\documentclass[journal]{IEEEtran}
\usepackage{caption}
\usepackage{graphicx}
\graphicspath{ {figures/} }  % folder containing graphics

\usepackage[font=small]{caption}
\usepackage[cmex10]{amsmath}
\usepackage{amsfonts} 
\usepackage{amssymb}
\usepackage{array}
\usepackage{mdwmath}
\usepackage{mdwtab}
\usepackage{eqparbox}
\usepackage{url}
\usepackage{comment}
\usepackage{amsthm}
\usepackage{algorithm}% http://ctan.org/pkg/algorithms
\usepackage{algpseudocode}% http://ctan.org/pkg/algorithmicx
\usepackage{graphicx}
\usepackage{float}
\usepackage{subfigure}
\usepackage{xcolor}
 \usepackage{cite} 
\newtheorem{Remark}{Remark}

  {\proof}{\proofend}
\newtheorem{proposition}{Proposition}

\ifCLASSINFOpdf
\else
\fi
\begin{document}

\title{Anti-Malicious ISAC: How to Jointly Monitor and Disrupt Your Foes?\\ }


\author{Zonghan~Wang,
        Zahra~Mobini,~\IEEEmembership{Senior Member,~IEEE,}
        Hien~Quoc~Ngo,~\IEEEmembership{Fellow,~IEEE,} Hyundong~Shin,~\IEEEmembership{Fellow,~IEEE,} 
                and Michail~Matthaiou,~\IEEEmembership{Fellow,~IEEE}
        }


\markboth{}%
{Shell \MakeLowercase{\textit{et al.}}: Anti-Malicious ISAC:}



\maketitle
\begin{abstract}
Integrated sensing and communication (ISAC) systems are key enablers of future networks but raise significant security concerns. In this realm, the emergence of malicious ISAC systems has amplified the need for authorized parties to legitimately monitor suspicious communication links and protect legitimate targets from potential detection or exploitation by malicious foes. In this paper, we propose a new wireless proactive monitoring paradigm, where a legitimate monitor intercepts a suspicious communication link while performing cognitive jamming to enhance the monitoring success probability (MSP) and simultaneously safeguard the target. To this end, we derive closed-form expressions of the signal-to-interference-plus-noise-ratio (SINR) at the user (UE), sensing access points (S-APs), and an approximating expression of the  SINR at the proactive monitor.  Moreover, we propose an optimization technique under which the legitimate monitor minimizes the success detection probability (SDP) of the legitimate target, by optimizing the jamming power allocation over both communication and sensing channels subject to total power constraints and monitoring performance  requirement. To enhance the monitor's longevity and reduce the risk of detection by malicious ISAC systems, we further propose an adaptive power allocation scheme aimed at minimizing the total transmit power at the monitor while meeting a pre-selected sensing SINR threshold and ensuring successful monitoring. Our numerical results show that the proposed algorithm
significantly compromises the sensing and communication performance of malicious ISAC.

\let\thefootnote\relax\footnotetext{

This work was supported by the U.K. Engineering and Physical Sciences Research Council (EPSRC) grant (EP/X04047X/2) for TITAN Telecoms Hub. The work of H.~Q.~Ngo was supported by the U.K. Research and Innovation Future Leaders Fellowships under Grant MR/X010635/1. The work of H. Q. Ngo and M. Matthaiou was also supported by a research grant from
the Department for the Economy Northern Ireland under the US-Ireland R$\&$D
Partnership Programme. The work of  H. Shin was supported by the National Research Foundation of Korea (NRF) grant funded by the Korean government (MSIT) (RS-2025-00556064 and RS-2025-25442355) and by the MSIT (Ministry of Science and ICT), Korea, under the ITRC (Information Technology Research Center) support program (IITP-2025-RS-2021-II212046) supervised by the IITP (Institute for Information \& Communications Technology Planning \& Evaluation). The work of M. Matthaiou was supported by the European Research Council (ERC) under the European Union’s Horizon 2020 research and innovation programme (grant agreement No. 101001331). 
\textit{(Corresponding authors: H. Shin and M. Matthaiou).}

Z. Wang,  H. Q. Ngo, and M. Matthaiou are with the Centre for Wireless Innovation (CWI), Queen's University Belfast, BT3 9DT Belfast, U.K. (email:\{zwang95,   hien.ngo, m.matthaiou\}@qub.ac.uk). M. Matthaiou is also with the Department of Electronic Engineering, Kyung Hee University, Yongin-si, Gyeonggi-do 17104, Republic of Korea.

Z.~Mobini is with the Department of Electrical and Electronic Engineering, The University of Manchester, Manchester M13 9PL, U.K., and was also with the Centre for Wireless Innovation (CWI), Queen’s University Belfast, BT3 9DT Belfast, U.K.  (e-mail: zahra.mobini@manchester.ac.uk). 

H.~Shin is with the Department of Electronics and Information Convergence Engineering, Kyung Hee University, Yongin-si, Gyeonggi-do
17104, Republic of Korea (e-mail: hshin@khu.ac.kr).

%This work was supported by the U.K. Engineering and Physical Sciences Research Council (EPSRC) (grants No. EP/X04047X/1 and EP/X040569/1). The work of Z. Mobini and H. Q. Ngo was supported by the U.K. Research and Innovation Future Leaders Fellowships under Grant MR/X010635/1. The work of M. Matthaiou was supported by the European Research Council (ERC) under the European Union’s Horizon 2020 research and innovation programme (grant agreement No. 101001331). 

Parts of this paper were presented at the 2024 IEEE GLOBECOM~\cite{ref:Wang2024AntiMaliciousIU}.}
\end{abstract}


% \fi
\IEEEpeerreviewmaketitle

\vspace{-1em}
\section{Introduction}
The fusion of sensing functions into communication systems is expected to be a key component of future networks~\cite{ref:Fan_JSAC,ref:Fan_coexist_to_joint}. By leveraging infrastructure and resources for both communication and sensing in a cooperative manner, integrated sensing and communication (ISAC) systems aim to efficiently enhance the performance of both counterparts. While this emerging concept has garnered increasing research interest, prior works on ISAC have primarily focused on signal processing and waveform design aspects within single-cell (cellular) networks.

Meanwhile, for large-scale networks, a shared infrastructure coordination system has been developed via cell-free massive multiple-input multiple-output (CF-mMIMO), where numerous access points (APs), each equipped with multiple antennas, are distributed across a given area and operate coherently to serve user equipment (UEs) \cite{ref:Hien_small_cell} \cite{ref:M_NetworkAssisted_cf}. In particular, precoded signals and channel state information (CSI) are shared among the APs via a backhaul and central processing unit (CPU) \cite{ref:Hien_total_energy}. This distributed system has also been extensively studied in the radar domain, where multiple transmitters and receivers are used to enhance performance in target detection and localization. Building on these two concepts, researchers have expanded ISAC to cell-free systems~\cite{ref:Zinat,ref:Zinat_WCL,ref:ISAC_mode_selection,ref:CaoYu_CL,ref:Secure_max_sens_SINR}, in which the APs can collaboratively serve UEs while simultaneously transmitting probing signals and directing beams toward targets for sensing purposes. A CF-mMIMO ISAC system can support both communication and sensing tasks simultaneously, offering significant improvements in performance, diversity gain, and adaptability compared to traditional cellular ISAC. In~\cite{ref:Zinat}, the authors proposed a CF-mMIMO ISAC system in which each AP serves UEs in the downlink while steering beams toward targets for detection. The concept of sensing spectral efficiency (SE) in a CF-mMIMO ISAC system was introduced as a key performance metric. By optimizing the precoding vectors and power allocation, the system can maximize the sensing signal-to-interference-plus-noise ratio (SINR) to achieve a high detection probability while meeting minimum communication requirements. Furthermore, the integration of ISAC into CF-mMIMO networks for multiple-target detection has been explored in~\cite{ref:ISAC_mode_selection}, where each AP can adaptively function as either a radar or communication AP. A dynamic AP operation mode selection strategy and optimal power allocation were proposed to maximize the minimum SE of the UEs and meet sensing requirements within multiple sensing zones.

In parallel, wireless security has gained significant attention from both academia and industry, leading to the adoption of various approaches aimed at enhancing the security of wireless systems~\cite{ref:Fred_Rusek_Magz}.   {One of the widely adopted
approaches is based on the idea of artificial noise (AN) injection. With the assistance of some advanced AN elimination techniques, AN can be widely adopted to disturb suspicious links while avoiding the interception on the legitimate communication link \cite{refRevised:ANelim_perspEav,refRevised:ANelim_noCSI,refRevised:A_Survey_on_AN}.} In particular, \cite{refRevised:AN_GSM} studied AN-aided generalized spatial modulation systems, and \cite{refRevised:SIM_SISO} introduced a stacked intelligent metasurface-aided communication system for enhanced secure transmission. In addition, there have been many studies focusing on the physical-layer security (PLS) of ISAC systems. 
In~\cite{ref4}, the authors studied the  PLS  of MIMO communication and radar systems in a single-user and target scenario, where artificial jamming was introduced to achieve secure transmission by either maximizing the secrecy rate of users or ensuring a specific target-detection criterion. A dual-functional single  AP  was proposed in~\cite{ref:SuNanchi}, which identifies the locations of eavesdroppers with the assistance of cooperative users' information. A weighted optimization problem was introduced to minimize the Cramér-Rao bound of eavesdroppers, subject to beamwidth constraints and transmit power limits. This optimal power control was used to demonstrate the achievable rate at the  UE  versus the eavesdropper's pilot power, while evaluating the information leakage to an active eavesdropper. The work in~\cite{ref:Yassen} studied the secrecy performance in a CF-mMIMO system under active eavesdropping attacks, introducing protective partial zero-forcing precoding and an AP selection method to enhance the secrecy SE, meet SINR requirements for legitimate users, and incorporate an eavesdropper detection method in the network. Later, \cite{ref:Secure_max_sens_SINR} considered a secure CF-mMIMO ISAC system, focusing on the trade-off between the secrecy rate of the UEs and the sensing SINR of the target. In~\cite{ref:sensing_eavesdropper}, the authors explored a scenario involving both communication data eavesdroppers and sensing data eavesdroppers. They proposed a power allocation scheme at the APs to meet secure transmission requirements while adaptively preserving sensing privacy.

Despite their immense potential, ISAC wireless systems also entail significant risks, as adversaries could exploit the technology for malicious or illegal purposes~\cite{ref:XuJie_surveillance}. For instance, attackers might use the sensing functionality to track legitimate targets and share this information among malicious UEs~\cite{ref:robustSecure}. While previous studies have examined scenarios in which reconfigurable intelligent surfaces (RISs) are employed to impair communication performance~\cite{refRevised:RIS_passiveBFAttack},~\cite{refRevised:maliciousRIS_Emil}, none have specifically investigated the misuse of ISAC systems for illicit activities. 
Therefore, it is crucial for authorized parties to develop security measures to counteract the threats posed by malicious ISAC systems. To address this, we draw inspiration from proactive eavesdropping security methods in the field of PLS~\cite{ref:Xu_Jie_TWC}, where a monitor intentionally sends jamming signals to degrade the malicious communication rate, thereby enhancing monitoring efficiency. Unlike traditional wireless PLS, which focuses on preventing information leakage to illegal eavesdroppers~\cite{Mobini:TIFS:2019}, our proposed approach deploys a legitimate monitor as part of the system, playing a critical role in ensuring public safety. 

Proactive monitors, also known as surveillance or proactive eavesdropping systems, enable authorized parties, such as government agencies, to legally monitor and disrupt suspicious communication links~\cite{ref:XuJie_surveillance},~\cite{ref:Xu_Jie_TWC},~\cite{ref:zahra_iot}. The rise of ISAC systems introduces new threats: unlike conventional communication-only systems, a malicious ISAC system can simultaneously serve malicious UEs and sense legitimate targets using radar capabilities, increasing the risk of sensitive information leakage. Existing works leave several gaps that motivate our study. First, prior studies focused on communication interception only~\cite{ref:XuJie_surveillance,ref:Xu_Jie_TWC,ref:battery_aided_surveillance,ref:EE_eavesdropper,ref:monitor_energy_efficiency,ref:zahra_iot}, neglecting the sensing capabilities of ISAC systems. Second, \cite{ref:XuJie_surveillance,ref:Xu_Jie_TWC,ref:battery_aided_surveillance} considered a single AP and/or a single suspicious link, oversimplifying real networks where multiple APs serve multiple UEs in cell-free architectures. Third, uplink training and channel estimation are often ignored, even though malicious ISAC systems use pilot-based channel estimation to precode signals, limiting passive monitoring opportunity. Against this background, we hereafter enable the proactive monitor to perform pilot spoofing during uplink training, enhancing its received SINR and ensuring reliable decoding of the information transmitted to the malicious UE.

To the best of our knowledge, no research has focused on jointly monitoring and disrupting malicious CF-mMIMO ISAC systems.

Motivated by the above discussion, for the first time, we consider a dual-functional proactive monitor within a malicious CF-mMIMO ISAC system. The monitor aims at protecting the legitimate target from being detected by the malicious system, while monitoring the malicious UE.\footnote{ While proactive monitoring can safeguard legitimate UEs and/or targets, improper use may constitute an invasion of privacy. Therefore, future deployments must comply with established frameworks. In this context, the U.S. National Security Agency (NSA) launched the Terrorist Surveillance Program in 2001 to legitimately monitor wireless devices for public safety~\cite{ref:XuJie_surveillance}, while 3GPP Technical Specification TR~33.854 defines requirements for UTM operators to legally monitor autonomous airbone vehicle (UAV) operations, including scenarios where unauthorized entities may spoof, track, or access sensitive flight information~\cite{refRevised:LowAltitudeEconomy}.}  The power allocated at the monitor to the target and to the malicious UE is carefully designed based on the various scenarios and the objectives to achieve effective jamming. Our specific contributions are the following:

\begin{itemize}
    \item We propose an anti-malicious CF-mMIMO ISAC design that utilizes proactive monitoring. The malicious ISAC system comprises multiple communication APs (C-APs) serving multiple UEs, with one UE suspected of engaging in illegal activities, and multiple sensing APs (S-APs) attempting to illicitly sense a legitimate target. In  our anti-malicious  design, the monitor has dual functionalities: it intercepts the transmissions of the suspicious UE and emits a jamming signal to disrupt the communication links between the APs and the suspicious UE. Concurrently, the monitor generates a precoded jamming signal directed at the legitimate target, thereby reducing the probability of successful sensing by the malicious ISAC system.
    \item We provide a detailed theoretic performance analysis of the proposed anti-malicious CF-mMIMO ISAC system and derive closed-form expressions for the SINR  at the  UEs and S-APs, and a closed-form approximation of the SINR at the proactive monitor. These closed-form expressions facilitate the subsequent system optimization and can shed useful insights into the system performance.
    \item We formulate two optimization problems, ($\mathbf{P}_1$) and ($\mathbf{P}_2$), with two different objectives: ($\mathbf{P}_1$) the malicious ISAC success detection probability (SDP) minimization and ($\mathbf{P}_2$) the proactive monitor's power consumption minimization. The formulated problems are under total power constraints and a successful monitoring requirement. Note that a common assumption in proactive monitoring literature is that the monitor is continuously powered by conventional energy sources. However, this assumption may be impractical, as power may only be supplied by batteries with limited capacity. A lack of jamming energy could result in the failure to intercept suspicious links, ultimately limiting the monitor's performance~\cite{ref:EE_eavesdropper}. In addition, in practical surveillance systems, the monitor is often located near the malicious UEs. Frequent replacement of the monitor's energy source increases the risk of exposure to these UEs. Therefore, efficient energy utilization is crucial for power-constrained monitors, and it is essential to extend their operational lifetime by minimizing the monitor's power consumption~\cite{ref:battery_aided_surveillance}, \cite{ref:monitor_energy_efficiency}.     
  
    \item Numerical results show that our proactive monitoring  effectively reduces the SDP of the malicious CF-mMIMO ISAC system while providing successful monitoring performance. Compared to equal power allocation (EPA) scenarios,  the optimization approach ($\mathbf{P}_1$) yields a significant decrease in the SDP. The simulation results also confirm that the optimization approach ($\mathbf{P}_2$) achieves a notable jamming power saving of $43.6\%$, while meeting both the successful monitoring and SDP requirements.
\end{itemize}

\emph{Notation}: We use lower and upper case letters to denote vectors and matrices. The superscripts $^{\mathtt{c}}$ and $^{\mathtt{s}}$ indicate communication and sensing functionalities. Italic footers indicate device indices. Fixed system entities are given in typewriter font footers. The superscripts $(\cdot)^{H}$, $(\cdot)^{\ast}$ and $(\cdot)^{T}$ stand for the Hermitian, conjugate and transpose operators; $||\cdot||$ denotes the Euclidean norm; $\mathbf{I}_N$ stands for the $N \times N$ identity matrix. A  circular symmetric, complex Gaussian distribution with variance $\sigma^2$ is denoted by $\mathcal{CN} (0,\sigma^2)$. Moreover, $\mathbb{E}\left \{ \cdot \right \}$ denotes the  expectation, while $\mathrm{Tr}(\cdot)$ denotes the trace of a matrix. 

%%%%%%%%%%%%%%%%%%%%% System Model %%%%%%%%%%%%%%%%%%%%%%%%%%%%%%%%%%
\section{System Model}
\begin{figure}[t]
  \centering
\includegraphics[width=0.85\linewidth,height=0.55\linewidth]{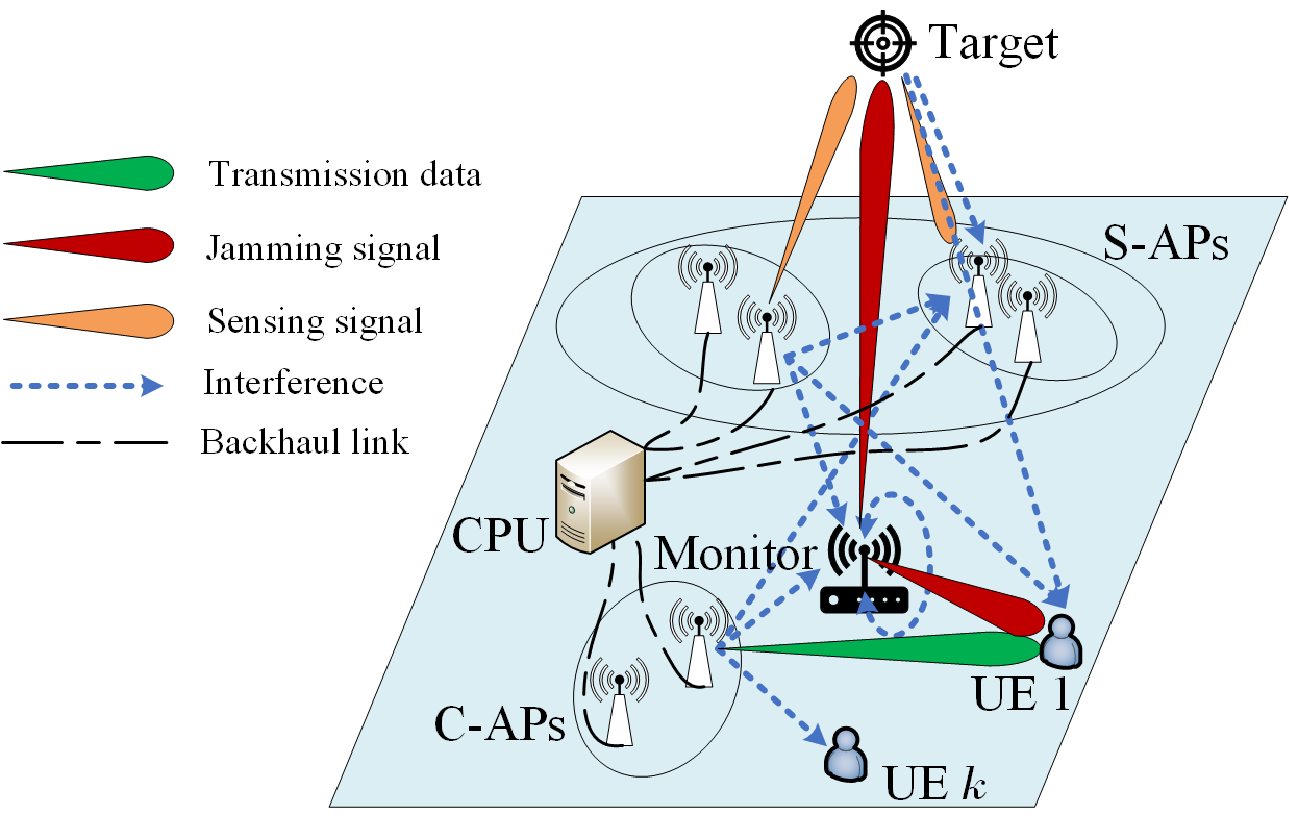}
  \caption{Anti-malicious CF-mMIMO ISAC design using a FD proactive monitor.}
  \label{fig:system model}
\end{figure}

\begin{table}[t]
\caption{List of notations.}\label{Table}
    \centering
    \begin{tabular}{|l|l|}
    \hline
        Parameter & Definition  \\ \hline
        $\mathbf{g}_{m,k}$ & Channel between the $m$-th C-AP and the $k$-th UE \\ \hline
        $\mathbf{g}_{\mathtt{pm},k}$ & Channel between the proactive monitor and the $k$-th UE \\ \hline
        $\mathbf{g}_{m',k}$ &  Channel between the $m'$-th S-AP and the $k$-th UE \\ \hline
        $h_{\mathtt{t},k}$ & LoS channel between the target and the $k$-th UE \\ \hline
        $\mathbf{h}_{t,m''}$ & LoS channel between the $m''$-th S-AP and target \\ \hline
        $\mathbf{h}_{\mathtt{pm},\mathtt{t}}$  & LoS channel between the proactive monitor and the target \\ \hline
        $\mathbf{h}_{m\!,\mathtt{t}}$  & LoS channel between the $m$-th C-AP and target \\ \hline
        $\mathbf{h}_{m'\!,\mathtt{t}}$  & LoS channel between the $m'$-th S-AP and target \\ \hline
        $\mathbf{w}_{\mathtt{pm},\mathtt{t}}^{\mathtt{s}} $ & MR precoding from the proactive monitor to target \\ \hline
        $\mathbf{w}_{\mathtt{pm},1}^{\mathtt{c}} $ & MR precoding from the proactive monitor to $1$-st UE \\ \hline
        $\mathbf{w}_{m'\!,\mathtt{t}} ^{\mathtt{s}}$ & MR precoding from the $m'$-th S-AP to target \\ \hline
        $\mathbf{w}_{m,k} ^{\mathtt{c}}$ &  MR precoding from the $m$-th C-AP to $k$-th UE \\ \hline
        $ \mathbf{w}_{\mathtt{comb},\mathtt{pm}}$ & Combining vector at the proactive monitor \\ \hline
        $ \mathbf{w}_{\mathtt{comb},m''}$ & Combining vector at the $m''$-th S-AP \\ \hline
        $\mathbf{G}_{\mathtt{pm},\mathtt{pm}}$ & Channel between Tx and Rx at the proactive monitor \\ \hline
        $\mathbf{G}_{m,\mathtt{pm}}$ & Channel between the $m$-the C-AP and proactive monitor \\ \hline
        $\mathbf{G}_{\mathtt{pm},m''}$ & Channel between the proactive monitor and $m''$-th AP \\ \hline
        $\mathbf{G}_{m,m''}$ & Channel between the $m$-th C-AP and $m''$-th S-AP \\ \hline
        $\mathbf{G}_{m',m''}$ & Channel between the $m'$-th S-AP and $m''$-th S-AP \\ \hline
        $\phi_{k}$ & Uplink pilot sequence of  UE $k$ \\ \hline
        $\boldsymbol{\Phi}_{\mathtt{pm}}$ & Uplink pilot sent by the proactive monitor \\ \hline
        $\eta_{m,k}$ & Power control coefficient at C-AP $m$ for  UE $k$ \\ \hline
        $\eta_{m',\mathtt{t}}$ & Power control coefficient at S-AP $m'$ for the target \\ \hline
        $\eta_{\mathtt{pm},1}$ & Power control coefficient at the proactive monitor for UE 1 \\
         \hline
        $\eta_{\mathtt{pm},\mathtt{t}}$ & Power control coefficient at the proactive monitor for the\\
        &  target\\ \hline
        $N$ & Number of antennas at APs \\ \hline
        $N_{\mathtt{pm}}$ & Number of antennas at the proactive monitor \\ \hline
        $\mathcal{M}_{\mathtt{c}}$ & Malicious C-AP set \\ \hline
        $\mathcal{M}_{\mathtt{s}}$ & Malicious S-AP set \\ \hline
        $\mathbf{N}_{\mathtt{p},m}$ & AWGN matrix at the $m$-th AP during the uplink training phase \\ \hline
        $\tilde{\mathbf{n}}_{\mathtt{p},m}$ & Effective noise vector at  AP $m$ after  \\
        & minimum-mean-square-error (MMSE) estimation\\ \hline
    \end{tabular}
    \vspace{-0.6em}
\end{table}
Let us consider a malicious CF-mMIMO ISAC system  operating under time division duplex (TDD) mode that involves $M$  APs and $K$  UEs, as shown in Fig.
\ref{fig:system model}. All UEs are untrusted. The APs are divided into two disjoint sets: i) C-AP set, denoted by $ \mathcal{M}_{\mathtt{c}}$, which is used to serve untrusted UEs and ii) S-AP set, $ \mathcal{M}_{\mathtt{s}}$, is used for detecting a legitimate target, where $\mathcal{M}_{\mathtt{c}} \cap \mathcal{M}_{\mathtt{s}}=\emptyset$. 
Furthermore, a multi-static sensing approach is considered, involving multiple transmit and receive S-APs subsets, denoted by $\mathcal{M}_{\mathtt{s},\mathtt{t}}$ and $\mathcal{M}_{\mathtt{s},\mathtt{r}}$, respectively, where $\mathcal{M}_{\mathtt{s}}=\mathcal{M}_{\mathtt{s},\mathtt{t}} \cup  \mathcal{M}_{\mathtt{s},\mathtt{r}}$ and $\mathcal{M}_{\mathtt{s},\mathtt{t}} \cap \mathcal{M}_{\mathtt{s},\mathtt{r}}=\emptyset$. Against this malicious CF-mMIMO ISAC system, we consider a full-duplex (FD) proactive monitor in the system, which is deployed to monitor and simultaneously send a jamming signal to interfere with the reception of a malicious UE  and S-APs in $\mathcal{M}_{\mathtt{s},\mathtt{r}}$. Without loss of generality, we assume that the proactive monitor aims to monitor UE $1$ among the $K$ malicious UEs. \ref{fig:system model}. Monitoring a specific UE (here UE $1$) is implemented in a specific snapshot in time and frequency band. Other UEs could be monitored in different time/frequency resources. To be more general, we assume an aerial legitimate target located in $3$ dimensional ($3$D) space with height $h$ m above the ground. The roles of nodes in the system are listed in Table \ref{table:roles of nodes}. Moreover,

\begin{itemize}
    \item We assume that each UE is equipped with a single antenna, while each AP is equipped with $N$ antennas, and the proactive monitor is equipped with $ N_{\mathtt{pm}}$ antennas.
The ground-to-ground channel between the $m$-th AP ($m \in\mathcal{M}_{\mathtt{c}}$) and the $k$-th UE is modeled as
\begin{align}
    \mathbf{g}_{m,k}=\beta_{m,k} ^{1/2}\mathbf {g}_{m,k}' ,
\end{align}
where $\mathbf{g}_{m,k}'\in \mathbb{C} ^{N\times 1}$ is the small-scale fading vector whose entries are independent and identically distributed (i.i.d.) $\mathcal{CN}(0,1)$. In addition, $\beta_{m,k}$ is the large-scale fading coefficient. The channel vectors $\mathbf{g}_{\mathtt{pm},k}$ and $\mathbf{g}_{m',k}$, can be defined similarly with appropriate modifications as shown in \text{Table}~\ref{Table}.
\item With regard to the channel between S-APs and  the target, it is reasonable to assume that the ground-to-air (air-to-ground) channels are line-of-sight (LoS) \cite{ref:Zinat}. In particular, the channel between the $m'$-th S-AP, $m'\in\mathcal{M}_{\mathtt{s},\mathtt{t}}$, and the target, $\mathbf{h}_{m'\!,\mathtt{t}}$, can be written as
\begin{equation}~\label{eq:gta_channel}
    \mathbf{h}_{m'\!,\mathtt{t}}=\sqrt{\zeta_{m'\!,\mathtt{t}} }\boldsymbol{\alpha } _{t}\left ( \phi_{m'\!,\mathtt{t}}^{a}, \phi_{m'\!,\mathtt{t}}^{e}\right ) ,
\end{equation}
%--------------
where $\zeta_{m'\!,\mathtt{t}} = (\frac{\lambda}{4\pi d_{m'\!,\mathtt{t}}} )^{L}$ is the free-space path loss,  ${L}$ is the path loss exponent, $\lambda$ is the wavelength and $d_{m'\!,\mathtt{t}}$ is the distance between the $m'$-th AP $(x_{m'},y_{m'},0)$ and the target $(x_{t},y_{t},h)$ in a 3D Euclidean space, which can be given by $d_{m'\!,\mathtt{t}}=\sqrt{(x_{m'}-x_{t})^2+(y_{m'}-y_{t})^2+h^2} $.
Moreover, $\boldsymbol{\alpha } _{t}\left ( \phi_{m'\!,\mathtt{t}}^{a}, \phi_{m'\!,\mathtt{t}}^{e}\right )$ is the steering vector, where $ \phi_{m'\!,\mathtt{t}}^{a}$ and $ \phi_{m'\!,\mathtt{t}}^{e}$ denote the azimuth and elevation angle of  departure (AoD) from the $m'$-th AP to the target, respectively \cite{ref:Emil_book}. The same steps can be followed to model the channel between the target and the $m''$-th S-AP, $m''\in\mathcal{M}_{\mathtt{s},\mathtt{r}}$, denoted by $\mathbf{h}_{t,m''}$.

\item  The ground-to-air channels between the target and the monitor, $\mathbf{h}_{\mathtt{pm},\mathtt{t}}\in \mathbb{C} ^{ N_{\mathtt{pm}} \times 1}$. In our model, the legitimate target and the proactive monitor cooperate and are located at fixed positions, thus the proactive monitor has prior knowledge of the legitimate target’s information, i.e. true location and radar cross-section (RCS). Under these assumptions,  the channel between the monitor and the target can be accurately modeled as a LoS link \cite{refRevised:Kaitao_NetworkISAC}.
The air-to-ground channel between the target and UE $k$, $h_{\mathtt{t},k}$, can be modeled using~\eqref{eq:gta_channel} with proper changes.
\end{itemize}

\begin{Remark} Although the classification of malicious ISAC systems is not the primary focus of this work, recognizing potential malicious behavior is crucial for the design of the proactive monitor. Feasible approaches include: (i) Sensitive content analysis: detecting unauthorized or confidential transmissions via physical-layer decoding and advanced data analysis techniques, such as text mining or multimedia analysis~\cite{ref:XuJie_surveillance, refRevised:MinChen_media}; (ii) Abnormal UE behavior: identifying UEs with atypical communication patterns using mobility profiling, social network analysis, or anomaly detection~\cite{ref:XuJie_surveillance, refRevised:OutlyingSequenceDetection}; and (iii) Feedback from legitimate UEs: monitoring communication quality and service disruptions to infer malicious activity and guide adaptive jamming.
 \end{Remark}

\begin{Remark} The proactive monitor does not serve untrusted UEs; it monitors UEs associated with a malicious ISAC system while protecting the legitimate target from potential detection. This occurs when a system appears legitimate but is considered malicious by the defender. Examples include military operations where enemy ISAC systems serve their users while sensing legitimate targets, law enforcement monitoring criminal networks, enterprise cyber-security where rogue or insider devices exfiltrate confidential data, and public safety scenarios involving unauthorized drones or Internet-of-Things (IoT) devices threatening privacy or infrastructure.
\end{Remark}

 {\begin{Remark}   The proactive monitor steers its jamming beam at the legitimate target rather than at the malicious S-APs. Since the target cooperates with the monitor, its location and CSI are known accurately, enabling precise LoS beamforming. The resulting interference corrupts the target echo received by the malicious S-APs, thereby degrading the sensing performance of the adversarial ISAC system. By contrast, the S-APs are non-cooperative: their positions and channels are unknown to the monitor, so forming effective jamming beams toward them is infeasible — particularly given the monitor’s limited spatial degrees of freedom.
\end{Remark}}

\begin{table}[t]
\caption{ Roles of nodes.\label{table:roles of nodes}}
    \centering
    \begin{tabular}{|l|l|}
    \hline
        Node & Role  \\ \hline
        UE & Malicious user in the system\\ \hline
        Target & Aerial legitimate target \\ \hline
        C-AP & Communication AP which serves untrusted UEs \\ \hline
        S-AP &  1) Transmit S-AP: sending probing signal to the \\
        & legitimate target\\
        &2) Receive S-AP: receiving reflected signal from \\
        &the legitimate target\\ \hline
        Proactive monitor & Jointly monitor untrusted UEs and disrupt   sensing  \\
        &the legitimate target\\ \hline
        CPU & All malicious APs cooperate through the CPU, \\
        &which processes echoes to detect the legitimate \\
        &target and coordinates the service of malicious UEs  \\ \hline
    \end{tabular}
    \vspace{0em}
\end{table}
\subsection{Uplink Training}
In  the uplink training phase of the malicious ISAC system, UE $k$ sends the pilot sequence $\boldsymbol{\varphi }_{k} \in \mathbb{C} ^{ \tau_{\mathtt{p}}\times 1}$ to the APs for the channel estimation, where $\tau_{\mathtt{p}}$ denotes the length of the uplink training phase.  The proactive monitor launches a pilot spoofing attack, i.e., sends the same pilot as  UE $1$ to the APs,  to enhance its overhearing performance. 
Following \cite{ref:HoangTiepM}, we define a matrix $\boldsymbol{\Phi}_{\mathtt{pm}} =
[ \boldsymbol{\varphi }_{1},  \boldsymbol{\varphi }_{1}, ...  \boldsymbol{\varphi }_{1}]^{H}\in \mathbb{C} ^{N_{\mathtt{pm}} \times\tau_{\mathtt{p}}}$ as the pilot sent by the monitor, where $\boldsymbol{\varphi }_{1} \in \mathbb{C} ^{ \tau_{\mathtt{p}}\times 1}$. The received pilot signal at the $m$-th AP can be written as
\begin{equation}
    \mathbf {Y}_{\mathtt{p},m} =  \sqrt{\tau_{\mathtt{p}}\rho_{\mathtt{p}}} \sum_{k=1}^{K} \mathbf{g}_{m,k}\boldsymbol{\varphi }_{k} ^{H} +\sqrt{\tau_{\mathtt{p}}\rho_{\mathtt{p},\mathtt{pm}}}  \mathbf{G}_{m,\mathtt{pm}}\boldsymbol{\Phi}_{\mathtt{pm}}+\mathbf{N}_{\mathtt{p},m},
    \label{Y_p,m}
\end{equation}
where $\rho_{\mathtt{p}}$ and $\rho_{\mathtt{p},\mathtt{pm}}$  are the transmit signal-to-noise ratios (SNRs) for pilot transmission at the UEs and monitor, respectively. Under the assumption of orthogonal pilot
sequences, we obtain $\check{\mathbf{y}}_{\mathtt{p},m}=\mathbf{Y}_{\mathtt{p},m} \boldsymbol{\varphi }_{k} $ as
\begin{equation}
     \check{\mathbf{y}}_{\mathtt{p},m}=\sqrt{\tau_{\mathtt{p}}\rho_{\mathtt{p}}} \mathbf{g}_{m,k}+\sqrt{\tau_{\mathtt{p}}\rho_{\mathtt{p},\mathtt{pm}}} \mathbf{G}_{m,\mathtt{pm}}\mathbf{u}_{N_{\mathtt{pm}}}+\tilde{\mathbf{n}}_{\mathtt{p},m},
\end{equation}
where $\mathbf{u}_{N_{\mathtt{pm}}}\in \mathbb{C} ^{N_{\mathtt{pm}} \times 1}$ is an all-one vector.
Then, given $\check{\mathbf{y}}_{\mathtt{p},m}$, the  MMSE  estimate of ${\mathbf{g} }_{m,k}$  is
\begin{equation}
\begin{split}
        \hat{\mathbf{g} }_{m,k}&=\mathbb{E}\left \{ \mathbf{g}_{m,k} \check{\mathbf{y}}_{\mathtt{p},m}^{H}\right \} (\mathbb{E}\left \{ \check{\mathbf{y}}_{\mathtt{p},m} \check{\mathbf{y}}_{\mathtt{p},m}^{H}\right \})^{-1}\check{\mathbf{y}}_{\mathtt{p},m}\\
        &=\frac{ \sqrt{\tau_{\mathtt{p}}\rho_{\mathtt{p}}} \beta _{m,k}}{\tau_{\mathtt{p}}\rho_{\mathtt{p}} \beta _{m,k}+\tau_{\mathtt{p}}\rho_{\mathtt{p},\mathtt{pm}} \beta _{m,\mathtt{pm}}N_{\mathtt{pm}}+1} \check{\mathbf{y}}_{\mathtt{p},m}.
        \label{eq:g_est_mk}
\end{split}
\end{equation}
From \eqref{eq:g_est_mk}, we can see that $\hat{\mathbf{g} }_{m,k}\sim \mathcal{CN} (\mathbf{0},\gamma_{m,k}\mathbf{I}_N)$, where 
\begin{align}\label{gamma_mk}
        \gamma_{m,k}
        =\left\{\begin{matrix}
\frac{ \tau_{\mathtt{p}}\rho_{\mathtt{p}} \beta _{m,1}^{2}}{\tau_{\mathtt{p}}\rho_{\mathtt{p}} \beta _{m,1}+\tau_{\mathtt{p}}\rho_{\mathtt{p},\mathtt{pm}} \beta _{m,\mathtt{pm}}N_{\mathtt{pm}}+1}, & k=1, \\
\frac{ \tau_{\mathtt{p}}\rho_{\mathtt{p}} \beta_{m,k}^{2}}{\tau_{\mathtt{p}}\rho_{\mathtt{p}} \beta_{m,k}+1}, &k\ne 1.
\end{matrix}\right.
\end{align}
%=====
\subsection{Downlink Transmission}
Let $s_{k}$ be the symbol intended for UE $k$ with $\mathbb{E}\left \{|s_{k}|^2\right \}=1$. Then, the signal transmitted by the $m$-th C-AP becomes
\begin{align}\label{eq:bold_x_m}
\mathbf {x}_{m}=\sum\nolimits_{k=1}^{K}\sqrt{\eta _{m,k}\rho _{\mathtt{c}}} \mathbf {w} _{m,k}^{\mathtt{c}}s_{k},
\end{align}
where $\mathbf {w} _{m,k}^{\mathtt{c}}\in \mathbb{C} ^{N\times 1}$ is the precoding vector generated by the $m$-th C-AP to UE $k$, $\eta_{m,k}$ is the power control coefficient chosen to satisfy the power constraint at the C-APs, and  $\rho_c$ is the normalized downlink SNR. The probing signal sent by the $m'$-th S-AP to the target is given by
\begin{align}\label{eq:bold_x_m't}
    \mathbf {x}_{m',\mathtt{t}}= \sqrt{\eta _{m'\!,\mathtt{t}}\rho _{\mathtt{s}}} \mathbf{w}_{m'\!,\mathtt{t}}^{\mathtt{s}}s_{\mathtt{t}},
\end{align}
where $\mathbf{w}_{m'\!,\mathtt{t}}^{\mathtt{s}}\in \mathbb{C} ^{N\times 1}$ denotes the precoding vector for sensing. Moreover, $\eta_{m'\!,\mathtt{t}}$ and  $\rho_s$ are the power control coefficient and normalized downlink SNR of the sensing signal, respectively. Additionally, $s_{\mathtt{t}}$ is the radar sensing symbol for the target with $\mathbb{E}\left \{|s_{\mathtt{t}}|^2\right \}=1$.

In the downlink communication phase, the monitor sends jamming signals  to disrupt the sensing performance of the malicious ISAC system and to interfere with the transmission links to  UE $1$ using a conjugate beamforming approach.\footnote{In our analysis, we assume perfect CSI of all suspicious links at the proactive monitor to study fundamental performance limits. In practice, the monitor can estimate these channels by overhearing pilot signals transmitted by the UEs~\cite{refRevised:Isabella_TWC}. Under imperfect or partial CSI, estimation errors can be modeled as bounded uncertainties within a given error radius~\cite{refRevised:Imperfect_CSI_errorbound}, which may degrade the effectiveness of jamming toward both UE $1$ and the target, potentially reducing the suppression of the sensing SINR.}  
The transmitted signal at the monitor is  
\begin{align}\label{eq:bold_x_pm}
    \mathbf {x}_{\mathtt{pm}}= \sqrt{\eta _{\mathtt{pm},\mathtt{t}}\rho _{\mathtt{pm}}} \mathbf{w}_{\mathtt{pm},\mathtt{t}}^{\mathtt{s}}s_{\mathtt{pm},\mathtt{t}}+\sqrt{\eta _{\mathtt{pm},1}\rho _{\mathtt{pm}}} \mathbf{w}_{\mathtt{pm},1}^{\mathtt{c}}s_{\mathtt{pm},1},
\end{align}
where
$s_{\mathtt{pm},\mathtt{t}}$ and $s_{\mathtt{pm},1}$ denote the transmit  jamming signal to the target and  UE $1$, respectively. The precoding vectors constructed at the monitor  for the target and   UE $1$  can be given by $\mathbf{w}_{\mathtt{pm},\mathtt{t}} ^{\mathtt{s}}=  \mathbf{h}_{\mathtt{pm},\mathtt{t}}^{\ast }$ and $\mathbf{w}_{\mathtt{pm},1} ^{\mathtt{c}}=  \mathbf{g}_{\mathtt{pm},1} ^{\ast }$, respectively. Moreover, we consider the conjugate scheme for precoding the probing signal at the $m'$-th S-AP to the target, $m' \in \mathcal{M}_{\mathtt{s},\mathtt{t}}$, and at the $m$-th C-AP to UE $k$, $m \in \mathcal{M}_{\mathtt{c}}$, such that $\mathbf{w}_{m'\!,\mathtt{t}} ^{\mathtt{s}}=  \mathbf{h}_{m'\!,\mathtt{t}} ^{\ast}$ and $\mathbf{w}_{m,k} ^{\mathtt{c}}=  \hat{\mathbf{g}}_{m,k} ^{\ast }$, respectively.
% Considering that the APs transmit at full power, the power coefficient at C-APs and S-APs can be expressed by $ \eta_{m,k} = \frac{1}{{N\sum_{k=1}^{K} \gamma_{m,k}}}$ and $\eta _{m'\!,\mathtt{t}} = \frac{1}{N\zeta_{m'\!,\mathtt{t}}}$, respectively~\cite{ref:ISAC_mode_selection}. We first consider the equal power allocation scheme at the monitor, thus the power allocation coefficients can be expressed by $\eta _{\mathtt{pm},\mathtt{t}} =\frac{1}{2 N_{\mathtt{pm}}\zeta_{\mathtt{pm},\mathtt{t}}}$ and $\eta _{\mathtt{pm},1}=\frac{1}{2 N_{\mathtt{pm}}\beta_{\mathtt{pm},1}}$.

\subsubsection{Received SINR at UE $k$}\label{subsubsection:Received SINR at UE}
The received signal at the $k$-th UE can be represented as
\begin{align}
\label{eq:y_k_xmk}
    &y_{k}  =\sum\nolimits_{m \in\mathcal{M}_{\mathtt{c}} } \mathbf{g}_{m,k}^{T}\mathbf {x}_{m} +\sum\nolimits_{m'\in\mathcal{M}_{\mathtt{s},\mathtt{t}} } \mathfrak{h}_{m',k}\mathbf {x}_{m',\mathtt{t}} \notag\\
    &\qquad+\mathfrak{h}_{\mathtt{pm},k}\mathbf {x}_{\mathtt{pm}} +n_{k},
\end{align}
where $n_{k}$ represents the additive white Gaussian noise (AWGN) with $n_{k}\sim \mathcal{CN}(0,1) $. It is worth noting that $\mathfrak{h}_{m',k}$ represents the effective channel between the $m'$-th S-AP, $m' \in \mathcal{M}_{\mathtt{s},\mathtt{t}}$, and the $k$-th UE. This channel   includes both the direct link and the reflected channel through the target, and it can be modeled as \cite{refRevised:ISAC_AAV, refRevised:FullDuplexISAC}  
\begin{align}              
\mathfrak{h}_{m',k}&=\mathbf{g}_{m',k}^{T}+\sqrt{\frac{\lambda^2\sigma_\mathtt{RCS}}{(4\pi)^3 d_{m',\mathtt{t}}^2d_{\mathtt{t},k}^2}}\boldsymbol{\alpha }^{T} _{t}\left ( \phi_{m'\!,\mathtt{t}}^{a}, \phi_{m'\!,\mathtt{t}}^{e}\right ) \notag\\
&\!=\!\mathbf{g}_{m',k}^{T}\!\!+\!\!\sqrt{\alpha}\!\sqrt{\frac{\lambda^2}{(4\pi)^2 d^2_{\mathtt{t},k}}}\!\sqrt{\frac{\lambda^2}{(4\pi)^2 d^2_{m'\!,\mathtt{t}}}}\boldsymbol{\alpha }^{T} _{t}\left ( \phi_{m'\!,\mathtt{t}}^{a}, \phi_{m'\!,\mathtt{t}}^{e}\right )\notag\\
&=\mathbf{g}_{m',k}^{T}+\sqrt{\alpha}h_{\mathtt{t},k}\mathbf{h}_{m'\!,\mathtt{t}}^{T},
\end{align}
where $\alpha$ is the target reflection gain that depends on the transmission and reflection coefficient, center frequency and a nonfluctuating RCS of the target,  given by $\alpha = 4\pi \sigma_{\mathtt{RCS}} / \lambda^2$, where $\sigma_{\mathtt{RCS}}$ is the RCS of the target.
Moreover, the effective channel between the proactive monitor and the $k$-th UE can be defined using a similar modeling approach, i.e., 
\begin{align}
\mathfrak{h}_{\mathtt{pm},k}=\mathbf{g}_{\mathtt{pm},k}^{T}+\sqrt{\alpha}h_{\mathtt{t},k}\mathbf{h}_{\mathtt{pm},\mathtt{t}}^{T}. 
\end{align}
Submitting (\ref{eq:bold_x_m}), (\ref{eq:bold_x_m't}) and (\ref{eq:bold_x_pm}) into (\ref{eq:y_k_xmk}), we can obtain:
\begin{align}
       & y_{k} =\mathrm {DS}_{k}s_{k}+\mathrm {BU}_{k}s_{k}+\sum\nolimits_{{k}'\ne k}^{K}\mathrm{IU}_{k',k}s_{k'}+\mathrm{IS}_{k}s_{\mathtt{t}} \notag\\
       &\qquad+\mathrm {JS}_{\mathtt{s},k}s_{\mathtt{pm},\mathtt{t}}+\mathrm {JS}_{\mathtt{c},k}s_{\mathtt{pm},1}+\!n_{k},
    \label{yk_rewrite}
\end{align}
%------------
where $\mathrm {DS}_{k}$, $\mathrm{BU}_{k}$, $\mathrm{IU}_{k',k}$, $\mathrm{IS}_{k}$, $\mathrm {JS}_{\mathtt{s},k}$, $\mathrm {JS}_{\mathtt{c},k}$ and $n_{k}$ denote the desired signal, beamforming uncertainty, inter-UE interference, interference from the S-APs, jamming signal to target, jamming signal to UE $k$ and noise respectively, given by
%-------------------
\begin{align}
    \mathrm {DS}_{k}&\triangleq\mathbb{E}\left \{ \sum\nolimits_{m \in\mathcal{M}_{\mathtt{c}} } \sqrt{\eta _{m,k}\rho _{\mathtt{c}}} \mathbf{g}_{m,k}^{T}\mathbf {w} _{m,k}^{\mathtt{c}}\right \},\\
    %------------
    \mathrm {BU}_{k}&\triangleq \sqrt{\rho} _{\mathtt{c}}\Big(\sum\nolimits_{m \in\mathcal{M}_{\mathtt{c}} }\!\sqrt{\eta _{m,k}}{\mathbf {g}}_{m,k}^{T}\hat{\mathbf {g}}_{m,k} ^{\ast }\!\notag\\
    &\hspace{0.4cm}-\!\mathbb{E} \Big \{ \sum\nolimits_{m \in\mathcal{M}_{\mathtt{c}} }\sqrt{\eta _{m,k}}\mathbf {g}_{m,k}^{T} \hat{\mathbf {g}}_{m,k} ^{\ast } \Big \}\!\Big),\\
    %------------
    \mathrm{IU}_{k',k}&\triangleq\sum\nolimits_{m \in\mathcal{M}_{\mathtt{c}} }  \sqrt{\eta _{m,{k}'}\rho _{\mathtt{c}}}\mathbf{g}_{m,k}^{T}\mathbf {w} _{m,{k}'}^{\mathtt{c}},\\
    %------------
\mathrm{IS}_{k}&\triangleq\sum\nolimits_{m'\in\mathcal{M}_{\mathtt{s},\mathtt{t}} } \sqrt{\eta _{m'\!,\mathtt{t}}\rho _{\mathtt{s}}}\mathfrak{h}_{m',k}\mathbf{w}_{m'\!,\mathtt{t}}^{\mathtt{s}},\\
    \mathrm {JS}_{\mathtt{s},k}&\triangleq \sqrt{\eta _{\mathtt{pm},\mathtt{t}}\rho _{\mathtt{pm}}}\mathfrak{h}_{\mathtt{pm},k}\mathbf {w}_{\mathtt{pm},\mathtt{t}} ^{\mathtt{s}},\\
    %------------
    \mathrm {JS}_{\mathtt{c},k}&\triangleq \sqrt{\eta _{\mathtt{pm},k}\rho _{\mathtt{pm}}}\mathfrak{h}_{\mathtt{pm},k}\mathbf {w}_{\mathtt{pm},1} ^{\mathtt{c}}.
    \label{eq:DS,IU...}
\end{align}
%--------------------

\begin{proposition}\label{Theorem1}
The effective SINR of the $k$-th UE is given by (\ref{eq:SINR_UE}), shown at the top of the next page, where
%=========================
\begin{figure*}
\begin{equation}
\mathrm{SINR}_{k}=\frac{\left |  \mathrm {DS}_{k}  \right |^2 }{\mathbb{E}\left \{ |\mathrm{BU}_{k}|^{2} \right \}+ \sum\nolimits_{{k}'\ne 1}^{K}\mathbb{E}\left \{|\mathrm{IU}_{k',k}|^{2}\right \}+\mathbb{E}\left \{ |\mathrm{IS}_{k}|^{2} \right \}+\mathbb{E}\left \{ |\mathrm {JS}_{\mathtt{s},k}|^2 \right\}+\mathbb{E}\left \{ |\mathrm {JS}_{\mathtt{c},k}|^2 \right\}+1},
\label{eq:SINR_UE}
\end{equation}
\hrulefill
\vspace{-1.5em}
\end{figure*}
%=========================
%--------------------
%\begin{subequations}
\begin{align}\label{eq:each part SINR_k}
    \mathrm {DS}_{k}  & =  \sum\nolimits_{m \in\mathcal{M}_{\mathtt{c}} }\sqrt{\eta _{m,k}\rho _{\mathtt{c}}}N\gamma _{m,k},\\
    %-------------
    \mathbb{E}\left \{ |\mathrm{BU}_{k}|^{2} \right \} 
    & = \sum\nolimits_{m \in\mathcal{M}_{\mathtt{c}} }\rho _{\mathtt{c}}N\eta _{m,k}\gamma _{m,k}\beta_{m,k},\\
    %-----------------
    \mathbb{E}\left \{ |\mathrm{IU}_{k',k}|^{2} \right \}& =  \!\sum\nolimits_{m \in\mathcal{M}_{\mathtt{c}} }\eta _{m,k'}\rho _{\mathtt{c}}N\gamma_{m,k'}\beta_{m,k}, \\
    %-------------
    \mathbb{E}\left \{ |\mathrm{IS}_{k}|^{2} \right \}  & =\!\!\!\!\!\!\sum_{m'\in\mathcal{M}_{\mathtt{s},\mathtt{t}} }\!\!\sqrt{\eta _{m'\!,\mathtt{t}}}\rho _{\mathtt{s}}\zeta_{m'\!,\mathtt{t}}N\bigg(\!\alpha \sqrt{\eta _{m'\!,\mathtt{t}}}N\zeta_{m'\!,\mathtt{t}}\zeta_{\mathtt{t},k}\!\notag\\
    &\hspace{-3.3em}+\sqrt{\eta _{m'\!,\mathtt{t}}}\beta_{m',k}
    \!+\!\!\!\!\!\!\!\!\!\!\sum\limits_{\tilde{m}'\in\mathcal{M}_{\mathtt{s},\mathtt{t}}, \tilde{m}' \ne m'}\!\!\!\!\sqrt{\eta _{\tilde{m}'\!,t}}\zeta_{\tilde{m}',t}\zeta_{\mathtt{t},k}\alpha N\!\bigg), \\
    %-------------
    \mathbb{E}\left \{ |\mathrm {JS}_{\mathtt{s},k}|^2 \right\} &=\!\eta _{\mathtt{pm},\mathtt{t}}\rho _{\mathtt{pm}} \zeta_{\mathtt{pm},\mathtt{t}}N_{\mathtt{pm}}(\beta_{\mathtt{pm},k}\!\!+\!\!\alpha\zeta_{\mathtt{t},k}N_{\mathtt{pm}}\zeta_{\mathtt{pm},\mathtt{t}}), \\
    %-------------
     \mathbb{E}\left \{ |\mathrm {JS}_{\mathtt{c},k}|^2 \right\}&\!=\!
    \eta_{\mathtt{pm},\!1}\rho _{\mathtt{pm}}N_{\mathtt{pm}}\beta_{\mathtt{pm},1}\!(\!N_{\mathtt{pm}}\beta_{\mathtt{pm},1}\!+\!\!\beta_{\mathtt{pm}\!,1}\!\!+\!\alpha\zeta_{t\!,k}\zeta_{\mathtt{pm},\mathtt{t}}).
\end{align}    
%\end{subequations}
%--------------
Proof: See Appendix \ref{ProofTheorem1}.
\end{proposition}

It is observed that the numerator scales with the square of the total number of service antennas across all APs, which is due to the array gain provided by cell-free massive MIMO technology. This implies that a malicious system can enhance the SE by adding more service antennas. However, as shown in (14), the denominator scales with $N_{\mathtt{pm}}^2$. Therefore, if the monitoring system can deploy more antennas (at least with the same order of the total number of AP antennas),  it can limit the performance of the malicious system.

We now consider the asymptotic scenario where the number of antennas at the proactive monitor approaches infinity $N_{\mathtt{pm}} \to \infty$. For simplicity, we assume that the APs transmit at full power, i.e., the power coefficient at C-APs and S-APs can be expressed by $ \eta_{m,k} = \frac{1}{{N\sum_{k=1}^{K} \gamma_{m,k}}}$ and $\eta _{m'\!,\mathtt{t}} = \frac{1}{N\zeta_{m'\!,\mathtt{t}}}$, respectively~\cite{ref:ISAC_mode_selection},  and EPA scheme at the monitor, i.e., the power allocation coefficients can be expressed by $\eta _{\mathtt{pm},\mathtt{t}} =\frac{1}{2 N_{\mathtt{pm}}\zeta_{\mathtt{pm},\mathtt{t}}}$ and $\eta _{\mathtt{pm},1}=\frac{1}{2 N_{\mathtt{pm}}\beta_{\mathtt{pm},1}}$. We also assume that the transmit power at the proactive monitor scales as $\rho _{\mathtt{pm}}=\frac{P_{\mathtt{pm}}}{N _{\mathtt{pm}}}$, where $P_{\mathtt{pm}}$ is a fixed value. Under these conditions, as $N_{\mathtt{pm}} \to \infty$, the desired signal at UE $k$ remains independent of $N_{\mathtt{pm}}$, while the dominant interference terms in the denominator are the jamming components,  $\mathbb{E}\left \{ |\mathrm{JS}_{\mathtt{s},k}|^2 \right\}$ and $\mathbb{E}\left \{ |\mathrm{JS}_{\mathtt{c},k}|^2 \right\}$, both of which scale quadratically with $N_{\mathtt{pm}}$. Considering the transmit power scaling and power control coefficients, when $N_{\mathtt{pm}}$ grows infinity, we can obtain $\mathbb{E}\left \{ |\mathrm {JS}_{\mathtt{s},k}|^2 \right\}\xrightarrow{N_{\mathtt{pm}}\to\infty}\frac{1}{2}P_{\mathtt{pm}} \alpha\zeta_{\mathtt{t},k}\zeta_{\mathtt{pm},\mathtt{t}}$ and $\mathbb{E}\left \{ |\mathrm {JS}_{\mathtt{c},k}|^2 \right\}\xrightarrow{N_{\mathtt{pm}}\to\infty}\frac{1}{2}P_{\mathtt{pm}}\beta_{\mathtt{pm},1}$.
 Therefore, the $\mathrm{SINR}_{k}$ converges to a constant value. The result implies that even if the transmit power at the proactive monitor is scaled down by 
$\frac{P_{\mathtt{pm}}}{N _{\mathtt{pm}}}$, the monitor can still affect the performance of the malicious users by varying $P_{\mathtt{pm}}$. 

\vspace{-2em}
\subsubsection{Received SINR for UE $1$ at the Proactive Monitor}\label{subsubsection:Received signal at PM}
%==============
The received signal at the monitor  can be written as
%===========
\begin{align}\label{eq:yj_xmk}
        &\mathbf{y}_{\mathtt{pm}}= \sum\nolimits_{m \in\mathcal{M}_{\mathtt{c}} }\mathbf{G}_{m,\mathtt{pm}}^{T}\mathbf {x} _{m} + \sum\nolimits_{m'\in\mathcal{M}_{\mathtt{s},\mathtt{t}} } \boldsymbol{\Lambda}_{m',\mathtt{pm}}\mathbf{x}_{m'\!,\mathtt{t}}  \notag \\
        &~\qquad+\boldsymbol{\Lambda}_{\mathtt{pm},\mathtt{pm}}\mathbf {x}_{\mathtt{pm}}+\boldsymbol{n}_{\mathtt{pm}},
\end{align}
%-----
where $\boldsymbol{\Lambda}_{m',\mathtt{pm}}\in \mathbb{C} ^{ N_{\mathtt{pm}} \times N}$ is the effective channel between the $m'$-th S-AP and the proactive monitor. Note that $\boldsymbol{\Lambda}_{\mathtt{pm},\mathtt{pm}}\in \mathbb{C} ^{ N_{\mathtt{pm}} \times N_{\mathtt{pm}}}$ is the effective channel between the transmitter and receiver of the proactive monitor. Hence, $\boldsymbol{\Lambda}_{m',\mathtt{pm}}$ and $\boldsymbol{\Lambda}_{\mathtt{pm},\mathtt{pm}}$ can be formulated as
\begin{align}    \label{eq:Lambda_channel}\boldsymbol{\Lambda}_{m',\mathtt{pm}}&=\mathbf{G}_{m',\mathtt{pm}}^{T}+\sqrt{\alpha}\mathbf{h}_{t,\mathtt{pm}}\mathbf{h}_{m'\!,\mathtt{t}}^{T},
\end{align}
\begin{align}
\boldsymbol{\Lambda}_{\mathtt{pm},\mathtt{pm}}&=\mathbf{G}_{\mathtt{pm},\mathtt{pm}}^{T}+\sqrt{\alpha}\mathbf{h}_{t,\mathtt{pm}}\mathbf{h}_{\mathtt{pm},\mathtt{t}}^{T},
\end{align}
%===
where $\mathbf{G}_{\mathtt{pm},\mathtt{pm}} \in \mathbb{C} ^{ N_{\mathtt{pm}} \times N_{\mathtt{pm}}}$ is the self-interference channel between the transmit and receive antennas at the FD monitor, which can be modeled through the Rayleigh fading model, and whose entries are i.i.d. $\mathcal{CN} (0,\sigma_{\mathtt{SI}}^2)$~\cite{Mohammad:2023:survey}, while $\mathbf{G}_{m',\mathtt{pm}}\in \mathbb{C} ^{N \times N_{\mathtt{pm}}}$ denotes the Rayleigh channel between the $m'$-th S-AP and the monitor. Moreover, $\boldsymbol{n}_{\mathtt{pm}}\in \mathbb{C} ^{ N_{\mathtt{pm}} \times 1}$ denotes an AWGN vector whose entries are i.i.d. $\mathcal{CN}(0,1)$. The monitor uses the combining vector
\begin{align}\label{eq:w_comb,pm}
    \mathbf{w}_{\mathtt{comb},\mathtt{pm}} = \Big(\sum\nolimits_{m \in\mathcal{M}_{\mathtt{c}} }\sqrt{\eta _{m,1}\rho _{\mathtt{c}}}\mathbf{G}_{m,\mathtt{pm}}^{T}\mathbf {w} _{m,1}^{\mathtt{c}}\Big) ^{\ast }
\end{align}
to overhear the signal of UE $1$. By substituting (\ref{eq:bold_x_m})--(\ref{eq:bold_x_pm}) and (\ref{eq:w_comb,pm}) into (\ref{eq:yj_xmk}), the received signal at the monitor becomes:
\begin{align}
       &z_{\mathtt{pm}} = \mathrm {DS}_{\mathtt{pm}}x_{1}+\mathrm{BU}_{\mathtt{pm}}s_{1}+\sum\nolimits_{{k}'\ne 1}^{K}\mathrm{IU}_{k',\mathtt{pm}}s_{k'}+\mathrm{IS}_{\mathtt{pm}}s_{\mathtt{t}}\notag\\
        &\qquad+\mathrm {JS}_{\mathtt{s},\mathtt{pm}}s_{\mathtt{pm},\mathtt{t}}+\mathrm {JS}_{\mathtt{c},\mathtt{pm}}s_{\mathtt{pm},1}+\mathrm{n}_{\mathtt{pm}},
    \label{zk_rewrite}
\end{align}
where $\mathrm {DS}_{\mathtt{pm}}$, $\mathrm{BU}_{\mathtt{pm}}$, $\mathrm{IU}_{{k}',\mathtt{pm}}$, $\mathrm{IS}_{\mathtt{pm}}$, $\mathrm{JS}_{\mathtt{s},\mathtt{pm}}$, $\mathrm{JS}_{\mathtt{c},\mathtt{pm}}$ and $\mathrm{n}_{\mathtt{pm}}$ are the desired signal, beamforming uncertainty, interference from the C-APs, interference from the S-APs, self-interference caused by the jamming signal and noise, respectively, given by 
\begin{align}
    \mathrm {DS}_{\mathtt{pm}} &\triangleq \mathbb{E}\left \{\mathbf{w}_{\mathtt{comb},\mathtt{pm}}^{T}\sum\nolimits_{m \in\mathcal{M}_{\mathtt{c}} }  \sqrt{\eta _{m,1}\rho _{\mathtt{c}}}\mathbf{G}_{m,\mathtt{pm}}^{T}\mathbf {w} _{m,1}^{\mathtt{c}}\right \},\\
    \mathrm {BU}_{\mathtt{pm}} &\triangleq \mathbf{w}_{\mathtt{comb},\mathtt{pm}}^{T}\!\sum\nolimits_{m \in\mathcal{M}_{\mathtt{c}} } \!\sqrt{\eta _{m,1}\rho _{\mathtt{c}}}\mathbf{G}_{m,\mathtt{pm}}^{T}\mathbf {w} _{m,1}^{\mathtt{c}} \notag\\
    &\hspace{0.2cm}- \mathbb{E}\bigg \{ \!\mathbf{w}_{\mathtt{comb},\mathtt{pm}}^{T}\!\sum\nolimits_{m \in\mathcal{M}_{\mathtt{c}} } \!\sqrt{\eta _{m,1}\rho _{\mathtt{c}}}\mathbf{G}_{m,\mathtt{pm}}^{T}\mathbf {w} _{m,1}^{\mathtt{c}}\bigg \}, \\
    \mathrm {IU}_{k',\mathtt{pm}} &\triangleq \mathbf{w}_{\mathtt{comb},\mathtt{pm}}^{T}\sum\nolimits_{m \in\mathcal{M}_{\mathtt{c}} }  \sqrt{\eta _{m,k'}\rho _{\mathtt{c}}}\mathbf{G}_{m,\mathtt{pm}}^{T}\mathbf {w} _{m,k'}^{\mathtt{c}},\\
    \mathrm {IS}_{\mathtt{pm}}& \triangleq\mathbf{w}_{\mathtt{comb},\mathtt{pm}}^{T}\sum\nolimits_{m'\in\mathcal{M}_{\mathtt{s},\mathtt{t}} } \sqrt{\eta _{m'\!,\mathtt{t}}\rho _{\mathtt{s}}}\boldsymbol{\Lambda}_{m',\mathtt{pm}}\mathbf{w}_{m'\!,\mathtt{t}}^{\mathtt{s}},
\end{align}
\begin{align}
    \mathrm{JS}_{\mathtt{s},\mathtt{pm}}& \triangleq\mathbf{w}_{\mathtt{comb},\mathtt{pm}}^{T}\sqrt{\eta _{\mathtt{pm},\mathtt{t}}\rho _{\mathtt{pm}}}\boldsymbol{\Lambda}_{\mathtt{pm},\mathtt{pm}}\mathbf {w}_{\mathtt{pm},\mathtt{t}}^{\mathtt{s}}, \\
    \mathrm{JS}_{\mathtt{c},\mathtt{pm}} &\triangleq\mathbf{w}_{\mathtt{comb},\mathtt{pm}}^{T}\sqrt{\eta _{\mathtt{pm},1}\rho _{\mathtt{pm}}}\boldsymbol{\Lambda}_{\mathtt{pm},\mathtt{pm}}\mathbf {w}_{\mathtt{pm},1}^{\mathtt{c}},\\
    \mathrm{n}_{\mathtt{pm}}& \triangleq \mathbf{w}_{\mathtt{comb},\mathtt{pm}}^{T}\boldsymbol{n}_{\mathtt{pm}}.
    \label{eq:DS_pm,IU_pm...}
\end{align}

\begin{proposition}\label{Theorem2} The received SINR for  UE $1$ at
the monitor can be calculated as (\ref{eq:SINR_J}), shown on the top of the next page, where
\begin{figure*}
\begin{equation}
    \mathrm{SINR}_{\mathtt{pm}} = \frac{|\mathrm {DS}_{\mathtt{pm}}|^{2}}{\mathbb{E}\left \{ |\mathrm{BU}_{\mathtt{pm}}|^{2} \right \}+\sum\nolimits_{{k}'\ne 1}^{K}\mathbb{E}\left \{ |\mathrm{IU}_{{k}',\mathtt{pm}}|^{2} \right \}+\mathbb{E}\left \{ |\mathrm{IS}_{\mathtt{pm}}|^{2} \right \}+\mathbb{E}\left \{ |\mathrm{JS}_{\mathtt{s},\mathtt{pm}}|^{2} \right \}+\mathbb{E}\left \{ |\mathrm{JS}_{\mathtt{c},\mathtt{pm}}|^{2} \right \}+\mathbb{E}\left \{ |\mathrm{n}_{\mathtt{pm}}|^{2} \right \}}, 
    \label{eq:SINR_J}
\end{equation}
\hrulefill
\vspace{-0.5em}
\end{figure*}
%====
\begin{align}\label{eq:each_part_SINR_pm}
   \mathrm {DS}_{\mathtt{pm}} &=\sum\nolimits_{m \in\mathcal{M}_{\mathtt{c}} } \sqrt{\eta _{m,1}\rho _{\mathtt{c}}}N_{\mathtt{pm}}\beta_{m,\mathtt{pm}}N\gamma_{m,1},\\
    \mathbb{E}\left \{ |\mathrm{BU}_{\mathtt{pm}}|^{2} \right \}\!&\approx\bigg(\sum\nolimits_{m \in\mathcal{M}_{\mathtt{c}} }\!\eta _{m,1}\rho _{\mathtt{c}}\beta_{m,\mathtt{pm}}\gamma_{m,1}N\bigg)^2N_{\mathtt{pm}},\\
    %----------------------
    \!\mathbb{E}\!\left \{\! |\mathrm{IU}_{{k}',\mathtt{pm}}\!|^{2} \!\right \}\!&\approx\!\sum\nolimits_{m \in\mathcal{M}_{\mathtt{c}} }\!\!\eta _{m,{k'}}\rho _{\mathtt{c}}^2N_{\mathtt{pm}}\!N\gamma_{m,k'}\beta_{m,\mathtt{pm}} \bigg [ \eta _{m,1}\notag\\
    \times(\!N_{\mathtt{pm}}\!+\!N)&\beta_{m,\mathtt{pm}}\gamma_{m,1}\!+\!\sum\nolimits_{\tilde{m} \ne m,\tilde{m} \in\mathcal{M}_{\mathtt{c}}}\!\eta _{\tilde{m},1}\!N\!\gamma_{\tilde{m},1}\beta_{\tilde{m},\mathtt{pm}} \!\bigg ], \\
    %---------------------------
    \mathbb{E}\left \{ |\mathrm{IS}_{\mathtt{pm}}|^{2} \right \}\!&=\!\sum\nolimits_{m\in\mathcal{M}_{\mathtt{c}} }\!\sum\nolimits_{m'\in\mathcal{M}_{\mathtt{s},\mathtt{t}} }\!\!\sqrt{\eta _{m'\!,\mathtt{t}}} \eta _{m,1}\rho _{\mathtt{s}}\rho _{\mathtt{c}}\beta_{m,\mathtt{pm}}\notag\\
    \times\gamma _{m,1}\zeta_{m'\!,\mathtt{t}}&N_\mathtt{pm}N^2\bigg ( \sqrt{\eta _{m',\mathtt{t}}} \beta_{m',\mathtt{pm}}+\!\sqrt{\eta _{m'\!,\mathtt{t}}} N\zeta_{\mathtt{pm},\mathtt{t}}\zeta_{m'\!,\mathtt{t}}\alpha\notag\\
    +\!&\sum\nolimits_{\tilde{m}'\in\mathcal{M}_{\mathtt{s},\mathtt{t}}, \tilde{m}' \ne m' } \!\sqrt{\eta _{\tilde{m}'\!,t}}N\zeta_{\mathtt{pm},\mathtt{t}}\zeta_{\tilde{m}',t}\alpha \bigg ),\\
    %------------------------------
    \!\mathbb{E}\left \{ |\mathrm{JS}_{\mathtt{s},\mathtt{pm}}|^{2} \right \}\!&=\!\sum\nolimits_{m\in\mathcal{M}_{\mathtt{s},\mathtt{t}} }\!\eta _{\mathtt{pm},\mathtt{t}}\rho _{\mathtt{pm}}\eta _{m,1}\rho _{\mathtt{c}}\zeta_{\mathtt{pm},\mathtt{t}}\beta_{m,\mathtt{pm}}N_{\mathtt{pm}}^2N\notag\\
    &\hspace{0.3cm}\times\gamma_{m,1}\big ( \beta_{\mathtt{pm},\mathtt{pm}}+\alpha N_{\mathtt{pm}}\zeta_{\mathtt{pm},\mathtt{t}}^2 \big ),
\end{align}
\begin{align}
    \mathbb{E}\left \{\! |\mathrm{JS}_{\mathtt{c},\mathtt{pm}}|^{2} \!\right \}\!&=\!\sum\nolimits_{m \in\mathcal{M}_{\mathtt{c}} }\eta _{\mathtt{pm},1}\eta _{m,1}\rho _{\mathtt{c}}\rho _{\mathtt{pm}}\gamma_{m,1}NN_{\mathtt{pm}}^2\beta_{m,\mathtt{pm}}\notag\\
    &\hspace{0.3cm}\times\beta_{\mathtt{pm},1}(\beta_{\mathtt{pm},\mathtt{pm}}+\alpha\zeta_{\mathtt{pm},\mathtt{t}}^2), \\
    \mathbb{E}\left \{ |\mathrm{n}_{\mathtt{pm}}|^{2} \right \}&=\sum\nolimits_{m\in\mathcal{M}_{\mathtt{c}} }\eta _{m,1}\rho _{\mathtt{c}}NN_{\mathtt{pm}}\beta_{m,\mathtt{pm}}\gamma_{m,1}.
\end{align}

Proof: See Appendix \ref{ProofTheorem2}.
\end{proposition}

We now analyze the impact of increasing the jamming power. From equation \eqref{eq:SINR_J}, we observe that $\rho_{\mathtt{pm}}$ appears in the denominator of $\mathrm{SINR}_{\mathtt{pm}}$. As a result, increasing the jamming power has two opposing effects: on one hand, it reduces $\mathrm{SINR}_{\mathtt{pm}}$ due to increased self-interference in the full-duplex proactive monitor. On the other hand, since MR precoding is specifically designed for UE $1$, the jamming signal has a stronger impact on reducing $\mathrm{SINR}_{1}$ than $\mathrm{SINR}_{\mathtt{pm}}$. Consequently, the overall monitoring performance improves.
Next, we consider a scenario where the number of antennas at the proactive monitor approaches infinity, $N_{\mathtt{pm}} \to \infty$, while the transmit power scales as $\rho _{\mathtt{pm}}=\frac{P_{\mathtt{pm}}}{N _{\mathtt{pm}}}$. From \eqref{eq:each_part_SINR_pm}, we  observe that the desired signal power $\left |\mathrm {DS}_{\mathtt{pm}} \right |^2 $ scales proportionally to $N_{\mathtt{pm}}^2$. Turning to the denominator of \eqref{eq:each_part_SINR_pm}, we find that the interference terms $\mathbb{E}\left\{|\mathrm{BU}_{\mathtt{pm}}|^{2}\right\}$, $\mathbb{E}\left \{ |\mathrm{IS}_{\mathtt{pm}}|^{2} \right \}$ and the noise $\mathbb{E}\left \{ |\mathrm{n}_{\mathtt{pm}}|^{2} \right \}$  grow linearly with $N_{\mathtt{pm}}$. The term $N_{\mathtt{pm}}^2$  in $\mathbb{E}\left \{ |\mathrm{JS}_{\mathtt{s},\mathtt{pm}}|^{2}\right \}$ and $\mathbb{E}\left \{ |\mathrm{JS}_{\mathtt{c},\mathtt{pm}}|^{2}\right \}$ can be eliminated by the power scaling factor and power control coefficient. Consequently, these terms become negligible compared to the numerator when $N_{\mathtt{pm}} \to \infty$. Furthermore, we find that the inter-UE interference term $\mathbb{E}\!\left \{\! |\mathrm{IU}_{{k}',\mathtt{pm}}\!|^{2} \!\right \}$ in the denominator contains $N_{\mathtt{pm}}^2$ terms, leading to the asymptotic result $\mathrm{SINR}_{\mathtt{pm}}\xrightarrow{N_{\mathtt{pm}}\to\infty}\frac{\big(\sum\nolimits_{m \in\mathcal{M}_{\mathtt{c}} } \sqrt{\eta _{m,1}}\beta_{m,\mathtt{pm}}N\gamma_{m,1}\big)^2}{\sum\nolimits_{\!m \in\mathcal{M}_{\mathtt{c}} }\!\!\eta _{m,{k'}}\eta _{m,1}\rho _{\mathtt{c}}N\gamma_{m,k'}\beta_{m,\mathtt{pm}}^2
\gamma_{m,1}}$. By increasing the number of monitor antennas, we can proportionally scale down its transmit power by a factor of  $\frac{1}{N _{\mathtt{pm}}}$, while maintaining  good monitoring performance.

\subsubsection{Sensing SINR}
The received signal at the $m''$-th S-AP, $m'' \in\mathcal{M}_{\mathtt{s},\mathtt{r}}$ can be expressed by 
\begin{align}\label{eq:y_m''_xmk}
    &\!\mathbf{y}_{m''} \!=\!\sum\nolimits_{m'\!\in\mathcal{M}_{\mathtt{s},\mathtt{t}} } \! (\mathbf{H}_{m',m''}\!+\!\mathbf{G}_{m',m''}^{T}\!)\mathbf{x}_{m'\!,\mathtt{t}}\notag\\
    &\qquad~+\sum\nolimits_{m\in\mathcal{M}_{\mathtt{c}} } \mathbf{G}_{m,m''}^{T}\mathbf{x} _{m}+\boldsymbol{\Lambda}_{\mathtt{pm},m''}\mathbf{x}_{\mathtt{pm}}+\boldsymbol{n}_{m''},
\end{align}
where $\boldsymbol{\Lambda}_{\mathtt{pm},m''}\in \mathbb{C} ^{ N \times N_\mathtt{pm}}$ denotes the effective channel between the proactive monitor and $m''$-th S-AP, which can be expressed as $\boldsymbol{\Lambda}_{\mathtt{pm},m''}=\mathbf{G}_{\mathtt{pm},m''}^{T}+\sqrt{\alpha}\mathbf{h}_{t,m''}\mathbf{h}_{\mathtt{pm},\mathtt{t}}^{T}$; $\mathbf{G}_{m,m''}\in \mathbb{C} ^{ N \times N}$ denotes the channel between the $m$-th C-AP and $m''$-th S-AP, while $\boldsymbol{n}_{m''}\in \mathbb{C} ^{ N \times 1}$ represents an AWGN vector whose entries are i.i.d. $\mathcal{CN}(0,1)$. We note that, since all APs cooperate and are connected to a central CPU, we consider the worst-case scenario for the monitoring side (and the best-case scenario for the malicious ISAC system). In this case, the AP-AP interference in the malicious ISAC system (the terms includes $\mathbf{G}_{m,m''}$ and $\mathbf{G}_{m',m''}$) can be canceled out in~\eqref{eq:y_m''_xmk}~\cite{ref:Secure_max_sens_SINR,ref:Tang_Bo_Sen_SINR}. Moreover, we define $\mathbf{H}_{m',m''}\in \mathbb{C} ^{ N \times N}$ and $\mathbf{H}_{m',m''}=\sqrt{\alpha}\mathbf{h}_{t,m''}\mathbf{h}_{m'\!,\mathtt{t}}^{T}$ as the reflected channel through the target between the $m'$-th and $m''$-th S-AP ($m' \in \mathcal{M}_{\mathtt{s},\mathtt{t}}$ and $m'' \in \mathcal{M}_{\mathtt{s},\mathtt{r}}$). 
Using the combining vector 
\begin{align}\label{eq:w_comb_m''}
    \!\mathbf{w}_{\mathtt{comb},m''} \!=\! \left(\!\sum\nolimits_{m'\in\mathcal{M}_{\mathtt{s},\mathtt{t}} } \!\sqrt{\eta _{m'\!,\mathtt{t}}\rho _{\mathtt{s}}}\mathbf{H}_{m',m''}\mathbf{w}_{m'\!,\mathtt{t}}^{\mathtt{s}}\!\right) ^{\ast}\!, 
\end{align}
at the $m''$-th S-AP, and substituting (\ref{eq:bold_x_m}), (\ref{eq:bold_x_m't}),  (\ref{eq:bold_x_pm}) and (\ref{eq:w_comb_m''}) into (\ref{eq:y_m''_xmk}), the received signal at the CPU of the malicious ISAC system to detect the legitimate target can be expressed by
\begin{align}
   z_{\mathtt{cpu}}  = &\mathrm {DS}_{\mathtt{cpu}}s_{\mathtt{t}}+\sum\nolimits_{k=1}^K \mathrm{IU}_{k,\mathtt{cpu}}s_{k}+\mathrm {JS}_{\mathtt{s},\mathtt{cpu}}s_{\mathtt{pm},\mathtt{t}} 
      \notag\\
        &+ \mathrm {JS}_{\mathtt{c},\mathtt{cpu}}s_{\mathtt{pm},1}+\mathrm{n}_{\mathtt{cpu}}, 
    \label{eq:zm''}
\end{align}
where $\mathrm {DS}_{\mathtt{cpu}}$, $\mathrm{IU}_{k,\mathtt{cpu}}$, $\mathrm {JS}_{\mathtt{s},\mathtt{cpu}}$, $ \mathrm {JS}_{\mathtt{c},\mathtt{cpu}}$ and $\mathrm{n}_{\mathtt{cpu}}$ are the desired signal, interference from C-APs, jamming signal to target, jamming signal to UE $1$ and noise, respectively, given by 
\begin{align}
    \!\!\mathrm {DS}_{\mathtt{cpu}}&\!\triangleq \!\!\sum_{m''\!\in\mathcal{M}_{\mathtt{s},\mathtt{r}} }\!\sum_{m'\!\in\mathcal{M}_{\mathtt{s},\mathtt{t}} } \!\!\sqrt{\eta _{m'\!,\mathtt{t}}\rho _{\mathtt{s}}}\!\mathbf{w}_{\mathtt{comb},m''}^{T}\mathbf{H}_{m',m''}\mathbf{w}_{m'\!,\mathtt{t}}^{\mathtt{s}},\\
    %\mathrm{BU}_{\mathtt{cpu}} & \triangleq \sum_{m''\!\in\mathcal{M}_{\mathtt{s},\mathtt{r}}}\!\sum_{m' \in\mathcal{M}_{\mathtt{s},\mathtt{t}} } \!\sqrt{\eta _{m,1}\rho _{\mathtt{c}}}\mathbf{w}_{\mathtt{comb},m''}^{T}\!\mathbf{H}_{m',m''}\mathbf{w}_{m'\!,\mathtt{t}}^{\mathtt{s}}\!\notag\\
    %\quad\!-\!\mathbb{E}&\bigg \{ \!\sum_{m''\!\in\mathcal{M}_{\mathtt{s},\mathtt{r}}}\!\mathbf{w}_{\mathtt{comb},m''}^{T}\!\sum_{m \in\mathcal{M}_{\mathtt{c}} } \sqrt{\eta _{m,1}\rho _{\mathtt{c}}}\mathbf{H}_{m',m''}\mathbf{w}_{m'\!,\mathtt{t}}^{\mathtt{s}}\bigg \}\notag\\
    \mathrm {IU}_{k,\mathtt{cpu}}\!& \triangleq \sum_{m''\in\mathcal{M}_{\mathtt{s},\mathtt{r}} }\!\sum_{m\in\mathcal{M}_{\mathtt{c}} } \sqrt{\eta _{m,k}\rho _{\mathtt{c}}}\mathbf{w}_{\mathtt{comb},m''}^{T}\mathbf{G}_{m,m''}^{T}\mathbf{w} _{m,k}^{\mathtt{c}},\\
    \mathrm{JS}_{\mathtt{s},\mathtt{cpu}} &\triangleq \sum_{m''\in\mathcal{M}_{\mathtt{s},\mathtt{r}} }\!\sqrt{\eta _{\mathtt{pm},\mathtt{t}}\rho _{\mathtt{pm}}}\mathbf{w}_{\mathtt{comb},m''}^{T}\boldsymbol{\Lambda}_{\mathtt{pm},m''}\mathbf{w}_{\mathtt{pm},\mathtt{t}}^{\mathtt{s}},\\
    \mathrm{JS}_{\mathtt{c},\mathtt{cpu}}& \triangleq\sum_{m''\in\mathcal{M}_{\mathtt{s},\mathtt{r}} }\!\sqrt{\eta _{\mathtt{pm},1}\rho _{\mathtt{pm}}}\!\mathbf{w}_{\mathtt{comb},m''}^{T}\boldsymbol{\Lambda}_{\mathtt{pm},m''}\mathbf{w}_{\mathtt{pm},1}^{\mathtt{c}},
\end{align}
\begin{align}
    \mathrm{n}_{\mathtt{cpu}} &\triangleq \sum\nolimits_{m''\in\mathcal{M}_{\mathtt{s},\mathtt{r}} }\!\mathbf{w}_{\mathtt{comb},m''}^{T}\boldsymbol{n}_{m''}.
    \label{eq:DS_cpu,IU_cpu...}
\end{align}

\begin{proposition}~\label{Theorem3}
The received $\mathrm{SINR}$ at the CPU for sensing the target can be defined as in \eqref{eq:SINR_m''} at the top of next page, where
\begin{figure*}
\begin{equation}
 \mathrm{SINR}_{\mathtt{cpu}}=\frac{|\mathrm {DS}_{\mathtt{cpu}}|^{2}}{\sum_{k=1}^K\mathbb{E}\left \{ |\mathrm{IU}_{k,\mathtt{cpu}}|^{2} \right \} +\mathbb{E}\left \{ |\mathrm{JS}_{\mathtt{s},\mathtt{cpu}}|^{2} \right \}+\mathbb{E}\left \{ |\mathrm{JS}_{\mathtt{c},\mathtt{cpu}}|^{2} \right \}+\mathbb{E}\left \{ |\mathrm{n}_{\mathtt{cpu}}|^{2} \right \}},
\label{eq:SINR_m''}
\end{equation}
\hrulefill
\end{figure*}
\begin{align}\label{eq:each_part_SINR_cpu}
    &\!\mathrm {DS}_{\mathtt{cpu}}\!= \!\sum\nolimits_{m''\!\in\mathcal{M}_{\mathtt{s},\mathtt{r}} }\sum\nolimits_{m'\in\mathcal{M}_{\mathtt{s},\mathtt{t}} } \sqrt{\eta _{m'\!,\mathtt{t}}}\rho_{\mathtt{s}} \zeta_{m'\!,\mathtt{t}}\zeta_{t,m''}\alpha N^3\notag\\
    &\times\Big ( \sqrt{\eta _{m'\!,\mathtt{t}}} \zeta_{m'\!,\mathtt{t}}+\sum\nolimits_{\tilde{m}'\in\mathcal{M}_{\mathtt{s},\mathtt{t}}, \tilde{m}' \ne m' } \sqrt{\eta _{\tilde{m}'\!,t}}\zeta_{\tilde{m}',t} \Big ), \\
    %--------------------------------
    &\mathbb{E}\left \{\! |\mathrm{IU}_{k,\mathtt{cpu}}|^{2}\! \right \}\!= \!\sum\nolimits_{m''\!\in\mathcal{M}_{\mathtt{s},\mathtt{r}} }\!\sum\nolimits_{m\in\mathcal{M}_{\mathtt{c}} }\!\sum\nolimits_{m'\!\in\mathcal{M}_{\mathtt{s},\mathtt{t}} }\!\!\eta _{m,k}\sqrt{\eta _{m'\!,\mathtt{t}}} \notag\\
    &\hspace{2.1cm}\times\rho _{\mathtt{c}}\rho _{\mathtt{s}}\gamma_{m,k}N^4\! \beta_{m,m''} \zeta_{t,m''}\zeta_{m'\!,\mathtt{t}}\Big( \sqrt{\eta _{m'\!,\mathtt{t}}}\zeta_{m'\!,\mathtt{t}}\notag\\
    &\hspace{2.1cm}+ \sum\nolimits_{ \tilde{m}'\in\mathcal{M}_{\mathtt{s},\mathtt{t}}, \tilde{m}' \ne m'}\sqrt{\eta _{\tilde{m}',t}}\zeta_{\tilde{m}',t}\Big ), \\
    %------------------------
    &\mathbb{E}\!\left \{\! |\mathrm{JS}_{\mathtt{s},\mathtt{cpu}}|^{2} \right \}\!=\!\sum\nolimits_{m''\!\in\mathcal{M}_{\mathtt{s},\mathtt{r}} }\!\eta _{\mathtt{pm},\mathtt{t}}\rho _{\mathtt{s}}\rho _{\mathtt{pm}}\alpha N^3N_{\mathtt{pm}}\zeta_{\mathtt{pm},\mathtt{t}}\zeta_{t,m''}\zeta_{m'\!,\mathtt{t}}\notag\\
    &\hspace{2.1cm}\times\beta_{\mathtt{pm},m''}\Big( \sum\nolimits_{m'\!\in\mathcal{M}_{\mathtt{s},\mathtt{t}} }\sqrt{\eta _{m'\!,\mathtt{t}}}\zeta_{m'\!,\mathtt{t}}\Big )^2\!+\!\eta _{\mathtt{pm},\mathtt{t}}\notag\\
    &\hspace{2.1cm}\times\!\rho _{\mathtt{pm}}\rho _{\mathtt{s}}\zeta_{\mathtt{pm},\mathtt{t}}^2N^4N_\mathtt{pm}^2\alpha^2\Big(\!\sum\nolimits_{m'\!\in\mathcal{M}_{\mathtt{s},\mathtt{t}} }\!\!\sum\nolimits_{m''\!\in\mathcal{M}_{\mathtt{s},\mathtt{r}} }\notag\\
    &\hspace{2.1cm}\sqrt{\eta _{m'\!,\mathtt{t}}}\zeta_{m'\!,\mathtt{t}}\zeta_{t,m''}\Big )^2,
    \\
    %----
    &\mathbb{E}\!\left \{\! |\mathrm{JS}_{\mathtt{c},\mathtt{cpu}}|^{2}\! \right \}\!=\!\sum\nolimits_{m''\!\in\mathcal{M}_{\mathtt{s},\mathtt{r}} }\eta _{\mathtt{pm},1}\rho _{\mathtt{pm}}\rho _{\mathtt{s}}\alpha \zeta_{t,m''}N_{\mathtt{pm}}N^3\beta_{\mathtt{pm},m''}\notag\\
    &\hspace{0.2cm}\times\beta_{\mathtt{pm},1}\Big(\sum\nolimits_{m'\!\in\mathcal{M}_{\mathtt{s},\mathtt{t}} }\sqrt{\eta_{m'\!,\mathtt{t}}}\zeta_{m'\!,\mathtt{t}}\Big)^2+\eta _{\mathtt{pm},1}\notag\\
    &\hspace{0.2cm}\times\rho _{\mathtt{pm}}\rho _{\mathtt{c}}\zeta_{\mathtt{pm},\mathtt{t}}\beta_{\mathtt{pm},1}N^4N_\mathtt{pm}\alpha^2\Big(\!\sum\nolimits_{m''\!\in\mathcal{M}_{\mathtt{s},\mathtt{r}} }\!\notag\\
    & \hspace{0.2cm}\sum\nolimits_{m'\in\mathcal{M}_{\mathtt{s},\mathtt{t}} }\sqrt{\eta _{m'\!,\mathtt{t}}}\zeta_{t,m''}\zeta_{m'\!,\mathtt{t}}\Big )^2,    
\end{align}
\begin{align}
    %---------------------------------
    \mathbb{E}\left \{ |\mathrm{n}_{\mathtt{cpu}}|^{2} \right \}\!&=\! \sum\nolimits_{m''\!\in\mathcal{M}_{\mathtt{s},\mathtt{r}} }\!\sum\nolimits_{m'\!\in\mathcal{M}_{\mathtt{s},\mathtt{t}} }\!\sqrt{\eta _{m'\!,\mathtt{t}}}\rho _{\mathtt{s}}\alpha \zeta_{t,m''}\zeta_{m'\!,\mathtt{t}} \notag\\
    \times N^3\Big(& \sqrt{\eta _{m'\!,\mathtt{t}}}\zeta_{m'\!,\mathtt{t}}\!+\! \sum\nolimits_{\tilde{m}'\!\in\mathcal{M}_{\mathtt{s},\mathtt{t}}, \tilde{m}' \ne m'}\!\!\sqrt{\eta _{\tilde{m}',t}}\zeta_{\tilde{m}',t}\Big ).
\end{align}
Proof: Follows a similar methodology as those used in the proof of Propositions 1 and 2.
\end{proposition}

We now consider the effect of increasing the jamming power and the number of antennas at the proactive monitor. From  \eqref{eq:SINR_m''}, we observe that increasing $\rho_{\mathtt{pm}}$ and $N_{\mathtt{pm}}$   amplifies  the strength of  jamming signals, while the desired signal and other interference components remain unchanged due to their independence from $\rho _{\mathtt{pm}}$ and $N_{\mathtt{pm}}$. Consequently, the sensing SINR at the malicious ISAC, $\mathrm{SINR}_{\mathtt{cpu}}$, decreases. In addition, we analyze the case where $N_{\mathtt{pm}} \to \infty$, under the transmit power constraint $\rho_ {\mathtt{pm}}=\frac{P_{\mathtt{pm}}}{N_{\mathtt{pm}}}$. We first observe that  $\mathbb{E}\!\left \{\! |\mathrm{JS}_{\mathtt{c},\mathtt{cpu}}|^{2} \right \}\xrightarrow{N_{\mathtt{pm}}\to\infty} 0$. In contrast, the second term in $\mathbb{E}\!\left \{\! |\mathrm{JS}_{\mathtt{s},\mathtt{cpu}}|^{2} \right \}$ contains an $N_{\mathtt{pm}}^2$ scaling factor, leading to  $\mathbb{E}\!\left \{\! |\mathrm{JS}_{\mathtt{s},\mathtt{cpu}}|^{2} \right \}\xrightarrow{N_{\mathtt{pm}}\to\infty}\frac{1}{2}P_{\mathtt{pm}}\rho _{\mathtt{s}}\zeta_{\mathtt{pm},\mathtt{t}}N^4\alpha^2\bigg(\!\sum\nolimits_{m'\!\in\mathcal{M}_{\mathtt{s},\mathtt{t}} }\!\sum\nolimits_{m''\!\in\mathcal{M}_{\mathtt{s},\mathtt{r}} }\!\!
\sqrt{\eta _{m'\!,\mathtt{t}}}\zeta_{m'\!,\mathtt{t}}\zeta_{t,m''}\bigg )^2$. Thus,  $\mathrm{SINR}_{\mathtt{cpu}}$ converges to a constant value. This analysis highlights the influence of $N_{\mathtt{pm}}$ and $\rho_{\mathtt{pm}}$ on $\mathrm{SINR}_{\mathtt{cpu}}$, as well as the importance of meticulously selecting the power allocation coefficients $\eta_{\mathtt{pm},\mathtt{t}}$ and $\eta_{\mathtt{pm},1}$.

%%%%%%%%%%%%%%%%%%%%%% Performance %%%%%%%%%%%%%%%%%%%%%%%%%%%%%%%%%%
\subsection{Performance Metrics}\label{sec:performance_metrics}
The goal of the proactive monitor is to fully recover the information that the malicious UE $1$ can decode. For a given modulation and coding scheme, the achievable data rate is a monotonically increasing function of SINR.  If $\mathrm{SINR}_{\mathtt{pm}} < \mathrm{SINR}_1$, the monitor's channel capacity is smaller than that of the malicious UE. As a result, the monitor may fail to decode some symbols or packets that are correctly received by UE $1$, leading to incomplete or erroneous monitoring \cite{ref:Xu_Jie_TWC}. By ensuring $\mathrm{SINR}_{\mathtt{pm}} \ge \mathrm{SINR}_1$, we guarantee that the monitor can support at least the same data rate as the malicious UE and thus reliably decode  message intended for UE 1 \cite{ref:zahra_iot,ref:Xu_Jie_TWC,ref:monitor_energy_efficiency}. 
To this end, the following indicator function can be considered for characterizing the event of successful monitoring at the monitor:
\begin{equation}
    X_1= \left\{\begin{matrix}
 1, & \mathrm{SINR}_{\mathtt{pm}}\ge \mathrm{SINR}_1,\\
 0, & \mathrm{SINR}_{\mathtt{pm}}<  \mathrm{SINR}_1.
\end{matrix}\right.
\label{eq:X_k}
\end{equation}
The expectation of the successful monitoring case: $\mathbb{E}\left \{ X_k \right \}$ can be written as $\mathbb{E}\left \{ X_k \right \} = \mathrm {Pr} \left \{ \mathrm{SINR}_{\mathtt{pm}}\ge \mathrm{SINR}_k \right \} $, and indicates the MSP \cite{ref:zahra_iot}.

The malicious ISAC system aims to detect the legitimate target from the echoes. To quantify this approach, we adopt the SDP, defined as the probability that the sensing SINR at a target exceeds a given threshold $\kappa$: $\mathrm{SDP} = \mathrm{Pr} \left\{ \mathrm{SINR}_{\mathtt{cpu}} \ge \kappa \right\}$~\cite{ref:jj_detection_probability, refRevised:Jiajun_ICC}. This metric is widely used in radar systems to evaluate whether a target can be detected and localized. A high SDP indicates a high likelihood that the target can be successfully detected by the malicious ISAC system. Since SDP is a monotonically increasing function of the SINR, minimizing SINR effectively degrades the target detection performance of the malicious system.

%%%%%%%%%%%%%%%%%%%%% Optimization %%%%%%%%%%%%%%%%%%%%%%%%%%%%%%%%%%
\section{Problem Formulations}\label{sec:Proposed Design Problems and Solutions}
In this section, we introduce power allocation algorithms for two scenarios: (i) the objective is to minimize the SDP while ensuring that $\mathrm{SINR}_{\mathtt{pm}} \ge \mathrm{SINR}_1$ for successful monitoring of the malicious UE; (ii) the goal is to extend the monitor's operational time and reduce the risk of exposure. The two optimization problems in our manuscript are motivated by different but complementary objectives, depending on the system objective and practical preference. In particular, the first problem  focuses on achieving the best possible performance ignoring power consumption minimization. This formulation prioritizes performance maximization and provides insight into the upper bound of effectiveness. In contrast, practical deployments may involve battery-powered or energy-limited monitors. Once the sensing SINR of the malicious ISAC system falls below a critical threshold, the target becomes effectively undetectable, and further SINR reduction requires extra jamming power but yields diminishing returns. Motivated by this, the second problem considers energy-efficient design, where the goal is to minimize the jamming power while maintaining successful monitoring and satisfying specific sensing SINR constraints. This ensures sustainable monitor operation.

% Prior to formulating and solving the problems we proposed, we need to impose power constraints for the proactive monitor.
% \begin{itemize}
%     \item We define $\theta_{\mathtt{pm},\mathtt{t}} \triangleq N_{\mathtt{pm}}\eta _{\mathtt{pm},\mathtt{t}}\zeta_{\mathtt{pm},\mathtt{t}}$, the power control coefficient constraint can be given by $ 0\le\theta_{\mathtt{pm},\mathtt{t}}\le 1$. Similarly, the power control coefficient for sensing the target is given by $\theta_{\mathtt{pm},1}\triangleq N_{\mathtt{pm}}\eta _{\mathtt{pm},1}\beta_{\mathtt{pm},1}$ and $ 0\le\theta_{\mathtt{pm},1}\le 1$.
%     \item We note that $\varsigma _{\mathtt{s}}=\rho_\mathtt{pm}\theta_{\mathtt{pm},\mathtt{t}}$   represents the normalized power allocated at the monitor to jam the target. Moreover, we have $\varsigma _{\mathtt{c}}=\rho_\mathtt{pm}\theta_{\mathtt{pm},1}$ represents the normalized power allocated at the monitor to jam the UE $1$.
%     \item With the power constraint at the proactive monitor expressed as $ 0\le\theta_{\mathtt{pm},\mathtt{t}}+\theta_{\mathtt{pm},1}\le 1$, showing the power control for jamming the target and UE $1$. The total transmit power constraint at the proactive monitor can be written as $0\le \varsigma _{\mathtt{s}}+\varsigma _{\mathtt{c}}\le \rho_\mathtt{pm}$. The transmit power in Watt can be calculated by multiplying with the noise power.
% \end{itemize}
\subsection{Minimize the SDP Performance at Malicious ISAC}
In this subsection, we seek to optimize the power control coefficients $\eta _{\mathtt{pm},\mathtt{t}}$ and $\eta _{\mathtt{pm},1}$ at the proactive monitor to minimize the $\mathrm{SINR}_{\mathtt{cpu}}$ of the malicious ISAC system, under the constraints on the successful monitoring of the malicious UE and total transmit power at the monitor.
More precisely, the optimization problem can be formulated as
\begin{subequations}\label{Problem1}
\begin{align}
\mathbf{(P_{1} )}:\min_{\eta _{\mathtt{pm},\mathtt{t}},\eta _{\mathtt{pm},1}} \quad & \mathrm{SINR}_{\mathtt{cpu}}  & \label{obj:opt1.1} \\
\mbox{s.t.}\quad
&\mathrm{SINR}_{\mathtt{pm}}\ge \mathrm{SINR}_1,  & \label{st:opt1.1 SINR_s>k}\\
&\eta _{\mathtt{pm},\mathtt{t}}\ge 0, & \label{st:opt1.2}\\
&\eta _{\mathtt{pm},1}\ge 0, & \label{st:opt1.3}\\
0\le \eta _{\mathtt{pm},\mathtt{t}}N_{\mathtt{pm}}&\zeta_{\mathtt{pm},\mathtt{t}}+\eta _{\mathtt{pm},1}N_{\mathtt{pm}}\beta_{\mathtt{pm},1}\le 1. & \label{st:opt1.4}
\end{align}
\end{subequations}
The constraint (\ref{st:opt1.1 SINR_s>k}) specifies that the received SINR at the proactive monitor, denoted as $\mathrm{SINR}_{\mathtt{pm}}$, must be consistently larger than the SINR at the UE 1, represented as $\mathrm{SINR}_1$, which serves as a fundamental condition to guarantee the success of the monitoring. Moreover, constraint (\ref{st:opt1.4}) represents the total transmit power.
For ease of description, let us denote $\theta_{\mathtt{pm},\mathtt{t}} \triangleq N_{\mathtt{pm}}\eta _{\mathtt{pm},\mathtt{t}}\zeta_{\mathtt{pm},\mathtt{t}}$ and $\theta_{\mathtt{pm},1}\triangleq N_{\mathtt{pm}}\eta _{\mathtt{pm},1}\beta_{\mathtt{pm},1}$ in the following steps. Accordingly, we  have
\begin{subequations}
\begin{align}
\mathbf{(P_{1} )}:\min_{\theta _{\mathtt{pm},\mathtt{t}},\theta _{\mathtt{pm},1}} \quad &\frac{q_9}{q_{10}\theta _{\mathtt{pm},\mathtt{t}}+q_{11}\theta _{\mathtt{pm},1}+q_{12}} & \label{obj:min_SINR} \\
\mbox{s.t.}\quad
& \frac{q_1q_6\theta _{\mathtt{pm},\mathtt{t}}+q_1q_7\theta _{\mathtt{pm},1}+q_1q_8}{q_2q_5\theta _{\mathtt{pm},\mathtt{t}}+q_3q_5\theta _{\mathtt{pm},1}+q_4q_5}\ge 1, & \\
&0\le \theta _{\mathtt{pm},\mathtt{t}}+\theta _{\mathtt{pm},1}\le 1, & \label{st:0<theta1+theta2<1} \\
&\theta _{\mathtt{pm},\mathtt{t}}\ge 0,& \label{st:0<theta1<1} \\
&\theta _{\mathtt{pm},1}\ge 0. & \label{st:0<theta2<1}
\end{align} 
\end{subequations}
To further simplify the expression, we introduce  symbol $q_\nu$ with various subscript $\nu$ from 1 to 12  to represent the desired signal and interference plus noise, which are independent of the power control coefficients at the monitor as follows:
\begin{align}
&q_{1}= |\mathrm {DS}_{\mathtt{pm}}|^2,\notag\\
&\!q_2\!\!=\!\!\sum\nolimits_{m\in\mathcal{M}_{\mathtt{s},\mathtt{t}} }\!\!\!\rho _{\mathtt{pm}}\eta _{m,1}\!\rho _{\mathtt{c}}\beta_{m,\mathtt{pm}}N_{\mathtt{pm}}N\gamma_{m,1}\big (\beta_{\mathtt{pm},\mathtt{pm}}\!+\!\alpha N_{\mathtt{pm}}\zeta_{\mathtt{pm},\mathtt{t}}^2 \big ),\notag
\end{align}
\begin{align}
&q_{3}=\sum\nolimits_{m \in\mathcal{M}_{\mathtt{c}} }\eta _{m,1}\rho _{\mathtt{c}}\rho _{\mathtt{pm}}\gamma_{m,1}N N_{\mathtt{pm}}\beta_{m,\mathtt{pm}}(\beta_{\mathtt{pm},\mathtt{pm}}+\alpha\zeta_{\mathtt{pm},\mathtt{t}}^2),\notag\\
&q_{4}=\mathbb{E}\left \{ |\mathrm{BU}_{\mathtt{pm}}|^{2} \right \}\!+\!\sum\nolimits_{{k}'\ne 1}^{K}\mathbb{E}\left \{\! |\mathrm{IU}_{{k}',\mathtt{pm}}\!|^{2} \!\right \}+\!\mathbb{E}\left \{ |\mathrm{IS}_{\mathtt{pm}}|^{2} \right \}\notag\\
&\hspace{0.7cm}+\!\mathbb{E}\left \{ |\mathrm{n}_{\mathtt{pm}}|^{2} \right \},\notag\\
&q_{5}=\!\left |  \mathrm {DS}_{k}  \right |^2,\notag\\ &q_{6}=\rho _{\mathtt{pm}} (\beta_{\mathtt{pm},k}+\alpha\zeta_{\mathtt{t},k}N_{\mathtt{pm}}\zeta_{\mathtt{pm},\mathtt{t}}),\notag\\
&q_{7}=\!\rho _{\mathtt{pm}}(N_{\mathtt{pm}}\beta_{\mathtt{pm},1}\!+\!\beta_{\mathtt{pm}\!,1}\!+\!\alpha\zeta_{t\!,k}\zeta_{\mathtt{pm},\mathtt{t}}),\notag\\
&q_{8}=\mathbb{E}\left \{ |\mathrm{BU}_{k}|^{2} \right \}+\sum\nolimits_{{k}'\ne 1}^{K}\mathbb{E}\!\left \{\! |\mathrm{IU}_{{k}',k}\!|^{2} \!\right \}+\mathbb{E}\left \{ |\mathrm{IS}_{k}|^{2} \right \}+1,\notag\\
&q_{9}= |\mathrm {DS}_{\mathtt{cpu}}|^2,\notag\\ 
%-------------
&q_{10}=\sum\nolimits_{m''\in\mathcal{M}_{\mathtt{s},\mathtt{r}} }\!\rho _{\mathtt{s}}\rho _{\mathtt{pm}}\zeta_{t,m''}\zeta_{m'\!,\mathtt{t}}
\beta_{\mathtt{pm},m''}N^3\alpha\notag\\
&\hspace{0.7cm}\times  \Big( \sum\nolimits_{m'\!\in\mathcal{M}_{\mathtt{s},\mathtt{t}} }\sqrt{\eta _{m'\!,\mathtt{t}}}\zeta_{m'\!,\mathtt{t}}\Big )^2\!+\!\rho _{\mathtt{pm}}
\rho _{\mathtt{s}}\zeta_{\mathtt{pm},\mathtt{t}}N^4N_\mathtt{pm}\notag\\
&\hspace{0.7cm}\times\alpha^2\Big(\!\sum\nolimits_{m'\!\in\mathcal{M}_{\mathtt{s},\mathtt{t}} }\!\sum\nolimits_{m''\!\in\mathcal{M}_{\mathtt{s},\mathtt{r}} }
\sqrt{\eta _{m'\!,\mathtt{t}}}\zeta_{m'\!,\mathtt{t}}\zeta_{t,m''}\Big )^2,\notag
\end{align}
\begin{align}
&q_{11}=\sum\nolimits_{m''\in\mathcal{M}_{\mathtt{s},\mathtt{r}} }\rho _{\mathtt{pm}}\rho _{\mathtt{s}}\alpha \zeta_{t,m''}N^3\beta_{\mathtt{pm},m''}\notag\\
&\hspace{0.7cm}\times
\Big(\sum\nolimits_{m'\!\in\mathcal{M}_{\mathtt{s},\mathtt{t}} }\sqrt{\eta_{m'\!,\mathtt{t}}}\zeta_{m'\!,\mathtt{t}}\Big)^2+
\rho _{\mathtt{pm}}\rho _{\mathtt{c}}\zeta_{\mathtt{pm},\mathtt{t}}N^4\alpha^2\notag\\
&\hspace{0.7cm}\times\Big(\!\sum\nolimits_{m''\!\in\mathcal{M}_{\mathtt{s},\mathtt{r}} }\!
\sum\nolimits_{m'\in\mathcal{M}_{\mathtt{s},\mathtt{t}} }\sqrt{\eta _{m'\!,\mathtt{t}}}\zeta_{t,m''}\zeta_{m'\!,\mathtt{t}}\Big )^2,\notag\\
&q_{12}=\sum\nolimits_{k=1}^K\mathbb{E}\!\left \{\! |\mathrm{IU}_{k,\mathtt{cpu}}\!|^{2} \!\right \}+\mathbb{E}\left \{ |\mathrm{n}_{\mathtt{cpu}}|^{2} \right \}.
\end{align}

The expression (\ref{obj:min_SINR}) is quasi-convex and the constraints of
the problem $\mathbf{(P_{1} )}$ are linear functions of the variables $\theta _{\mathtt{pm},\mathtt{t}}$ and $\theta _{\mathtt{pm},1}$. By introducing a nonnegative auxiliary variable $t$, the optimization problem can be equivalently reformulated as follows: 
\begin{subequations}\label{opt1:reformP1}
\begin{align}
\mathbf{(P_{1} )}:\max_{\theta _{\mathtt{pm},\mathtt{t}},\theta _{\mathtt{pm},1}} \quad & t &  \\
\mbox{s.t.}\quad
& q_{10}\theta _{\mathtt{pm},\mathtt{t}}+q_{11}\theta _{\mathtt{pm},1}+q_{12}\ge t q_{9}, & \\
& q_1q_6\theta _{\mathtt{pm},\mathtt{t}}+q_1q_7\theta _{\mathtt{pm},1}+q_1q_8\ge \notag \\
&q_2q_5\theta _{\mathtt{pm},\mathtt{t}}+q_3q_5\theta _{\mathtt{pm},1}+q_4q_5,  & \\
&(\ref{st:0<theta1+theta2<1})-(\ref{st:0<theta2<1}). &\notag
\end{align} 
\end{subequations}
%======
Now, for a fixed $t$, all inequalities involved in the problem $\mathbf{(P_{1} )}$ are linear, hence the optimum solution to the problem can be obtained yielding a line-search to find the maximal value of $t$ while satisfying all constraints. We use bisection search method to find the optimized solution, where  in each step we solve a sequence of linear feasibility problem. The corresponding bisection-based search algorithm is shown in \textbf{Algorithm \ref{alg:bisection}}.

The total number of iterations required in the bi-section algorithm can be calculated by  $\log_{2}\Big(\frac{t_{\rm max}-t_{\rm min}}{\epsilon}\Big)$. Furthermore, the computational complexity of the optimization problem $\mathbf{(P_{1})}$ 
involves $C_{\mathtt{l}}=4$ linear constraints and $C_{\mathtt{v}}=2$ real-valued scalar variables. 
Hence, solving $\mathbf{(P_{1})}$ via bisection requires 
%$\mathcal{O}\!\Bigg(
% \Big\lceil \log_{2}\!\Big(\frac{t_{\rm max}-t_{\rm min}}{\epsilon}\Big)\Big\rceil
% \cdot C_{\mathtt{v}}^{2} \sqrt{C_{\mathtt{l}}} (C_{\mathtt{v}}+C_{\mathtt{l}})
% \Bigg) = \mathcal{O}\!\Big(\log_2\!\Big(\frac{t_{\rm max}-t_{\rm min}}{\epsilon}\Big)\Big)$,
\begin{align}
&\mathcal{O}\!\Bigg(
\Big\lceil \log_{2}\!\Big(\frac{t_{\rm max}-t_{\rm min}}{\epsilon}\Big)\Big\rceil
\cdot C_{\mathtt{v}}^{2} \sqrt{C_{\mathtt{l}}} (C_{\mathtt{v}}+C_{\mathtt{l}})
\Bigg) \notag\\
&= \mathcal{O}\!\Big(\log_2\!\Big(\frac{t_{\rm max}-t_{\rm min}}{\epsilon}\Big)\Big),
\end{align}
since $C_{\mathtt{l}}$ and $C_{\mathtt{v}}$ are constants in our case.

\begin{algorithm}[t]
    \caption{Bisection algorithm for solving optimization problem $\mathbf{(P_{1} )}$ }
\begin{itemize}
   \item[(1)] {\emph{Initialization}}: Choose the initial values of $t_{\rm max}$ and $t_{\rm min}$, where $t_{\rm max}$ and $t_{\rm min}$ define a range of objective function values. Set tolerance $\epsilon >0$.
   \item[(2)] Set $t:=\frac{t_{\rm max}+t_{\rm min}}{2}$ and solve the following convex feasibility problem:
   \begin{align}
   \begin{cases}
&\mathrm{SINR}_{\mathtt{pm}}\ge \mathrm{SINR}_1,   \\
&0\le \eta _{\mathtt{pm},\mathtt{t}}N\zeta_{\mathtt{pm},\mathtt{t}}\le 1, \\
&0\le \eta _{\mathtt{pm},\mathtt{t}}N\beta_{\mathtt{pm},1}\le 1, \\
&0\le \eta _{\mathtt{pm},\mathtt{t}}N_{\mathtt{pm}}\zeta_{\mathtt{pm},\mathtt{t}}+\eta _{\mathtt{pm},1}N_{\mathtt{pm}}\beta_{\mathtt{pm},1}\le 1. 
   \end{cases}
   \end{align}
   \item[(3)] If the problem in Step ($2$) is feasible, set $t_{\min}:=t$; else set $t_{\rm max}:=t$.
   \item[(5)] Stop if $t_{\rm max}-t_{\rm min}< \epsilon$. Otherwise, go to Step $2$.
\end{itemize}
\label{alg:bisection}
\end{algorithm}
%%%%%%%%%%% 2nd OPtimization %%%%%%%%%%%%

\subsection{Minimize Total Transmit Power at the Proactive Monitor}
Let us introduce the additional nonnegative variables $\varsigma _{\mathtt{s}}=\theta _{\mathtt{pm},\mathtt{t}}\rho_{\mathtt{pm}}$ and $\varsigma _{\mathtt{c}}=\theta _{\mathtt{pm},k}\rho_{\mathtt{pm}}$. Therefore, the minimum total transmit power required at the proactive monitor is $\varsigma _{\mathtt{s}}+\varsigma _{\mathtt{c}}$. Accordingly, we formulate the following optimization problem to minimize the total transmit power at the proactive monitor while ensuring successful monitoring performance and maintaining the SINR for detection below a threshold at the malicious ISAC:
\begin{subequations}\label{Problem2}
\begin{align}
\mathbf{(P_{2} )}:\min_{\varsigma _{\mathtt{s}},\varsigma _{\mathtt{c}}} \quad & (\varsigma _{\mathtt{s}}+\varsigma _{\mathtt{c}}) & \label{obj:opt2 min_pow} \\
\mbox{s.t.}\quad
&\mathrm{SINR}_{\mathtt{cpu}}\le\kappa, \label{st:opt2.1 SINR_s>k} & \\
&\mathrm{SINR}_{\mathtt{pm}}\ge \mathrm{SINR}_{1}, \label{st:opt2.2 SINR_pm>SINR_1}  \\
&\varsigma _{\mathtt{s}}\ge 0, & \label{st:opt2.3 power constraint}\\
&\varsigma _{\mathtt{c}}\ge 0, & \label{st:opt2.4 power constraint}\\
&0\le \varsigma _{\mathtt{s}}+\varsigma _{\mathtt{c}}\le \rho_\mathtt{pm}. &\label{st:opt2.5 power constraint}
\end{align}
\end{subequations}
%=========
We can reformulate $\mathbf{(P_{2} )}$ into the following problem by introducing a new auxiliary variable $\varsigma _{\mathtt{s}}+\varsigma _{\mathtt{c}} \le \frac{1}{\varsigma}$ as
\begin{subequations}
\begin{align}
\mathbf{(P_{2} )}:\max_{\varsigma _{\mathtt{s}},\varsigma _{\mathtt{c}}} \quad & \varsigma &  \\
\mbox{s.t.}\quad
& 1\ge \varsigma(\varsigma _{\mathtt{s}}+\varsigma _{\mathtt{c}}),& \\
& q_{10}'\varsigma _{\mathtt{s}}+ q_{11}'\varsigma _{\mathtt{c}}+q_{12}\ge \kappa' q_{9}, & \\
& q_{1}q_{6}'\varsigma _{\mathtt{s}}+q_{1}q_{7}'\varsigma _{\mathtt{c}}+q_1q_8\ge \notag \\
&q_{2}'q_5\varsigma _{\mathtt{s}}+q_{3}'q_5\varsigma _{\mathtt{c}}+q_4q_5,  & \\
&(\ref{st:opt2.3 power constraint})-(\ref{st:opt2.5 power constraint}). &
\end{align} 
\end{subequations}
where $q_{2}'$, $q_{3}'$, $q_{6}'$, $q_{7}'$, $q_{10}'$ and $q_{11}'$ can be obtained by dividing the corresponding $q_{2}$, $q_{3}$, $q_{6}$, $q_{7}$, $q_{10}$ and $q_{11}$ with the normalized power $\rho_{\mathtt{pm}}$. We note that $\mathbf{(P_{2})}$ can be solved using the same methodology as $\mathbf{(P_{1})}$, and therefore, we omit the details for brevity.

\begin{Remark} It is notable that for a given  energy storage capacity $E_{\mathtt{pm}}^{\max}$, the operational lifetime of proactive monitor can be computed as\cite{refRevised:SleepingStrategy}
\begin{align}\label{eq:Epm2}
    T_{\mathtt{serv}} = \frac{E_{\mathtt{pm}}^{\max}}{P_{\mathtt{sta}} + \frac{(\theta_{\mathtt{pm},\mathtt{t}} + \theta_{\mathtt{pm},1}) P_{\mathtt{pm}}}{\eta_{\mathtt{amp}}}},
\end{align}
where $P_{\mathtt{sta}}$ is the constant static power consumption and $\eta_{\mathtt{amp}}$ is the power amplifier efficiency.
Accordingly, reducing the power consumption at proactive monitor (the second term in the denominator of~\eqref{eq:Epm2}) directly increases $T_{\mathtt{serv}}$.
\end{Remark}

\section{Numerical Results}
\begin{figure}[t]
  \centering
  \includegraphics[width=0.85\linewidth,height=0.65\linewidth]{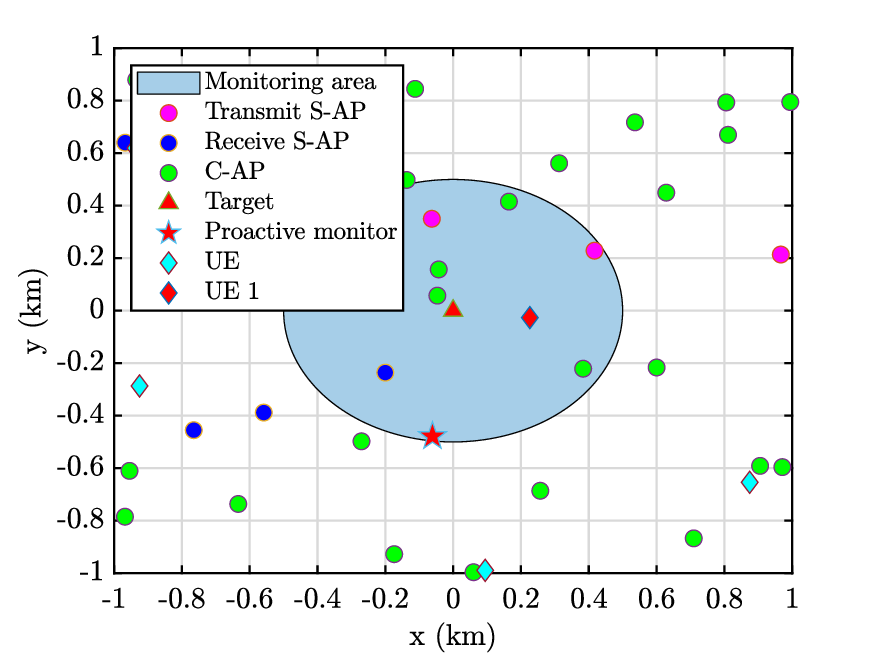}
  \caption{ The 2D locations of the system, where $M_{\mathtt{c}}=32$, $M_{\mathtt{s},\mathtt{t}}=M_{\mathtt{s},\mathtt{r}}=4$, $K=5$, $r=500$ m.}
  \label{fig:2D location}
\end{figure}

In this section, we evaluate the performance of anti-malicious ISAC
system and validate the impact of key system parameters. 
To begin with, we consider 32 C-APs for transmitting communication signals, 4 S-APs for transmitting and receiving sensing signals respectively, and $K$ UEs in the malicious ISAC system. The APs and UEs are randomly distributed in an area of $2\times 2$ $\mathrm{km}^2$ having wrapped around edges to reduce the boundary effects. We assume that each AP is equipped with a uniform linear array (ULA) antenna. The $n$-th element of the steering vector $\boldsymbol{\alpha }_{t}\left( \phi_{m'\!,\mathtt{t}}^{a}, \phi_{m'\!,\mathtt{t}}^{e}\right) \in \mathbb{C}^{N\times 1}$ is
$\left [\boldsymbol{\alpha } _{t}\left ( \phi_{m'\!,\mathtt{t}}^{a}, \phi_{m'\!,\mathtt{t}}^{e}\right ) \right ]_{n}\!=\!{\exp\left [ j2\pi\frac{\Delta d}{\boldsymbol{\Lambda} }(n-1) \sin \phi_{m'\!,\mathtt{t}}^{a}\cos \phi_{m'\!,\mathtt{t}}^{e} \right ] }$, where $\Delta d= \frac{\boldsymbol{\Lambda}}{2}$ is the distance between any two adjacent antennas. The legitimate monitor is positioned randomly inside a circle centred around the target with a radius of $r$, while the target is located in the center of the area at the altitude of $h$ m above the ground. The 2D locations of all APs, UEs and target in a typical realization
are illustrated in Fig. \ref{fig:2D location}.

\begin{align}\label{eq:3PL}
    \mathrm {PL}_{m,k}\!=\!\begin{cases}
        \!-\!L\!-\!15\log_{10}{(d_{1})}-20\log_{10}{(d_{0})},~$if$~d\le d_{0},\\
        \!-\!L\!-\!15\log\!_{10}{(d_{1})}\!-\!20\!\log_{10}{(d)},~$if$~ d_{0}\!<\!d\!\le \!d_{1},\\
        \!-\!L\!-\!35\log_{10}{(d)},~$if$~d>d_{1}.
    \end{cases}
\end{align}

The large-scale fading coefficient can be calculated by $\beta_{m,k} = \mathrm {PL}_{mk} \cdot 10^{\frac{\sigma _{sh}z_{mk}}{10} }  $, where $\sigma_{sh}=9$dB and $z_{mk}\sim \mathcal{CN} (0,1)$, $\mathrm {PL}_{mk}$ denotes the path-loss, which is modeled as \eqref{eq:3PL}. We further consider that the noise figure is $8$ dB, $d_{0}=10$ m and $d_{1}=50$ m \cite{ref:Hien_small_cell}. We assume that the power for downlink data transmission and sensing is $P _{\mathtt{c}} = P_{\mathtt{s}} = 1$ W, the transmit power for sending the pilot sequences is $P _{\mathtt{p}} =0.2$ W and the power allocated at the monitor is denoted by $P_{\mathtt{pm}}$. 
Moreover, we assume that $ \eta_{m,k} = \frac{1}{{N\sum_{k=1}^{K} \gamma_{m,k}}}$ and $\eta _{m'\!,\mathtt{t}} = \frac{1}{N\zeta_{m'\!,\mathtt{t}}}$.

Now, we compare our optimized proactive monitoring  approach with two baselines: (i) passive monitoring, where the monitor remains silent and only overhears suspicious links, and (ii) proactive monitoring with constant-power jamming, where the jamming power is fixed. We highlight that our monitor employs a reactive jamming strategy with optimized power allocation. Specifically, two separate power allocation coefficients are used: one controlling the jamming power toward the malicious UE 1, and the other controlling the jamming power toward the legitimate target to reduce its SDP as perceived by the malicious ISAC system. Both coefficients can adaptively become zero when jamming is unnecessary. They are optimized to balance between successful monitoring, target protection,  and  energy efficiency, thereby realizing a fully reactive and adaptive jamming strategy.

\begin{figure}[t]
  \centering
  \includegraphics[width=0.9\linewidth,height=0.65\linewidth]{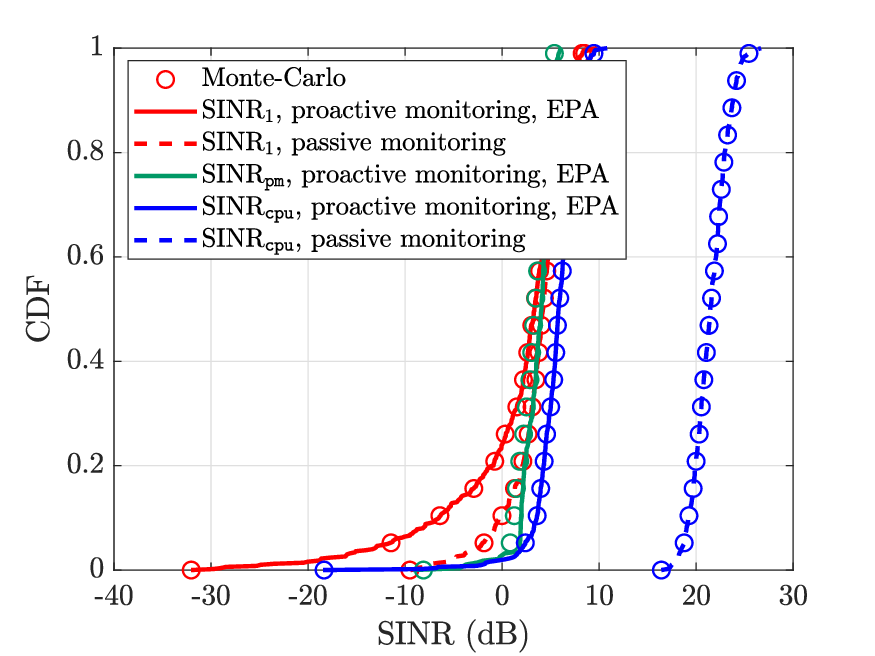}
  \caption{ CDF of the anti-malicious system, $P_\mathtt{pm}=3W$, $N_\mathtt{pm}=32$, $h=500$~m, $r=300$~m.}
  \label{fig:CDF_5_nojam}
\end{figure}

We first consider proactive monitoring with constant-power jamming, where the monitor uses an  EPA  scheme with $\theta_{\mathtt{pm},\mathtt{t}} = \theta_{\mathtt{pm},1} = \frac{1}{2}$. 
To verify the accuracy of the closed-form expressions presented in Propositions \ref{Theorem1}-\ref{Theorem3}, we compare their cumulative distribution functions (CDFs) with Monte-Carlo simulation results, where the simulations are averaged over $500$ random channel realizations. We also show the CDFs for passive monitoring scheme. Our numerical results in Fig. \ref{fig:CDF_5_nojam} lead to the
following conclusions: i) the analytical results (solid and dash curves) match tightly with the simulation results (markers); ii) the proactive monitor can effectively degrade both $\mathrm{SINR}_\mathtt{cpu}$ and $\mathrm{SINR}_1$ at the malicious ISAC system by transmitting jamming signals.

\begin{figure}[t]
  \centering
  \includegraphics[width=0.9\linewidth,height=0.65\linewidth]{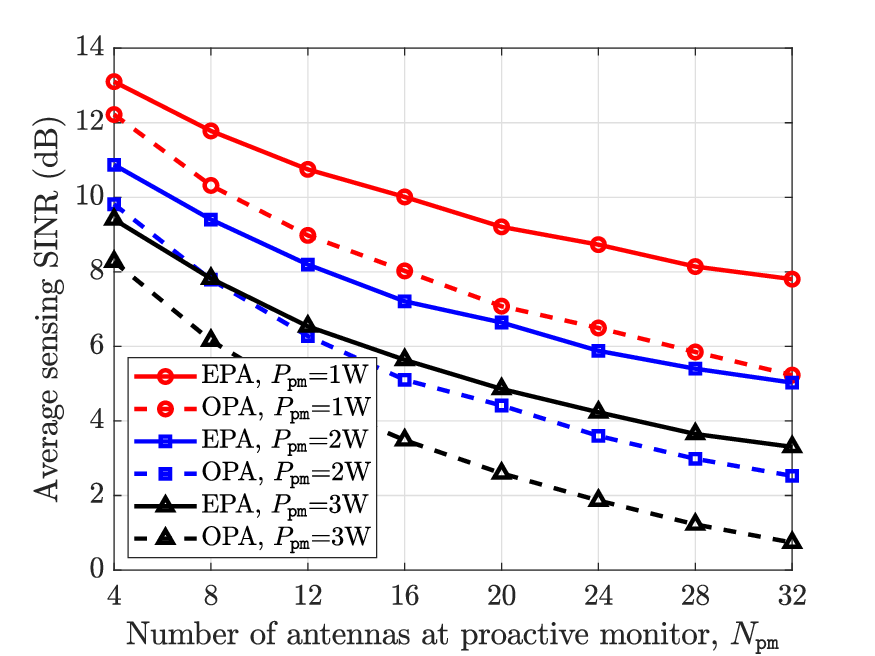}
  \caption{$\mathrm{SINR}_\mathtt{cpu}$ versus $N_\mathtt{pm}$, where $h=500$~m, $r=300$~m}
  \label{fig:SINR_Nj}
\end{figure}

Figure~\ref{fig:SINR_Nj} compares the performance of proactive monitoring using the optimized power allocation (OPA) scheme $\mathbf{(P_{1})}$, which minimizes the malicious sensing $\mathrm{SINR}_\mathtt{cpu}$, with that of constant-power jamming using  EPA.
 The comparison is evaluated against the number of antennas at the proactive monitor, with the average sensing SINR computed over $1,000$ random channel realizations. The results demonstrate that the proactive monitor's performance significantly improves with an increased number of antennas and higher jamming power. Specifically, the optimized power allocation scheme consistently degrades the sensing performance of the malicious ISAC system more effectively than the baseline scheme. For instance, with $N_{\mathtt{pm}}=32$, the monitor achieves an approximate $3$ dB reduction in the average $\mathrm{SINR}_\mathtt{cpu}$, highlighting the advantage of employing a larger antenna array at the monitor.
\begin{figure}[t]
  \centering
  \includegraphics[width=0.9\linewidth,height=0.65\linewidth]{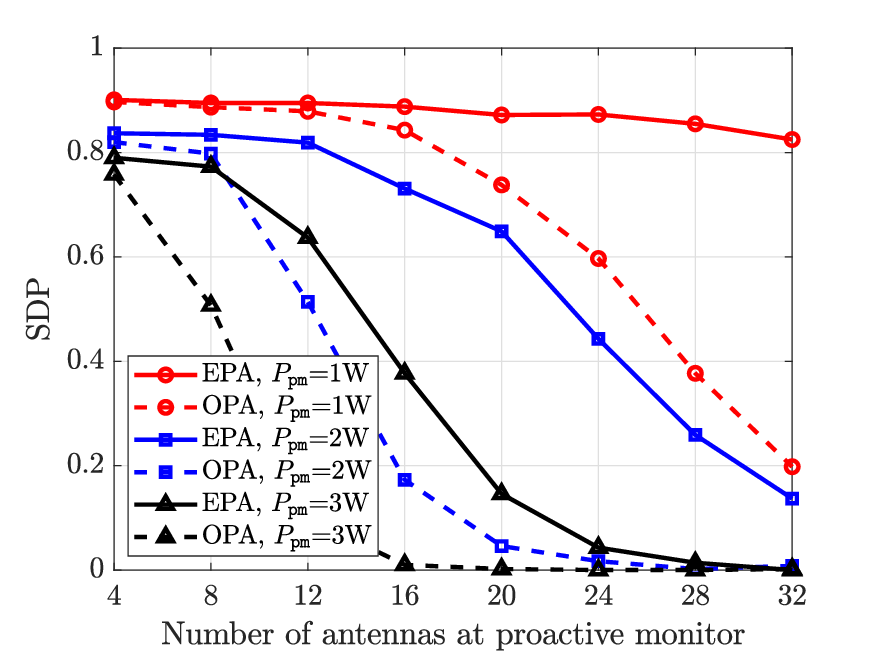}
  \caption{SDP versus $N_\mathtt{pm}$, where $\kappa=8$~dB, $h=500$~m, $r=300$~m.}
  \label{fig:SDP_Nj}
\end{figure}

Figure~\ref{fig:SDP_Nj} depicts the SDP as a function of the number of monitor antennas, evaluated under varying jamming power levels and a fixed detection threshold of $\kappa = 8$ dB. For scenarios with lower jamming power (e.g., $P_\mathtt{pm} = 1$ W), increasing the number of monitor antennas results in only a slight reduction in the SDP; for instance, an increase from $4$ to $32$ antennas yields just a $3\%$ improvement. However, when our proposed optimization approach is employed for power allocation, the SDP decreases significantly—by nearly $70\%$—with $32$ antennas. 
In contrast, under higher jamming power scenarios, the optimization algorithm demonstrates remarkable efficiency. With just $16$ antennas, the SDP is reduced from $40\%$ to below $5\%$, showcasing the capability of the optimization approach to achieve substantial performance improvements with minimal hardware resources. This demonstrates the adaptability and efficiency of our method in balancing performance, hardware constraints, and energy efficiency.

\begin{figure}[t]
  \centering
  \includegraphics[width=0.9\linewidth,height=0.7\linewidth]{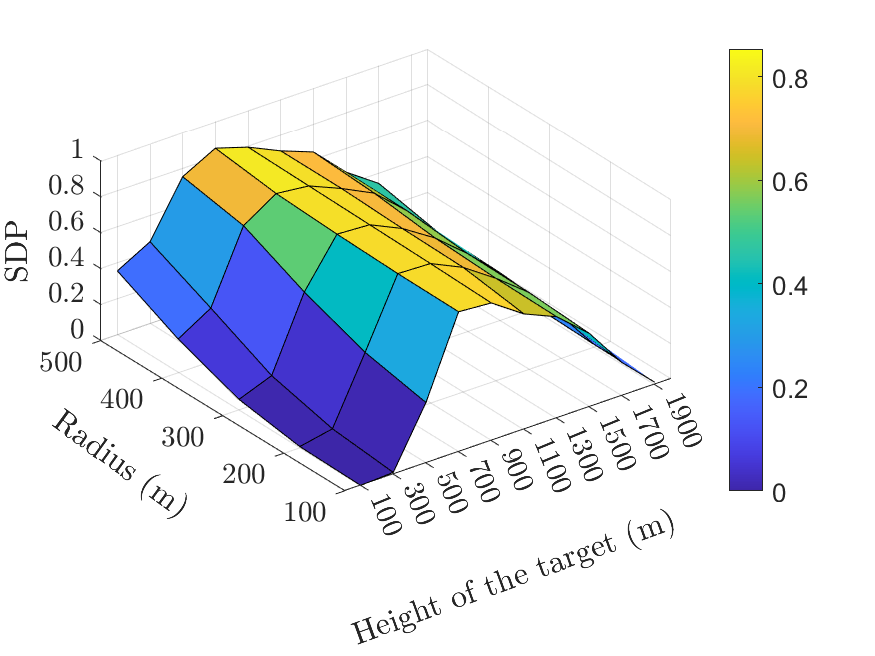}
  \caption{SDP versus $h$ and $r$, where $\kappa=8$~dB, $P_\mathtt{pm}=1$ W, $N_\mathtt{pm}=32$.}
  \label{fig:SDP_h_radius}
\end{figure}
In Fig. \ref{fig:SDP_h_radius}, we analyze the performance of the proactive monitor with an optimized power allocation scheme under varying height and radius of the target location. As shown, increasing the radius directly affects the monitor's effectiveness. This is because a larger distance leads to greater path loss, weakening the impact of AN in the jamming signal. Consequently, as the distance increases, the likelihood of the target being exposed to the malicious ISAC system also increases. Additionally, we evaluate the performance under different target height conditions. On  one hand, increasing the target's height reduces the power of the desired signal received by the malicious ISAC, thereby decreasing the sensing SINR. On the other hand, as the height increases, the reflected jamming signal weakens, leading to reduced interference and a higher sensing SINR at the malicious ISAC. As a result, in the height range of 100-700 m, the jamming signal's interference has a more dominant effect on $\mathrm{SINR}_\mathtt{cpu}$. However, once the target's height exceeds 700 m, there is a noticeable decrease in SDP. This reduction occurs because the reflected sensing signals of the malicious ISAC become significantly weaker due to increased propagation losses. Consequently, the reduced signal levels lead to a noticeable drop in the malicious ISAC system's ability to sense the target. This relationship between target height and SDP highlights the optimal operational range of the proactive monitor’s jamming strategy.

\begin{figure}[t]
  \centering
  \includegraphics[width=0.9\linewidth,height=0.65\linewidth]{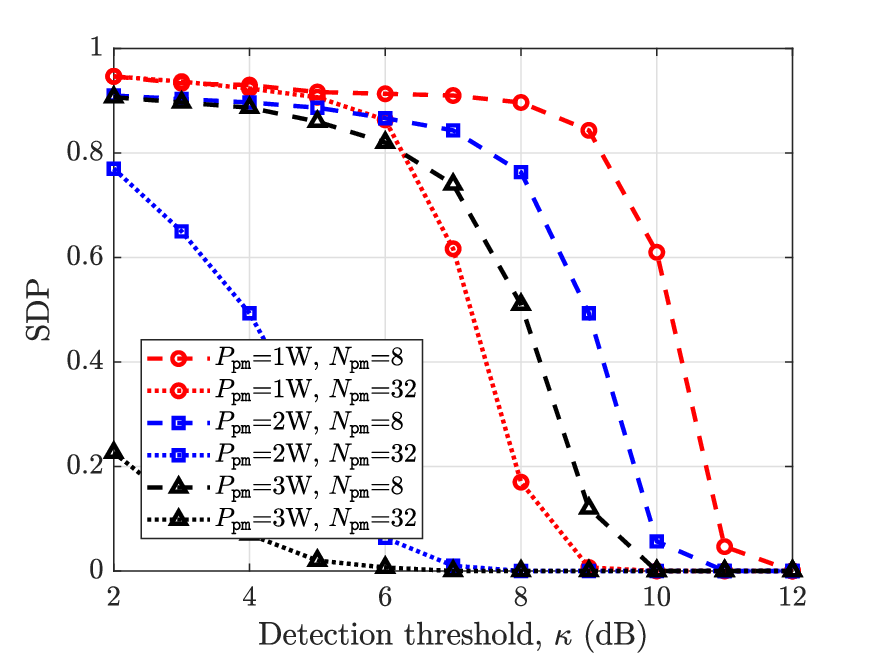}
  \caption{SDP versus the detection threshold, $\kappa$.}
  \label{fig:SDP_kappa}
\end{figure}

The SDP at the malicious ISAC receiver is shown in Fig. \ref{fig:SDP_kappa} as a function of the detection threshold under optimal power allocation. First, it is readily seen that SDP decreases as the detection threshold $\kappa$ increases. Additionally, higher values of jamming power and number of antennas speed up the point at which SDP drops to zero. Notably, across all detection threshold values, increasing the number of antennas achieves a more notable effect than increasing jamming power. For instance, at $\kappa = 8$ dB, raising the jamming power from 1W to 2W and 3W results in a 10\% and 42\% decrease in the SDP, respectively, while increasing the number of antennas reduces SDP by more than 75\%. The effect is especially notable in low $\kappa$ cases with higher number of antennas: at $\kappa = 2$ dB and $N_\mathtt{pm} = 8$, increasing jamming power from 2W to 3W reduces SDP by less than 10\%; however, with $N_\mathtt{pm} = 32$, this increase in jamming power yields an SDP reduction of more than 70\%.

\begin{figure}[t]
  \centering
  \includegraphics[width=0.9\linewidth,height=0.7\linewidth]{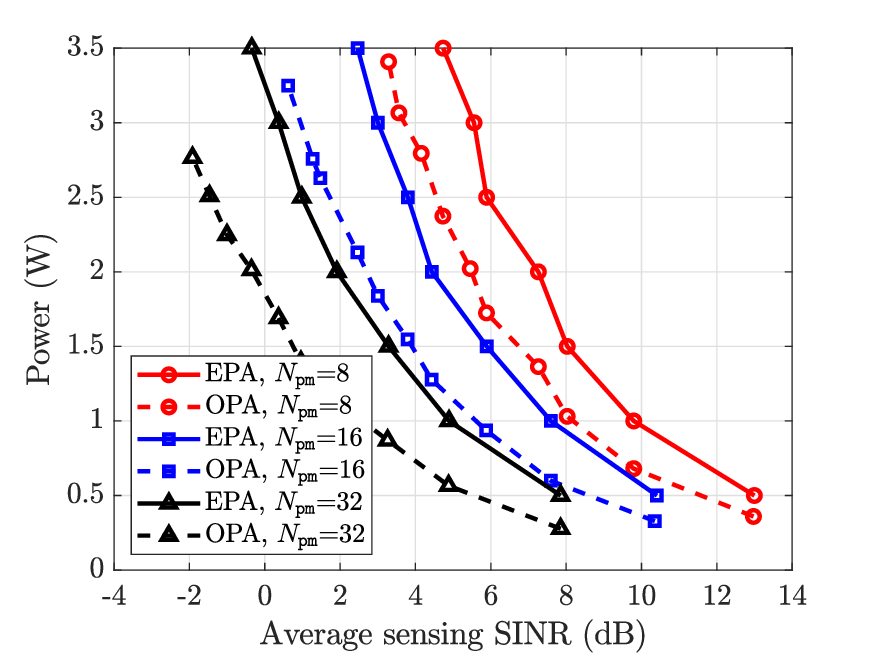}
  \caption{Total transmit power (W) versus $\mathrm{SINR}_\mathtt{cpu}$ (dB).}
  \label{fig:minpow_Npm}
\end{figure}

Figure~\ref{fig:minpow_Npm} compares the  monitor power consumption for proactive monitoring using EPA scheme and  proactive monitoring using the OPA scheme  $\mathbf{P_{2}}$   versus the average sensing $\mathrm{SINR}_{\mathtt{cpu}}$ under varying numbers of monitor antennas.  The average sensing $\mathrm{SINR}_{\mathtt{cpu}}$ is calculated by averaging over $1,000$ network realizations. For each realization, the threshold $\kappa$ is equal to the value of $\mathrm{SINR}_{\mathtt{cpu}}$ under equal power allocation. It is observed that the OPA results in an average jamming power savings of $43.6\%$ compared to the EPA scheme when $N_\mathtt{pm}=32$, while ensuring successful monitoring performance. Similarly, for $8$ and $16$ antennas the average power savings are $30.8\%$ and $37.1\%$, respectively. These results demonstrate the efficiency of our optimization approach ($\mathbf{P_{2}}$), which intelligently distributes power to minimize the overall consumption while maintaining the required monitoring and target detection performance. The significant reduction in the power consumption highlights the potential for extended operational lifetimes and reduced exposure risks for the proactive monitor in practical surveillance scenarios.

\section{Conclusion}
This paper introduced a proactive monitor for intercepting a suspicious communication link and protecting a legitimate target in a malicious CF-mMIMO ISAC system. We derived closed-form expressions for the effective SINR at the malicious UE, sensing SINR at the malicious ISAC, and the approximation of SINR at the proactive monitor. Two optimization problems were formulated: the first aimed to minimize the sensing SINR by optimizing jamming power allocation to the target and malicious UE, leading to a significant reduction in the SDP. The second focused on minimizing the proactive monitor’s power consumption while maintaining successful monitoring and sensing SINR degradation. Numerical results confirmed that 1) the proposed monitoring approach significantly degrades the sensing and communication performance of the malicious ISAC system; 2) the second power allocation scheme reduced the jamming power consumption by more than one-third, while maintaining comparable monitoring performance. The results also highlighted the impact of factors such as antenna count, jamming power, target height, monitoring radius, and detection threshold on the system performance.

% or


\appendices

\section{Proof of Proposition 1}~\label{ProofTheorem1}
1) Compute $\mathrm {DS}_{k}$:
\begin{align}
    \mathrm {DS}_{k}  &= \sum\nolimits_{m \in\mathcal{M}_{\mathtt{c}} } \sqrt{\eta _{m,k}\rho _{\mathtt{c}}} \mathbb{E}\left \{(\hat{\mathbf {g}}_{m,k}+\tilde{\mathbf {g}}_{m,k}  )^{T}\hat{\mathbf {g}}_{m,k} ^{\ast } \right \} \notag\\
    &= \sum\nolimits_{m \in\mathcal{M}_{\mathtt{c}} }\sqrt{\eta _{m,k}\rho _{\mathtt{c}}}N\gamma _{m,k}.
\label{Eq:DS_k}
\end{align}

2) Compute $\mathbb{E}\left \{ |\mathrm{BU}_{k}|^{2} \right \}$:
\begin{align}
        &\mathbb{E}\left \{ |\mathrm{BU}_{k}|^{2} \right \} 
        \!=\! \!\sum_{m \in\mathcal{M}_{\mathtt{c}} }\!\!\eta _{m,k}\rho _{\mathtt{c}} \mathbb{E}\! \left \{  |{\mathbf {g}}_{m,k}^{T}\hat{\mathbf {g}}_{m,k} ^{\ast }\!-\!\mathbb{E} \!\left \{ \mathbf {g}_{m,k}^{T} \hat{\mathbf {g}}_{m,k} ^{\ast }\!\right \} \! |^{2} \right \}\! \notag\\
        &=\rho _{\mathtt{c}}\!\sum_{m \in\mathcal{M}_{\mathtt{c}} }\!\!\eta _{m,k}\bigg(\mathbb{E} \left \{  |{\mathbf {g}}_{m,k}^{T}\hat{\mathbf {g}}_{m,k} ^{\ast } |^{2} \right \}\!-\!|\mathbb{E} \left \{ \mathbf {g}_{m,k}^{T} \hat{\mathbf {g}}_{m,k} ^{\ast }\right \}|^{2}\bigg)\notag\\
        &=\rho _{\mathtt{c}}\!\sum\nolimits_{m \in\mathcal{M}_{\mathtt{c}} }\!\!\eta _{m,k}\bigg (\mathbb{E} \left \{  ||{\hat{\mathbf {g}}_{m,k}}||^4 \right \} \!+\!\tilde{\mathbf {g}}_{m,k}^{T}\hat{\mathbf {g}}_{m,k} ^{\ast }\!-\!N^{2}\gamma_{m,k}^2 \bigg  )\notag\\
        &\!\overset{(a)}{=}\rho _{\mathtt{c}}N \sum\nolimits_{m \in\mathcal{M}_{\mathtt{c}} }\eta _{m,k}\gamma _{m,k}\beta_{m,k},
    \label{Eq:BU_k}
\end{align}
where $(a)$ follows from the fact that $\mathbb{E}\big \{ \left \| \hat{\mathbf{g}}_{m,k} \right \| ^4 \big \}=N(N+1)\gamma_{m,k}^2$ and $\tilde{\mathbf {g}}_{m,k}^{T}\hat{\mathbf {g}}_{m,k} ^{\ast }=\mathbf (\mathbf {g}_{m,k}^{T}-\hat{\mathbf {g}}_{m,k})^{T}\hat{\mathbf {g}}_{m,k} ^{\ast }=N(\beta_{m,k}-\gamma _{m,k})\gamma _{m,k}$.

3) Compute $\mathbb{E}\left \{ |\mathrm{IU}_{k',k}|^{2} \right \} $:
\begin{align}
        \mathbb{E}\left \{ |\mathrm{IU}_{k',k}|^{2} \right \} 
        &= \sum\nolimits_{m \in\mathcal{M}_{\mathtt{c}} }\!\eta _{m,k'}\rho _{\mathtt{c}} \mathbb{E} \left \{  | \mathbf{g}_{m,k}^{T}\hat{\mathbf{g}}_{m,k'}^{\ast} |^{2} \right \}  \notag \\
        &= \sum\nolimits_{m \in\mathcal{M}_{\mathtt{c}} }\eta _{m,k'}\rho _{\mathtt{c}}N\gamma_{m,k'}\beta_{m,k}.
    \label{Eq:IU_k'k}
\end{align}

4) Compute $\mathbb{E}\left \{ |\mathrm{IS}_{k}|^{2} \right \}$:
\begin{align}\label{Eq:IS_k}
    &\mathbb{E}\left \{ |\mathrm{IS}_{k}|^{2} \right \} =\mathbb{E}\Big \{ \big|\sum\nolimits_{m'\in\mathcal{M}_{\mathtt{s},\mathtt{t}} }\sqrt{\eta _{m'\!,\mathtt{t}}\rho _{\mathtt{s}}}\mathbf{g}_{m',k}^{T}\mathbf{h}_{m'\!,\mathtt{t}}\big|^2 \Big \}\notag\\
    &\hspace{0.4cm}+\mathbb{E}\Big \{ \big|\sum\nolimits_{m'\in\mathcal{M}_{\mathtt{s},\mathtt{t}} }\sqrt{\eta _{m'\!,\mathtt{t}}\rho _{\mathtt{s}}}\sqrt{\alpha}h_{\mathtt{t},k}N\zeta_{m'\!,\mathtt{t}}\big|^2 \Big \}\notag\\
    &\overset{(b)}{=}\!\sum\nolimits_{m'\in\mathcal{M}_{\mathtt{s},\mathtt{t}} }\!\eta _{m'\!,\mathtt{t}}\rho _{\mathtt{s}}\beta_{m',k} \mathrm {Tr} (   \mathbf{h}_{m'\!,\mathtt{t}}\mathbf{h}_{m'\!,\mathtt{t}}^{H}))\notag\\
    &\hspace{0.4cm}+ \sum\nolimits_{m'\in\mathcal{M}_{\mathtt{s},\mathtt{t}} }\eta _{m'\!,\mathtt{t}}\rho _{\mathtt{s}}\alpha\mathbb{E}\left \{ | h_{\mathtt{t},k}N\zeta_{m'\!,\mathtt{t}}|^2 \right \} +\rho _{\mathtt{s}}\alpha\zeta_{\mathtt{t},k} N^2\notag\\
    &\hspace{0.4cm}\times\!\sum\nolimits_{m'\in\mathcal{M}_{\mathtt{s},\mathtt{t}} }\!\sqrt{\eta _{m'\!,\mathtt{t}}}\zeta_{m'\!,\mathtt{t}}\sum\nolimits_{\tilde{m}'\in\mathcal{M}_{\mathtt{s},\mathtt{t}}, \tilde{m}' \ne m' }\sqrt{\eta _{\tilde{m}'\!,t}}\zeta_{\tilde{m}',t}\notag\\
    &=\sum\nolimits_{m'\in\mathcal{M}_{\mathtt{s},\mathtt{t}} }\sqrt{\eta _{m'\!,\mathtt{t}}}\rho _{\mathtt{s}}\zeta_{m'\!,\mathtt{t}}N(\sqrt{\eta _{m'\!,\mathtt{t}}}\alpha\zeta_{m'\!,\mathtt{t}}\zeta_{\mathtt{t},k}N\notag\\
    &\hspace{0.4cm}+\sqrt{\eta _{m'\!,\mathtt{t}}}\beta_{m',k}+\sum\nolimits_{\tilde{m}'\in\mathcal{M}_{\mathtt{s},\mathtt{t}}, \tilde{m}' \ne m' }\sqrt{\eta _{\tilde{m}'\!,t}}\zeta_{\tilde{m}',t}\zeta_{\mathtt{t},k}\alpha N),
\end{align}
where in $(b)$  we have used  \cite[Lemma 7] {ref:Zhikangda_TIT}. That is 
\begin{align}
    \mathbb{E} \left \{ \mathbf{X}\mathbf{W}\mathbf{X}^{H} \right \} =v_x\mathrm{Tr}(\mathbf{W})\mathbf{I}_{M},
\end{align}
where $\mathbf{W}\in \mathbb{C} ^{N\times N}$
is a deterministic matrix, while $\mathbf{X}\in \mathbb{C} ^{M\times N}$, whose
entries are i.i.d. random variables with zero mean  and $v_x$ variance.

5) Compute $\mathbb{E}\left \{ |\mathrm {JS}_{\mathtt{s},k}|^2 \right\}$:
\begin{align}\label{Eq:JS_sk}
    &\mathbb{E}\left \{ |\mathrm {JS}_{\mathtt{s},k}|^2 \right\} = \eta _{\mathtt{pm},\mathtt{t}}\rho _{\mathtt{pm}} \mathbb{E}\left \{|(\mathbf{g}_{\mathtt{pm},k}^{T}+\sqrt{\alpha}\mathbf{h}_{\mathtt{t},k}\mathbf{h}_{\mathtt{pm},\mathtt{t}}^{T})\mathbf {h}_{\mathtt{pm},\mathtt{t}} ^{\ast }|^{2} \right \} \notag\\
   %& = \sqrt{\eta _{m,1}\rho _{\mathtt{c}}}N_{\mathtt{pm}}\beta_{\mathtt{pm},1}
   &=\eta _{\mathtt{pm},\mathtt{t}}\rho _{\mathtt{pm}} \zeta_{\mathtt{pm},\mathtt{t}}N_{\mathtt{pm}}(\beta_{\mathtt{pm},k}+\alpha\zeta_{\mathtt{t},k}N_{\mathtt{pm}}\zeta_{\mathtt{pm},\mathtt{t}}).
\end{align}

6) Compute $\mathbb{E}\!\left \{ |\mathrm {JS}_{\mathtt{c},k}|^2 \!\right\}$: For $k=1$, following the same process as in \eqref{Eq:BU_k} and using the fact that $\mathbb{E}\big \{ \big \| \mathbf{g}_{\mathtt{pm},1} \big \| ^4 \big \}=\beta_{\mathtt{pm},1}^2 N_{\mathtt{pm}}(N_{\mathtt{pm}}\!+\!1)$, $\mathbb{E}\!\left \{ |\mathrm {JS}_{\mathtt{c},1}|^2 \!\right\} $ can be obtained as:
\begin{align}\label{Eq:JS_ck}
        &\!\mathbb{E}\!\left \{ |\mathrm {JS}_{\mathtt{c},1}|^2 \!\right\} 
       \!=\!\eta _{\mathtt{pm},1}\rho _{\mathtt{pm}}\mathbb{E}\!\left \{\!|(\mathbf{g}_{\mathtt{pm},1} ^{T}+\sqrt{\alpha}\mathbf{h}_{t,1}\mathbf{h}_{\mathtt{pm},\mathtt{t}}^{T})\mathbf{w}_{\mathtt{pm},1}^{\mathtt{c}}|^{2} \right \}\notag\\
       &=\!\eta _{\mathtt{pm},1}\rho _{\mathtt{pm}}\mathbb{E}\left \{ \left \| \mathbf{g}_{\mathtt{pm},1} \right \| ^4 \right \} \!+\!\eta _{\mathtt{pm},1}\rho _{\mathtt{pm}}\alpha\beta_{\mathtt{pm},1}\zeta_{t,1}N_{\mathtt{pm}}\zeta_{\mathtt{pm},\mathtt{t}}\notag\\
       &=\eta _{\mathtt{pm},1}\rho _{\mathtt{pm}}\beta_{\mathtt{pm},1}N_{\mathtt{pm}}((N_{\mathtt{pm}}\!+\!1)\beta_{\mathtt{pm},1}+\alpha\zeta_{t,1}\zeta_{\mathtt{pm},\mathtt{t}}).
\end{align}
Moreover, for $k\ne1$, $\mathbb{E}\!\left \{ |\mathrm {JS}_{\mathtt{c},k}|^2 \!\right\}$ can be obtained as:
\begin{align}\label{Eq:JS_ckno1}
        \!\mathbb{E}\!\left \{ |\mathrm {JS}_{\mathtt{c},k}|^2 \!\right\} 
        &=\!\eta _{\mathtt{pm},1}\rho _{\mathtt{pm}}\mathbb{E}\!\left \{\!|(\mathbf{g}_{\mathtt{pm},k} ^{T}+\sqrt{\alpha}\mathbf{h}_{\mathtt{t},k}\mathbf{h}_{\mathtt{pm},\mathtt{t}}^{T})\mathbf {g}_{\mathtt{pm},1} ^{\ast }|^{2} \right \}\notag\\
        &=\!\eta _{\mathtt{pm},1}\rho _{\mathtt{pm}}\beta _{\mathtt{pm},1}N_{\mathtt{pm}}(\beta _{\mathtt{pm},k}\!+\!\alpha\zeta_{\mathtt{t},k}\zeta_{\mathtt{pm},\mathtt{t}}).
\end{align}

\section{Proof of Proposition 2}\label{ProofTheorem2}
1) Compute $\mathrm {DS}_{\mathtt{pm}}$:
\begin{align}\label{eq:DS_J}
   \mathrm {DS}_{\mathtt{pm}}&= \sum\nolimits_{m \in\mathcal{M}_{\mathtt{c}} } \sqrt{\eta _{m,1}\rho _{\mathtt{c}}} \mathbb{E}\left \{\mathbf{w}_{\mathtt{comb},\mathtt{pm}}^{T}\mathbf{G}_{m,\mathtt{pm}}^{T}\mathbf {w} _{m,1}^{\mathtt{c}} \right \} \notag\\
    %&=\sum\nolimits_{m \in\mathcal{M}_{\mathtt{c}} } \sqrt{\eta _{m,1}\rho _{\mathtt{c}}} \mathbb{E}\left \{\mathbf{w}_{\mathtt{comb},\mathtt{pm}}^{T}\mathbf{G}_{m,\mathtt{pm}}^{T}\mathbf {w} _{m,1}^{\mathtt{c}} \right \}, \notag\\
    &=\sum\nolimits_{m \in\mathcal{M}_{\mathtt{c}} } \eta _{m,1}\rho _{\mathtt{c}}\mathbb{E}\left \{\hat{\mathbf{g}}_{m,1}^{T}\mathbf{G}_{m,\mathtt{pm}}^{\ast}\mathbf{G}_{m,\mathtt{pm}}^{T}\hat{\mathbf{g}}_{m,1}^{\ast} \right \}\notag\\
    &=\sum\nolimits_{m \in\mathcal{M}_{\mathtt{c}} } \eta _{m,1}\rho _{\mathtt{c}}N_{\mathtt{pm}}\beta_{m,\mathtt{pm}}N\gamma_{m,1}.
\end{align}

2) To compute $\mathbb{E}\left \{ |\mathrm{BU}_{\mathtt{pm}}|^{2} \right \}$, we first notice that for each $m\in\mathcal{M}_{\mathtt{c}}$, the 
product of the matrix $\mathbf{G}_{m,\mathtt{pm}}^{T}$ and the vector $\hat{\mathbf{g} }_{m,1} ^{\ast }$ results in a vector $\mathbf{y}^{(m)}\in \mathbb{C}^{N_{\mathtt{pm}}\times 1}$. For $i$-th entry $y_{i}^{(m)}$ in the vector $\mathbf{y}^{(m)}$, it can be shown that the entry can be approximated as zero-mean, with variance $\beta_{m,\mathtt{pm}}\gamma_{m,1}N$. We then proceed to compute $\mathbb{E}\left \{ |\mathrm{BU}_{\mathtt{pm}}|^{2} \right \}$:
\begin{align}
\label{eq:BU_J}
    \mathbb{E}\left \{ |\mathrm{BU}_{\mathtt{pm}}|^{2} \right \}\!
    &=\!\!\mathbb{E}\Big \{ \!\Big |\mathbf{w}_{\mathtt{comb},\mathtt{pm}}^{T}\!\sum\nolimits_{m \in\mathcal{M}_{\mathtt{c}} } \!\!\sqrt{\eta _{m,1}\rho _{\mathtt{c}}}\mathbf{G}_{m,\mathtt{pm}}^{T}\mathbf {w} _{m,1}^{\mathtt{c}}\Big|^2\!\Big \} \notag\\
    &\hspace{0.1cm}-\!\Big |  \mathbb{E}\Big \{ \!\mathbf{w}_{\mathtt{comb},\mathtt{pm}}^{T}\!\sum\nolimits_{m \in\mathcal{M}_{\mathtt{c}} }\!\!\sqrt{\eta _{m,1}\rho _{\mathtt{c}}}\mathbf{G}_{m,\mathtt{pm}}^{T}\mathbf {w} _{m,1}^{\mathtt{c}}\!\Big \}\Big |^2 \notag\\
    %&= \mathbb{E}\bigg \{ \bigg|\bigg(\sum\nolimits_{m \in\mathcal{M}_{\mathtt{c}} }\sqrt{\eta _{m,1}\rho _{\mathtt{c}}}\mathbf{G}_{m,\mathtt{pm}}^{T}\hat{\mathbf{g} }_{m,1} ^{\ast }\bigg) ^{H }\notag\\
    %&\bigg(\sum\nolimits_{m \in\mathcal{M}_{\mathtt{c}} }\sqrt{\rho _{\mathtt{c}}}\sqrt{\eta _{m,1}}\!\mathbf{G}_{m,\mathtt{pm}}^{T}\hat{\mathbf{g} }_{m,1} ^{\ast }\bigg)\bigg|^2\bigg \}-\left | \sum\nolimits_{m \in\mathcal{M}_{\mathtt{c}} }\!\!\eta _{m,1}\rho _{\mathtt{c}}N_{\mathtt{pm}}\beta_{m,\mathtt{pm}}N\gamma_{m,1}\right |^2, \notag\\
    &= \mathbb{E}\Big \{ \Big\| \sum\nolimits_{m \in\mathcal{M}_{\mathtt{c}} }\sqrt{\eta _{m,1}\rho _{\mathtt{c}}}\mathbf{G}_{m,\mathtt{pm}}^{T}\hat{\mathbf{g} }_{m,1} ^{\ast }\Big\|^4\Big \}\notag\\
    &\hspace{0.1cm}-\Big | \sum\nolimits_{m \in\mathcal{M}_{\mathtt{c}} } \!\eta _{m,1}\rho _{\mathtt{c}}N_{\mathtt{pm}}\beta_{m,\mathtt{pm}}N\gamma_{m,1}\Big |^2.
\end{align}
%========
The product $\mathbf{G}_{m,\mathtt{pm}}^{T}\hat{\mathbf{g} }_{m,1} ^{\ast }$ in~\eqref{eq:BU_J} does not inherently follow a Gaussian distribution. However,  we approximate the above term as Gaussian for analytical tractability. This approximation is motivated by the central limit theorem. That is when the number of communication APs, $\mathcal{M}_{\mathtt{c}}$,  is large, the sum of many independent random variables tends to follow a Gaussian distribution. In CF-mMIMO systems, where multiple APs contribute to the received signal, the aggregation of these terms behaves approximately as Gaussian, especially when the APs' channel estimates are uncorrelated. Furthermore, our simulation results validate this assumption, showing that the approximation is highly accurate, with an error of less than $3\%$. With a Gaussian approximation, we obtain
%======
\begin{align}
    \!\mathbb{E}\left \{ |\mathrm{BU}_{\mathtt{pm}}|^{2} \right \}\!&
       \!\approx\!\bigg(\!\sum\nolimits_{m \in\mathcal{M}_{\mathtt{c}} }\!\eta _{m,1}\rho _{\mathtt{c}}\beta_{m,\mathtt{pm}}\gamma_{m,1}N\!\bigg)^2\!\!N_{\mathtt{pm}}\!(N_{\mathtt{pm}}\!+\!1)\notag\\
    &\hspace{0.4cm}-\!\Big (\sum\nolimits_{m \in\mathcal{M}_{\mathtt{c}} } \eta _{m,1}\rho _{\mathtt{c}}N_{\mathtt{pm}}\beta_{m,\mathtt{pm}}N\gamma_{m,1}\Big )^2\notag\\
    &=\!\bigg(\sum\nolimits_{m \in\mathcal{M}_{\mathtt{c}} }\!\eta _{m,1}\rho _{\mathtt{c}}\beta_{m,\mathtt{pm}}\gamma_{m,1}N\bigg)^2N_{\mathtt{pm}}.
\end{align}

%======
3) Compute $\mathbb{E}\left \{ |\mathrm{IU}_{{k}',\mathtt{pm}}|^{2} \right \}$: We have
\begin{align}\label{eq:IC_k',J}
    &\mathbb{E}\big \{\! |\mathrm{IU}_{{k}',\mathtt{pm}}|^{2} \!\big \}
    %&\!=\!\mathbb{E}\bigg \{\bigg | \sum\nolimits_{m \in\mathcal{M}_{\mathtt{c}} }\!\sqrt{\eta _{m,1}\rho _{\mathtt{c}}}\hat{\mathbf{g} }_{m,1}^{T}\mathbf{G}_{m,\mathtt{pm}}^{*}\sum\nolimits_{m \in\mathcal{M}_{\mathtt{c}} }\!\sqrt{\eta _{m,{k'}}\!\rho _{\mathtt{c}}}\mathbf{G}_{m,\mathtt{pm}}^{T}\hat{\mathbf{g} }_{m,k'} ^{\ast } \bigg |^{2} \bigg \} \notag\\
    =\mathbb{E}\bigg \{\bigg | \sum\nolimits_{m \in\mathcal{M}_{\mathtt{c}} }\rho _{\mathtt{c}}\sqrt{\eta _{m,1}\eta _{m,{k'}}}\hat{\mathbf{g} }_{m,1}^{T}\mathbf{G}_{m,\mathtt{pm}}^{*} \notag\\
    &\quad\times \mathbf{G}_{m,\mathtt{pm}}^{T}\hat{\mathbf{g} }_{m,k'} ^{\ast }+ \sum\nolimits_{\tilde{m}\in\mathcal{M}_{\mathtt{c}}, \tilde{m}\ne m}\!\sum\nolimits_{m \in\mathcal{M}_{\mathtt{c}} }\!\rho _{\mathtt{c}}\sqrt{\eta _{\tilde{m},1}\eta _{m,{k'}}}\notag\\
    &\quad\times\hat{\mathbf{g} }_{\tilde{m},1}^{T}\mathbf{G}_{\tilde{m},\mathtt{pm}}^{*}\!\mathbf{G}_{m,\mathtt{pm}}^{T}\hat{\mathbf{g} }_{m,k'} ^{\ast } \bigg |^{2} \bigg \}\notag\\
    &\approx \mathbb{E}\bigg \{\bigg | \sum\nolimits_{m \in\mathcal{M}_{\mathtt{c}} }\rho _{\mathtt{c}}\sqrt{\eta _{m,1}\eta _{m,{k'}}}\hat{\mathbf{g} }_{m,1}^{T}\mathbf{G}_{m,\mathtt{pm}}^{*}\mathbf{G}_{m,\mathtt{pm}}^{T}\hat{\mathbf{g} }_{m,k'} ^{\ast } \bigg |^{2} \bigg \}\notag\\
    &\quad+\mathbb{E}\bigg \{\bigg | \sum\nolimits_{\tilde{m}\in\mathcal{M}_{\mathtt{c}}, \tilde{m} \ne m}\!\sum\nolimits_{m \in\mathcal{M}_{\mathtt{c}} }\!\rho _{\mathtt{c}}\sqrt{\eta _{\tilde{m},1}\eta _{m,{k'}}}\hat{\mathbf{g} }_{\tilde{m},1}^{T}\mathbf{G}_{\tilde{m},\mathtt{pm}}^{*}\notag\\
    &\quad\times\mathbf{G}_{m,\mathtt{pm}}^{T}\hat{\mathbf{g} }_{m,k'} ^{\ast } \bigg |^{2} \bigg \},
\end{align}
where the approximations applied here align with fundamental properties of large-scale communication systems. Specifically, the aggregate impact of off-diagonal terms becomes negligible compared to the on-diagonal terms as the system dimension grows, particularly under independent fading conditions. Our simulation results validate this approximation, showing an error of less than $3\%$, confirming its high accuracy in the evaluated scenarios.
Eq.~\eqref{eq:IC_k',J} can be further expressed as
\begin{align}\label{eq:IC_k',J2}
    &\mathbb{E}\big \{\! |\mathrm{IU}_{{k}',\mathtt{pm}}|^{2} \!\big \}\nonumber\\
    &=\sum\nolimits_{m \in\mathcal{M}_{\mathtt{c}} }\!\eta _{m,1}\eta _{m,{k'}}\rho _{\mathtt{c}}^2\gamma_{m,k'}\mathbb{E}\big \{ \hat{\mathbf{g} }_{m,1}^{T}\mathbb{E}\big \{ \big \|\mathbf{G}_{m,\mathtt{pm}}^{\ast }\big \|^4 \big \}\hat{\mathbf{g} }_{m,1}^{\ast } \big \} \notag\\
    &\quad+\!\sum\nolimits_{m \in\mathcal{M}_{\mathtt{c}} }\!\sum\nolimits_{\tilde{m} \ne m,\tilde{m} \in\mathcal{M}_{\mathtt{c}}}\eta _{m,{k'}}\eta _{\tilde{m},1}\rho _{\mathtt{c}}^2N\gamma_{m,k'}\beta_{m,\mathtt{pm}}\notag\\
    &\quad\times\mathbb{E}\left \{\left | \hat{\mathbf{g} }_{\tilde{m},1}^{T}\mathbf{G}_{\tilde{m},\mathtt{pm}}^{*} \right |^{2} \right \}\notag\\
    &\overset{(c)}= \!\sum\nolimits_{m \in\mathcal{M}_{\mathtt{c}} }\!\eta _{m,1}\eta _{m,{k'}}\rho _{\mathtt{c}}^2\gamma_{m,k'}\mathbb{E}\big \{ \hat{\mathbf{g} }_{m,1}^{T}\beta_{m,\mathtt{pm}}^2N_{\mathtt{pm}}(N\!+\!N_{\mathtt{pm}})\notag\\
    &\quad\times\mathbf{I}_{N}\hat{\mathbf{g} }_{m,1}^{\ast } \big \}\!+\!\sum\nolimits_{m \in\mathcal{M}_{\mathtt{c}} }\sum\nolimits_{\tilde{m} \ne m,\tilde{m} \in\mathcal{M}_{\mathtt{c}}}\!\!\eta _{m,{k'}}\eta _{\tilde{m},1}\rho_{\mathtt{c}}^2 N^2\notag\\
    &\quad\times N_{\mathtt{pm}}\gamma_{m,k'}\gamma_{\tilde{m},1}\beta_{m,\mathtt{pm}}\beta_{\tilde{m},\mathtt{pm}},\notag\\
    &=\!\sum\nolimits_{m \in\mathcal{M}_{\mathtt{c}} }\eta _{m,{k'}}\rho _{\mathtt{c}}^2N_{\mathtt{pm}}\!N\gamma_{m,k'}\beta_{m,\mathtt{pm}}\bigg [\eta _{m,1}\beta_{m,\mathtt{pm}}\gamma_{m,1} \notag\\
    &\quad\times(N_{\mathtt{pm}}\!+\!N)\!+\!\sum\nolimits_{\tilde{m} \ne m,\tilde{m} \in\mathcal{M}_{\mathtt{c}}}\!\eta _{\tilde{m},1}N\gamma_{\tilde{m},1}\beta_{\tilde{m},\mathtt{pm}} \bigg ],
\end{align}
where $\overset{(c)}= $ is obtained by first expanding $\mathbb{E}\big \{ \big \|\mathbf{G}_{m,\mathtt{pm}}^{\ast }\big \|^4 \big \}$ as:
$$\mathbb{E}\big \{ \big \|\mathbf{G}_{m,\mathtt{pm}}^{\ast }\big \|^4 \big \}=\mathbb{E}\big \{ \mathbf{G}_{m,\mathtt{pm}}^{\ast }\mathbf{G}_{m,\mathtt{pm}}^{T} \mathbf{G}_{m,\mathtt{pm}}^{\ast } \mathbf{G}_{m,\mathtt{pm}}^{T}\big \},$$ then, using \cite[Lemma 10] {ref:Zhikangda_TIT}, to obtain 
\begin{align}
    &\mathbb{E}\big \{ \mathbf{G}_{m,\mathtt{pm}}^{\ast }\mathbf{G}_{m,\mathtt{pm}}^{T} \mathbf{I}_{N}\mathbf{G}_{m,\mathtt{pm}}^{\ast } \mathbf{G}_{m,\mathtt{pm}}^{T}\big \}
    \nonumber\\
    &=\beta_{m,\mathtt{pm}}^2\big(N_{\mathtt{pm}}^2\!+\!N_{\mathtt{pm}}\mathrm {Tr}(\mathbf{I}_{N})\big) \mathbf{I}_{N}=\beta_{m,\mathtt{pm}}^2N_{\mathtt{pm}}(N\!+\!N_{\mathtt{pm}})\mathbf{I}_{N}.
\end{align}

4) Compute $\mathbb{E}\left \{ |\mathrm{IS}_{\mathtt{pm}}|^{2} \right \}$: 
\begin{align}\label{eq:IS_j}
        &\!\!\mathbb{E}\!\left \{ |\mathrm{IS}_{\mathtt{pm}}|^{2} \right \}\!\!=\!\! \mathbb{E}\!\Big \{ \!\Big |\mathbf{w}_{\mathtt{comb},\mathtt{pm}}^T\!\sum\nolimits_{m'\in\mathcal{M}_{\mathtt{s},\mathtt{t}} }\!\!\sqrt{\eta _{m'\!,\mathtt{t}}\rho _{\mathtt{s}}}\mathbf{G}_{m',\mathtt{pm}}^{T}\mathbf{h}_{m'\!,\mathtt{t}}^{\mathtt{\ast}} \!\Big |^2 \Big \}\notag\\
        &+\mathbb{E}\Big \{ \!\Big |\mathbf{w}_{\mathtt{comb},\mathtt{pm}}^T\sum\nolimits_{m'\in\mathcal{M}_{\mathtt{s},\mathtt{t}} } \sqrt{\eta _{m'\!,\mathtt{t}}\rho _{\mathtt{s}}}\sqrt{\alpha}\mathbf{h}_{\mathtt{pm},\mathtt{t}}\mathbf{h}_{m'\!,\mathtt{t}}^{T}\mathbf{h}_{m'\!,\mathtt{t}}^{\mathtt{\ast}}  \Big |^2 \Big \} \notag\\
       &\overset{(d)}{=}\!\sum\nolimits_{m\in\mathcal{M}_{\mathtt{c}} }\!\sum\nolimits_{m'\in\mathcal{M}_{\mathtt{s},\mathtt{t}} } \eta _{m'\!,\mathtt{t}}\rho _{\mathtt{s}}\mathbb{E}\big \{\hat{\mathbf{g}}_{m,1}^{T}\mathbf{G}_{m,\mathtt{pm}}^{\ast}\mathbb{E}\big \{|\mathbf{G}_{m',\mathtt{pm}}^{T}\notag\\
       &\hspace{0.4cm}\times\mathbf{h}_{m'\!,\mathtt{t}}^{\ast}|^2\big \}\mathbf{G}_{m,\mathtt{pm}}^{T}\hat{\mathbf{g}}_{m,1}^{\ast}\big \}+ \sum\nolimits_{m \in\mathcal{M}_{\mathtt{c}} }\sum\nolimits_{m'\in\mathcal{M}_{\mathtt{s},\mathtt{t}} }\eta _{m'\!,\mathtt{t}}\rho _{\mathtt{s}} \notag\\
       &\hspace{0.4cm}\times\!\eta _{m,1}\rho _{\mathtt{c}}N^2\zeta_{m'\!,\mathtt{t}}^2\alpha\mathbb{E}\!\left \{ \!\left |\! (\!\mathbf{G}_{m,\mathtt{pm}}^{T}\hat{\mathbf{g} }_{m,1} ^{\ast }\!) ^{H }\mathbf{h}_{\mathtt{pm},\mathtt{t}} \!\right |^2 \!\right \}\!\!+\!\!\sum\nolimits_{m \in\mathcal{M}_{\mathtt{c}} }\notag\\
       &\hspace{0.4cm}\sum\nolimits_{m'\!\in\mathcal{M}_{\mathtt{s},\mathtt{t}} }\!\eta _{m,1}\rho _{\mathtt{c}}\rho _{\mathtt{s}}N^2\alpha\zeta_{\mathtt{pm},\mathtt{t}}\sqrt{\eta _{m'\!,\mathtt{t}}}\zeta_{m'\!,\mathtt{t}}\!\notag\\
       &\hspace{0.4cm}\sum\nolimits_{\tilde{m}'\in\mathcal{M}_{\mathtt{s},\mathtt{t}}, \tilde{m}' \ne m' }\sqrt{\eta _{\tilde{m}'\!,t}} \zeta_{\tilde{m}',t}\mathbb{E}\!\Big \{  |\mathbf{G}_{m,\mathtt{pm}}^{T}\hat{\mathbf{g} }_{m,1} ^{\ast }|^2 \Big \}.
\end{align}
We first note that the second term on the left-hand side of $(d)$ follows from the property $\mathbf{h}_{\mathtt{pm},\mathtt{t}}^{T}\mathbf{h}_{\mathtt{pm},\mathtt{t}}^{\ast}=N_\mathtt{pm}\zeta_{\mathtt{pm},\mathtt{t}}$. Then, for computing $\mathbb{E}\Big \{ \Big |\sum\nolimits_{m'\in\mathcal{M}_{\mathtt{s},\mathtt{t}} }\! \sqrt{\eta _{m'\!,\mathtt{t}}}\mathbf{h}_{m'\!,\mathtt{t}}^{T}\mathbf{h}_{m'\!,\mathtt{t}}^{\mathtt{\ast}}\Big |^2 \Big \}$, we consider the square of the summation inside the expectation, which can be expanded into two distinct cases based on $m'$, the diagonal terms of the expansion $\mathbb{E}\Big \{\sum\nolimits_{m'\in\mathcal{M}_{\mathtt{s},\mathtt{t}} }\!\eta _{m'\!,\mathtt{t}}\mathbf{h}_{m'\!,\mathtt{t}}^{T}\mathbf{h}_{m'\!,\mathtt{t}}^{\mathtt{\ast}}\mathbf{h}_{m'\!,\mathtt{t}}^{T}\mathbf{h}_{m'\!,\mathtt{t}}^{\mathtt{\ast}} \Big \}$ and the off-diagonal terms of the expansion $\mathbb{E}\Big \{\sum\nolimits_{m'\in\mathcal{M}_{\mathtt{s},\mathtt{t}} }\!\sqrt{\eta _{m'\!,\mathtt{t}}}N^2\zeta_{m'\!,\mathtt{t}}\sum\nolimits_{\tilde{m}'\in\mathcal{M}_{\mathtt{s},\mathtt{t}}, \tilde{m}' \ne m' }\!\sqrt{\eta _{\tilde{m}'\!,t}}\zeta_{\tilde{m}',t} \Big \}$ for the deterministic vector $\mathbf{h}_{m'\!,\mathtt{t}}$, the result can be written as  $N^2\sum\nolimits_{m'\in\mathcal{M}_{\mathtt{s},\mathtt{t}} }\Big(\eta _{m'\!,\mathtt{t}}\zeta_{m'\!,\mathtt{t}}^2+\sum\nolimits_{\tilde{m}'\in\mathcal{M}_{\mathtt{s},\mathtt{t}}, \tilde{m}' \ne m' }\sqrt{\eta _{m'\!,\mathtt{t}}\eta _{\tilde{m}'\!,t}}\zeta_{m'\!,\mathtt{t}}\zeta_{\tilde{m}',t}\Big)$. Finally, we can obtain
\begin{align}\label{eq:IS_j2}
 &\mathbb{E}\!\left \{ |\mathrm{IS}_{\mathtt{pm}}|^{2} \right \}\nonumber\\
       &=\sum\nolimits_{m\in\mathcal{M}_{\mathtt{c}} }\sum\nolimits_{m'\in\mathcal{M}_{\mathtt{s},\mathtt{t}} } \eta _{m'\!,\mathtt{t}}\rho _{\mathtt{s}}\eta _{m,1}\rho _{\mathtt{c}}N\beta_{m',\mathtt{pm}}\zeta_{m'\!,\mathtt{t}}\notag\\
       &\hspace{0.4cm}\times\mathbb{E}\left \{\hat{\mathbf{g}}_{m,1}^{T}\mathbf{G}_{m,\mathtt{pm}}^{\ast}\mathbf{G}_{m,\mathtt{pm}}^{T}\hat{\mathbf{g}}_{m,1}^{\ast}\right \}\!+\!\sum\nolimits_{m \!\in\mathcal{M}_{\mathtt{c}} }\!\sum\nolimits_{m'\in\mathcal{M}_{\mathtt{s},\mathtt{t}} }\!\!\alpha \notag\\
       &\hspace{0.4cm}\times\eta _{m'\!,\mathtt{t}}\rho _{\mathtt{s}}\eta _{m,1}\rho _{\mathtt{c}}N^2N_{\mathtt{pm}}\zeta_{\mathtt{pm},\mathtt{t}}\zeta_{m'\!,\mathtt{t}}^2\beta_{m,\mathtt{pm}}\mathbb{E}\left \{ \left | \hat{\mathbf{g} }_{m,1} ^{T}\! \right |^2 \right \}\notag\\
       &\hspace{0.4cm}+\sum\nolimits_{m \in\mathcal{M}_{\mathtt{c}} }\sum\nolimits_{m'\in\mathcal{M}_{\mathtt{s},\mathtt{t}} }\!\alpha\sqrt{\eta _{m'\!,\mathtt{t}}}\rho _{\mathtt{s}} \eta _{m,1}\rho _{\mathtt{c}}N^3N_{\mathtt{pm}}\zeta_{\mathtt{pm},\mathtt{t}}\notag\\   &\hspace{0.4cm}\times\zeta_{m',\mathtt{t}}\beta_{m,\mathtt{pm}}\gamma_{m,1}\sum\nolimits_{\tilde{m}'\in\mathcal{M}_{\mathtt{s},\mathtt{t}}, \tilde{m}' \ne m' }\sqrt{\eta _{\tilde{m}'\!,t}}\zeta_{\tilde{m}',t}\notag\\
       &=\!\sum\nolimits_{m\in\mathcal{M}_{\mathtt{c}} }\!\sum\nolimits_{m'\in\mathcal{M}_{\mathtt{s},\mathtt{t}} } \!\!\sqrt{\eta _{m'\!,\mathtt{t}}} \eta _{m,1}\rho _{\mathtt{s}}\rho _{\mathtt{c}}\beta_{m,\mathtt{pm}}\gamma _{m,1}\zeta_{m'\!,\mathtt{t}} N_\mathtt{pm}\notag\\
        &\hspace{0.4cm}\times \!N^2\bigg ( \sqrt{\eta _{m'\!,\mathtt{t}}} N\zeta_{\mathtt{pm},\mathtt{t}}\zeta_{m'\!,\mathtt{t}}\alpha+\sum\nolimits_{\tilde{m}'\in\mathcal{M}_{\mathtt{s},\mathtt{t}}, \tilde{m}' \ne m' }\sqrt{\eta _{\tilde{m}'\!,t}}\notag\\
       &\hspace{0.4cm}\times N\zeta_{\mathtt{pm},\mathtt{t}}\zeta_{\tilde{m}',t}\alpha+\sqrt{\eta _{m'\!,\mathtt{t}}} \beta_{m',\mathtt{pm}} \bigg ).
\end{align}

5) Compute $\mathbb{E}\left \{ |\mathrm{JS}_{\mathtt{s},\mathtt{pm}}|^{2} \right \} $:
\begin{align}
   &\mathbb{E}\left \{ |\mathrm{JS}_{\mathtt{s},\mathtt{pm}}|^{2} \right \}=\mathbb{E}\left \{ \left |\mathbf{w}_{\mathtt{comb},\mathtt{pm}}^T\sqrt{\eta _{\mathtt{pm},\mathtt{t}}\rho _{\mathtt{pm}}}\mathbf{G}_{\mathtt{pm},\mathtt{pm}}^{T}\mathbf {h}_{\mathtt{pm},\mathtt{t}}^{\ast}\right |^2 \right \}\notag\\
   &\hspace{0.2cm}+\mathbb{E}\left \{ \left |\mathbf{w}_{\mathtt{comb},\mathtt{pm}}^T\sqrt{\eta _{\mathtt{pm},\mathtt{t}}\rho _{\mathtt{pm}}}\sqrt{\alpha}\mathbf{h}_{\mathtt{pm},\mathtt{t}}\mathbf{h}_{\mathtt{pm},\mathtt{t}}^{T}\mathbf {h}_{\mathtt{pm},\mathtt{t}}^{\ast}\right |^2 \right \}\notag\\
   &=\!\sum\nolimits_{m \in\mathcal{M}_{\mathtt{c}} }\!\eta _{m,1}\rho _{\mathtt{c}}\eta _{\mathtt{pm},\mathtt{t}}\rho _{\mathtt{pm}}N_{\mathtt{pm}}\beta_{\mathtt{pm},\mathtt{pm}}\zeta_{\mathtt{pm},\mathtt{t}}\mathbb{E}\left \{ \left |\hat{\mathbf{g} }_{m,1}^{T}\mathbf{G}_{m,\mathtt{pm}}^{\ast }\right |^2 \right \}\notag\\
   &\hspace{0.2cm}+\!\sum\nolimits_{m \in\mathcal{M}_{\mathtt{c}} }\!\!\eta _{m,1}\rho _{\mathtt{c}}\eta _{\mathtt{pm},\mathtt{t}}\rho _{\mathtt{pm}}\alpha N_{\mathtt{pm}}^2\zeta_{\mathtt{pm},\mathtt{t}}^2\mathbb{E}\!\left \{ \left |\hat{\mathbf{g} }_{m,1}^{T}\mathbf{G}_{m,\mathtt{pm}}^{*}\mathbf{h}_{\mathtt{pm},\mathtt{t}}\!\right |^2 \right \}\notag\\
   &\overset{(e)}{=}\sum\nolimits_{m\in\mathcal{M}_{\mathtt{c}} }\eta _{\mathtt{pm},\mathtt{t}}\rho _{\mathtt{pm}}\eta _{m,1}\rho _{\mathtt{c}}\beta_{\mathtt{pm},\mathtt{pm}}\zeta_{\mathtt{pm},\mathtt{t}}\beta_{m,\mathtt{pm}}N_{\mathtt{pm}}^2N\gamma_{m,1}\notag\\
   &\hspace{0.2cm}+\sum\nolimits_{m\in\mathcal{M}_{\mathtt{c}} }\eta _{\mathtt{pm},\mathtt{t}}\rho _{\mathtt{pm}}\eta _{m,1}\rho _{\mathtt{c}}\alpha\beta_{m,\mathtt{pm}}N_{\mathtt{pm}}^3N\gamma_{m,1}\zeta_{\mathtt{pm},\mathtt{t}}^3,\notag
\end{align}
\begin{comment}
\begin{align}
   &=\sum\nolimits_{m\in\mathcal{M}_{\mathtt{s}} }\eta _{\mathtt{pm},\mathtt{t}}\rho _{\mathtt{pm}}\eta _{m,1}\rho _{\mathtt{c}}\zeta_{\mathtt{pm},\mathtt{t}}\beta_{m,\mathtt{pm}}N_{\mathtt{pm}}^2N\gamma_{m,1}\notag\\
    &\hspace{0.4cm}\times\big ( \beta_{\mathtt{pm},\mathtt{pm}}+\alpha N_{\mathtt{pm}}\zeta_{\mathtt{pm},\mathtt{t}}^2 \big ),
\end{align}
\end{comment}
where $(e)$ follows \cite[Lemma 1] {ref:Zeping_cellfree}. More specifically,  $\mathbb{E}\!\Big \{ \Big |\hat{\mathbf{g} }_{m,1}^{T}\mathbf{G}_{m,\mathtt{pm}}^{*}\mathbf{h}_{\mathtt{pm},\mathtt{t}}\!\Big |^2 \Big \}=\mathbb{E}\!\Big \{ \hat{\mathbf{g} }_{m,1}^{T}\mathbb{E}\!\Big \{\mathbf{G}_{m,\mathtt{pm}}^{*}\mathbf{h}_{\mathtt{pm},\mathtt{t}}\mathbf{h}_{\mathtt{pm},\mathtt{t}}^{H}\times\\\mathbf{G}_{m,\mathtt{pm}}^{T}\Big \}\hat{\mathbf{g} }_{m,1}^{*}\Big \}$. By using \cite[Lemma 1] {ref:Zeping_cellfree} we can calculate
\begin{align}
    \mathbb{E}\!\left \{ \mathbf{G}_{m,\mathtt{pm}}^{*}\mathbf{h}_{\mathtt{pm},\mathtt{t}}\mathbf{h}_{\mathtt{pm},\mathtt{t}}^{H} \mathbf{G}_{m,\mathtt{pm}}^{T}\right \}&=\beta_{m,\mathtt{pm}}\mathrm{Tr}(\mathbf{h}_{\mathtt{pm},\mathtt{t}}\mathbf{h}_{\mathtt{pm},\mathtt{t}}^{H})\mathbf{I}_{N}\notag\\
    &=N_{\mathtt{pm}}\beta_{m,\mathtt{pm}}\zeta_{\mathtt{pm},\mathtt{t}}\mathbf{I}_{N}.
\end{align}

6) Compute $\mathbb{E}\left \{ |\mathrm{JS}_{\mathtt{c},\mathtt{pm}}|^{2} \right \} $:
\begin{align}
    &\mathbb{E}\left \{\! |\mathrm{JS}_{\mathtt{c},\mathtt{pm}}|\!^{2} \right \}=\mathbb{E}\left \{ \left |\mathbf{w}_{\mathtt{comb},\mathtt{pm}}^T\sqrt{\eta _{\mathtt{pm},1}\rho _{\mathtt{pm}}}\mathbf{G}_{\mathtt{pm},\mathtt{pm}}^{T}\mathbf{g}_{\mathtt{pm},1} ^{\ast }\right |^2 \right \} \notag \\
    &\hspace{0.2cm}+\mathbb{E}\left \{ \left |\mathbf{w}_{\mathtt{comb},\mathtt{pm}}^T\sqrt{\eta _{\mathtt{pm},\mathtt{t}}\rho _{\mathtt{pm}}}\sqrt{\alpha}\mathbf{h}_{\mathtt{pm},\mathtt{t}}\mathbf{h}_{\mathtt{pm},\mathtt{t}}^{T}\mathbf {g}_{\mathtt{pm},1}^{\ast}\right |^2 \right \}\notag\\
    &\!\!=\!\sum\nolimits_{m \in\mathcal{M}_{\mathtt{c}} }\!\!\eta _{\mathtt{pm},1}\eta _{m,1}\rho_{\mathtt{pm}}\rho _{\mathtt{c}}N_{\mathtt{pm}}\beta_{\mathtt{pm},\mathtt{pm}}\beta_{\mathtt{pm},1}\mathbb{E}\left \{ \!\left |\hat{\mathbf{g}}_{m,1}^{T}\mathbf{G}_{m,\mathtt{pm}}^{\ast}\right |^2 \!\right \}\notag\\
    &\hspace{0.2cm}+\!\sum\nolimits_{m \in\mathcal{M}_{\mathtt{c}} }\!\!\eta _{\mathtt{pm},1}\eta _{m,1}\rho_{\mathtt{pm}}\rho _{\mathtt{c}}\alpha N_{\mathtt{pm}}\beta_{\mathtt{pm},1}\zeta_{\mathtt{pm},\mathtt{t}}^2\mathbb{E}\left \{ \!\left |\hat{\mathbf{g}}_{m,1}^{T}\mathbf{G}_{m,\mathtt{pm}}^{\ast}\right |^2 \!\right \}\notag\\
    &=\sum\nolimits_{m \in\mathcal{M}_{\mathtt{c}} }\eta _{\mathtt{pm},1}\eta _{m,1}\rho _{\mathtt{c}}\rho _{\mathtt{pm}}\gamma_{m,1}N\beta_{\mathtt{pm},\mathtt{pm}}N_{\mathtt{pm}}^2\beta_{m,\mathtt{pm}}\beta_{\mathtt{pm},1}\notag\\
    &\hspace{0.2cm}+\!\sum\nolimits_{m \in\mathcal{M}_{\mathtt{c}} }\!\!\eta _{\mathtt{pm},1}\eta _{m,1}\rho_{\mathtt{pm}}\rho _{\mathtt{c}}\alpha N_{\mathtt{pm}}^2\beta_{\mathtt{pm},1}\zeta_{\mathtt{pm},\mathtt{t}}^2\beta_{m,\mathtt{pm}}N\gamma_{m,1}\notag\\
    &=\!\sum\nolimits_{m \in\mathcal{M}_{\mathtt{c}} }\eta _{\mathtt{pm},1}\eta _{m,1}\rho _{\mathtt{c}}\rho _{\mathtt{pm}}\gamma_{m,1}NN_{\mathtt{pm}}^2\beta_{m,\mathtt{pm}}\beta_{\mathtt{pm},1}\notag\\
    &\hspace{0.2cm}\times(\beta_{\mathtt{pm},\mathtt{pm}}+\alpha\zeta_{\mathtt{pm},\mathtt{t}}^2).
\end{align}

7) Compute $\mathbb{E}\left \{ |\mathrm{n}_{\mathtt{pm}}|^{2} \right \}$:
\begin{align}\label{eq:n_j}
        &\mathbb{E}\left \{ |\mathrm{n}_{\mathtt{pm}}|^{2} \right \}=\sum\nolimits_{m \in\mathcal{M}_{\mathtt{c}} } \eta _{m,1}\rho _{\mathtt{c}}\mathbb{E}\left \{ \hat{\mathbf{g}}_{m,1}^{T}\mathbf{G}_{m,\mathtt{pm}}^{\ast}    \mathbf{G}_{m,\mathtt{pm}}^{T} \hat{\mathbf{g}}_{m,1}^{\ast} \right \} \notag \\
        &\qquad\qquad~=\sum\nolimits_{m\in\mathcal{M}_{\mathtt{c}} }\eta _{m,1}\rho _{\mathtt{c}}NN_{\mathtt{pm}}\beta_{m,\mathtt{pm}}\gamma_{m,1}.    
\end{align}
\ifCLASSOPTIONcaptionsoff
  \newpage
\fi



%\bibliographystyle{IEEEtran}

\bibliographystyle{IEEEtran}
\bibliography{IEEEabrv}

\begin{IEEEbiography}[{\includegraphics[width=1in,height=1.25in,clip,keepaspectratio]{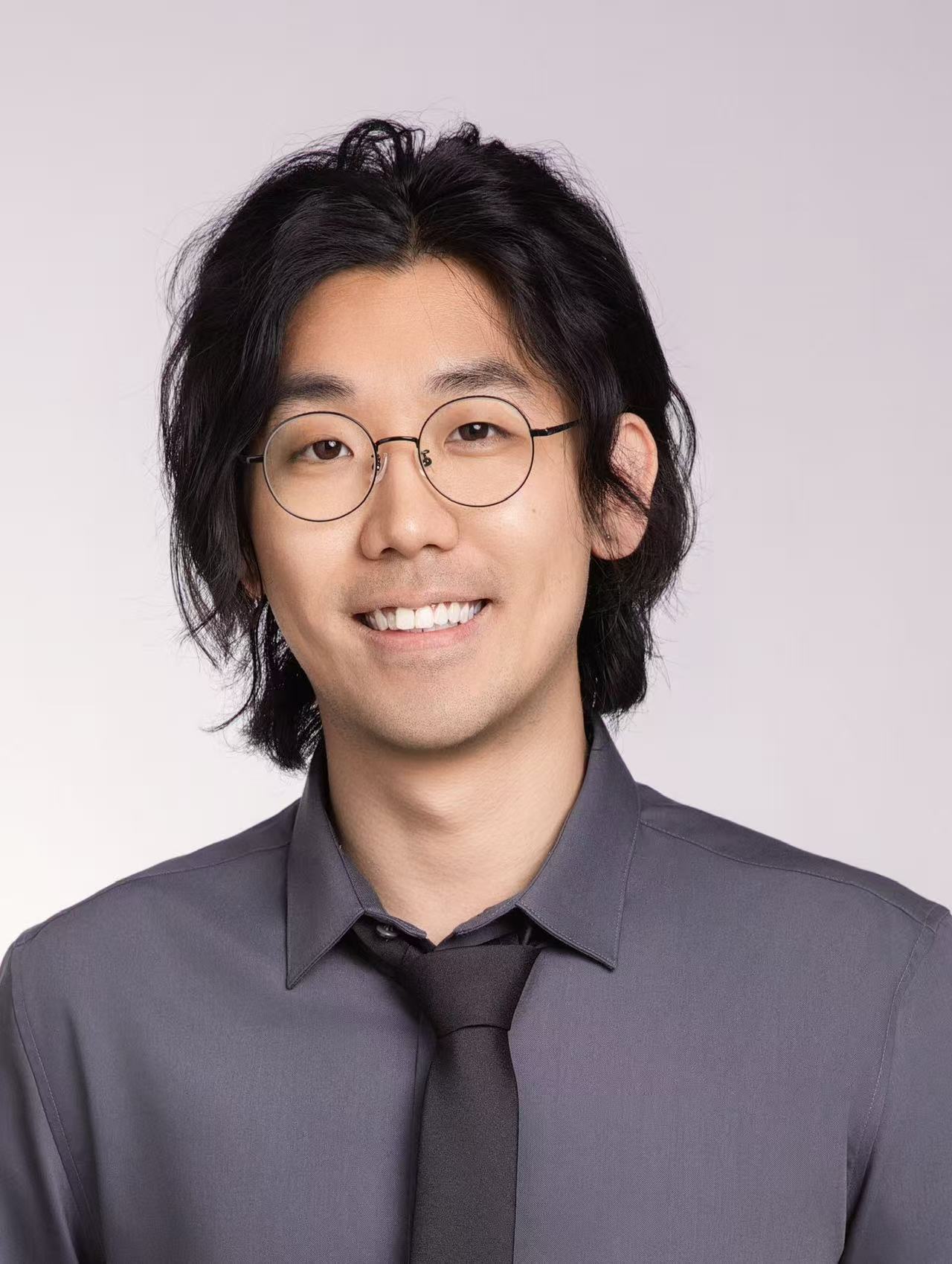}}]{Zonghan Wang} (Student Member, IEEE) received the B.S. degree in electronic and information engineering from Nanjing University of Science and Technology, China, and the M.S. degree in communication systems from Lund University, Sweden, in 2021 and 2023, respectively. He is currently pursuing the Ph.D. degree with the Centre for Wireless Innovation (CWI), Queen’s University, Belfast, U.K. His main research interests include massive MIMO systems, integrated sensing and communications, and physical-layer security.
\end{IEEEbiography}

\begin{IEEEbiography}[{\includegraphics[width=1in,height=1.25in,clip,keepaspectratio]{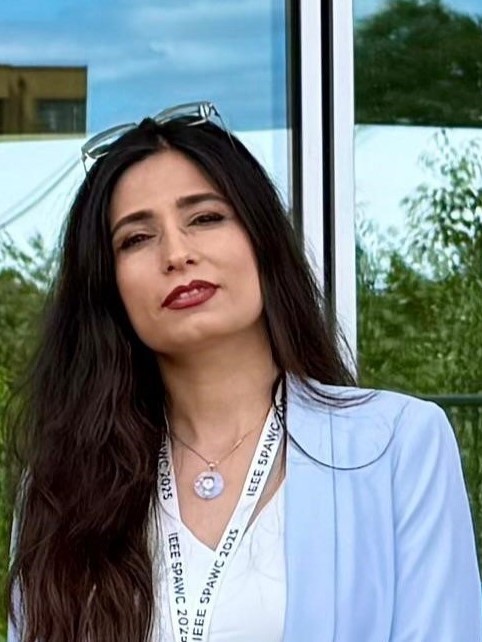}}]{Zahra Mobini} (Senior Member, IEEE) is currently an Assistant Professor in Communications Engineering and Signal Processing at the University of Manchester, U.K. She was a Post-Doctoral Research Fellow with Queen’s University Belfast, U.K., from  2021 to  2025.  From November 2010 to November 2011, she was a Visiting Researcher at the School of Engineering, Australian National University (ANU), Canberra, Australia. She received her Ph.D. degree from K. N. Toosi University of Technology, Tehran, Iran.
Her research interests include physical-layer security, cell-free massive MIMO,     integrated sensing and communications (ISAC), reconfigurable intelligent surfaces, and resource management and optimization. She has co-authored numerous research papers in wireless communications. She received a Commendation in the 2025 Postdoc Support Award from Queen’s University Belfast. She serves as the Editor for the IEEE Transactions on Communications, and the Physical Communication (Elsevier), and has served as  a Technical Program Committee member for prominent IEEE conferences such as ICC, GLOBECOM, and VTC.
\end{IEEEbiography}

\begin{IEEEbiography}[{\includegraphics[width=1in,height=1.25in,clip,keepaspectratio]{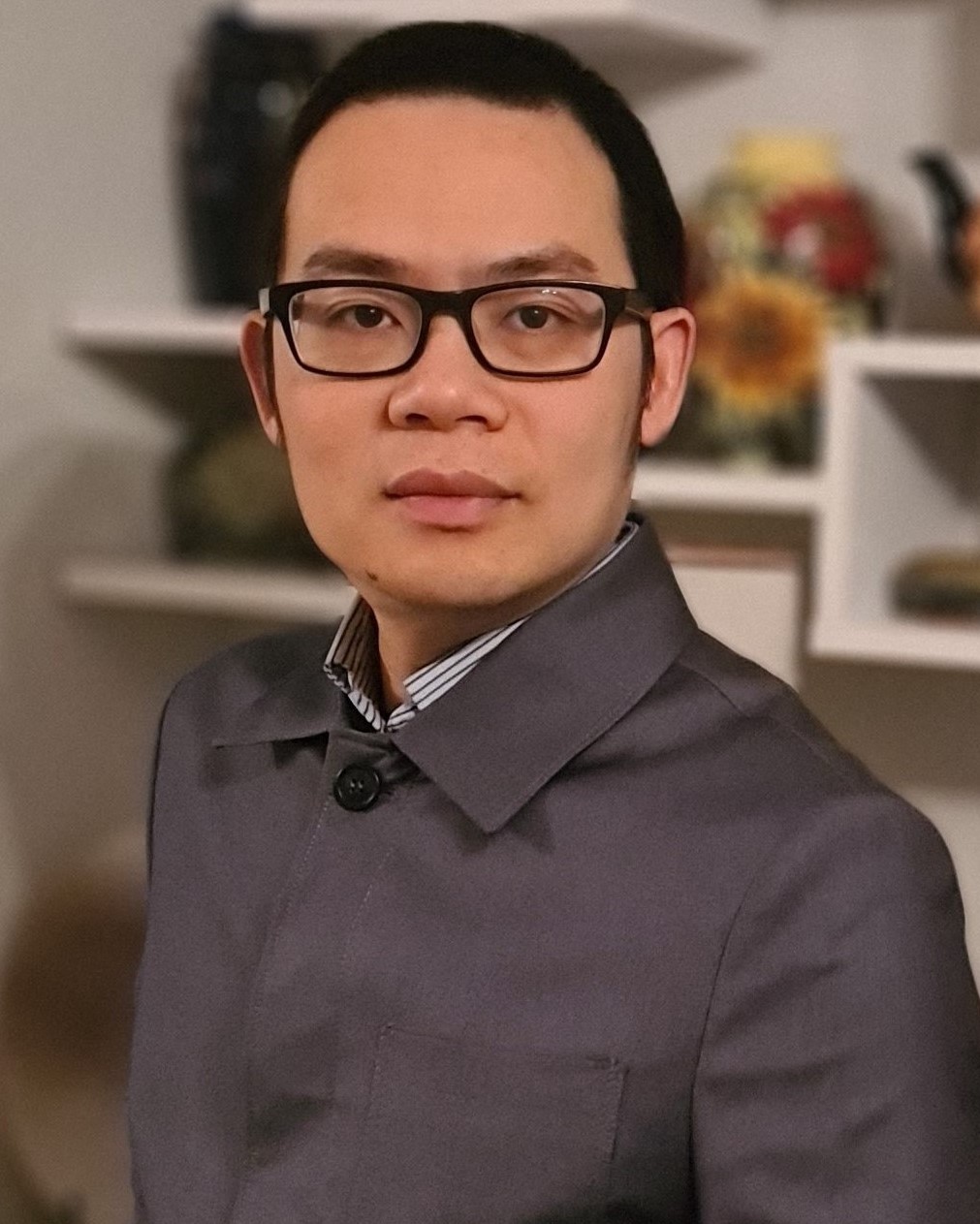}}]
{Hien Quoc Ngo} (Fellow, IEEE)  is currently a Professor with Queen's University Belfast, U.K. He is also an Eminent Scholar with Kyung Hee University, Republic of Korea. His main research interests include cellular/cell-free massive MIMO systems, integrated sensing and communications, reconfigurable intelligent surfaces, and physical layer security. He has co-authored many research papers in wireless communications and co-authored the Cambridge University Press textbook \emph{Fundamentals of Massive MIMO} (2016).

He received the IEEE ComSoc Stephen O. Rice Prize in 2015, the IEEE ComSoc Leonard G. Abraham Prize in 2017, the Best Ph.D. Award from EURASIP in 2018, and the IEEE CTTC Early Achievement Award in 2023. He also received the IEEE Sweden VT-COM-IT Joint Chapter Best Student Journal Paper Award in 2015. He was awarded the UKRI Future Leaders Fellowship in 2019. He serves as the Editor for the IEEE Transactions on Wireless Communications, and the IEEE Transactions on Communications. He was an Editor of the Digital Signal Processing, the Physical Communication (Elsevier), the IEEE Wireless Communications Letters, a Guest Editor of IET Communications, and a Guest Editor of IEEE ACCESS in 2017.
\end{IEEEbiography}

\begin{IEEEbiography}[
{\includegraphics[width=1in,height=1.25in,clip,keepaspectratio]{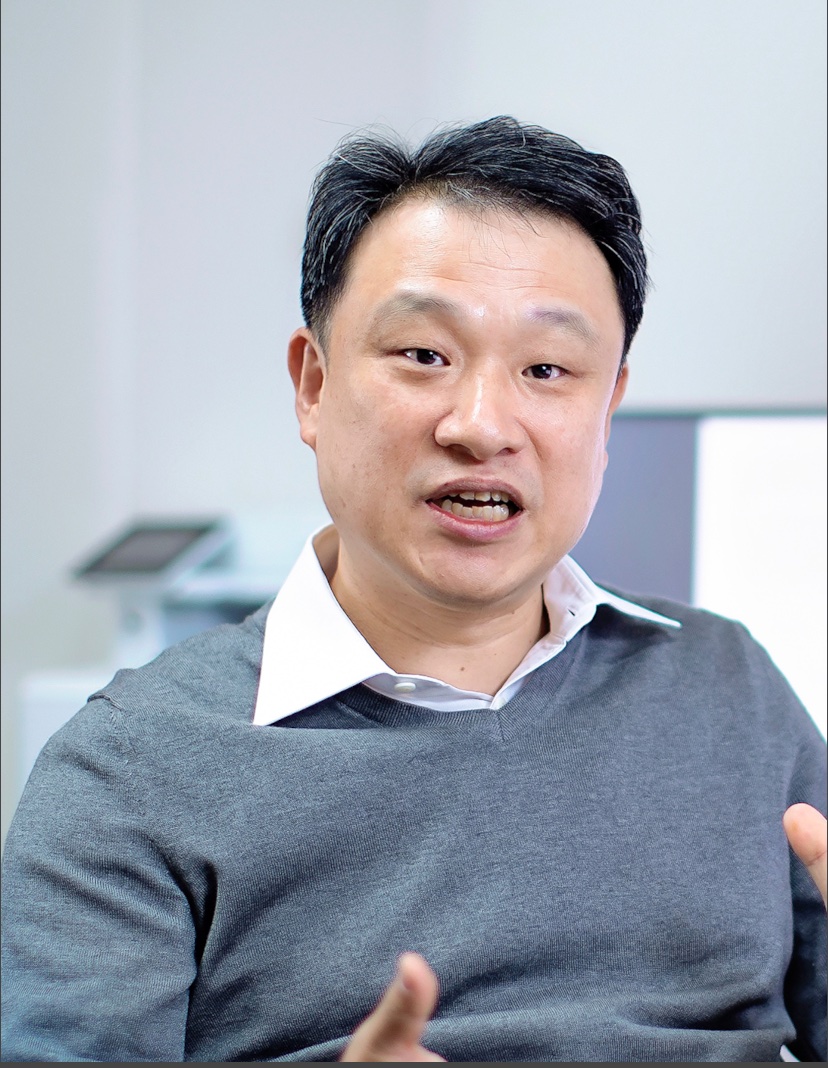}}
]{Hyundong Shin} (Fellow, IEEE)  
received the B.S. degree in Electronics Engineering from Kyung Hee University (KHU), Yongin-si, Korea, in 1999, and the M.S. and Ph.D. degrees in Electrical Engineering from Seoul National University, Seoul, Korea, in 2001 and 2004, respectively.
During his postdoctoral research at the Massachusetts Institute of Technology (MIT) from 2004 to 2006, he was with the Laboratory for Information Decision Systems (LIDS). In 2006, he joined the KHU, where he is currently a Professor in the Department of Electronic Engineering. His research interests include quantum information science, wireless communication, and machine intelligence.
Dr. Shin received the IEEE Communications Society’s Guglielmo Marconi Prize Paper Award and William R. Bennett Prize Paper Award. He served as the Publicity Co-Chair for the IEEE PIMRC and the Technical Program Co-Chair for the IEEE WCNC and the IEEE GLOBECOM. He was an Editor of \textsc{IEEE Transactions on Wireless Communications} and \textsc{IEEE Communications Letters}.
\end{IEEEbiography}

\begin{IEEEbiography}[{\includegraphics[width=1in,height=1.55in,clip,keepaspectratio]{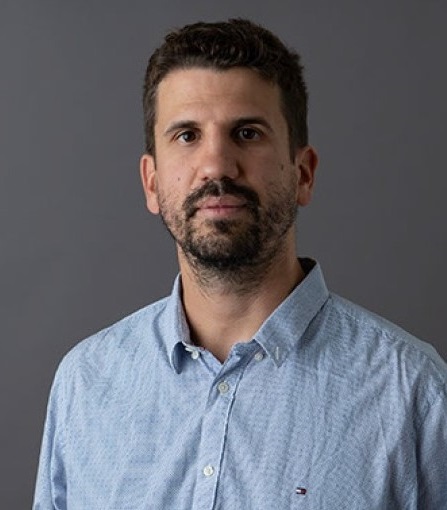}}]
{Michail Matthaiou}(Fellow, IEEE) obtained his Ph.D. degree from the University of Edinburgh, U.K. in 2008.
He is currently a Professor of Communications Engineering and Signal Processing and Deputy Director of the Centre for Wireless Innovation (CWI) at Queen’s University Belfast, U.K. He is also an Eminent Scholar at the Kyung Hee University, Republic of Korea. He has held research/faculty positions at Munich University of Technology (TUM), Germany and Chalmers University of Technology, Sweden. His research interests span signal processing for wireless communications, beyond massive MIMO, reflecting intelligent surfaces, mm-wave/THz systems and AI-empowered communications.

Dr. Matthaiou and his coauthors received the IEEE Communications Society (ComSoc) Leonard G. Abraham Prize in 2017. He currently holds the ERC Consolidator Grant BEATRICE (2021-2026) focused on the interface between information and electromagnetic theories. To date, he has received the prestigious 2023 Argo Network Innovation Award, the 2019 EURASIP Early Career Award and the 2018/2019 Royal Academy of Engineering/The Leverhulme Trust Senior Research Fellowship. His team was also the Grand Winner of the 2019 Mobile World Congress Challenge. He was the recipient of the 2011 IEEE ComSoc Best Young Researcher Award for the Europe, Middle East and Africa Region and a co-recipient of the 2006 IEEE Communications Chapter Project Prize for the best M.Sc. dissertation in the area of communications. He has co-authored papers that received best paper awards at the 2018 IEEE WCSP and 2014 IEEE ICC. In 2014, he received the Research Fund for International Young Scientists from the National Natural Science Foundation of China. He is currently the Editor-in-Chief of Elsevier Physical Communication, a Senior Editor for \textsc{IEEE Wireless Communications Letters} and \textsc{IEEE Signal Processing Magazine}, an Area Editor for \textsc{IEEE Transactions on Communications} and Editor-in-Large for \textsc{IEEE Open Journal of the Communications Society}. He is an IEEE and AAIA Fellow.
\end{IEEEbiography}

\begin{comment}
\begin{IEEEbiography}{Michael Shell}
Biography text here.
\end{IEEEbiography}

\begin{IEEEbiographynophoto}{John Doe}
Biography text here.
\end{IEEEbiographynophoto}

%\newpage

\begin{IEEEbiographynophoto}{Jane Doe}
Biography text here.
\end{IEEEbiographynophoto}


%\vfill

\end{comment}

\end{document}